\documentstyle[rmp,aps,amsfonts]{revtex}
\input epsfig.sty

\newcommand{\NP}{{\rm Nucl.\ Phys.\ }}

\newcommand{\PL}{{\rm Phys.\ Lett.\ }}

\newcommand{\PRP}{{\rm Phys.\ Rep.\ }}
\newcommand{\CMP}{{\rm Comm.\ Math.\ Phys.\ }}
\newcommand{\MPL}{{\rm Mod.\ Phys.\ Lett.\ }}
\newcommand{\PRL}{{\rm Phys.\ Rev.\ Lett.\ }}

\def\half{{1 \over 2}}

\newcommand{\be}{\begin{equation}}
\newcommand{\ee}{\end{equation}}
\newcommand{\bea}{\begin{eqnarray}}
\newcommand{\eea}{\end{eqnarray}}
\newcommand{\beas}{\begin{eqnarray*}}
\newcommand{\eeas}{\end{eqnarray*}}

\def\tr{{\rm Tr} \,}
\def\str{{\rm STr} \,}
\def\identity{{\rlap{1} \hskip 1.6pt \hbox{1}}}
\def\laplace{{\kern1pt\vbox{\hrule height 1.2pt\hbox{\vrule width 1.2pt\hskip
  3pt\vbox{\vskip 6pt}\hskip 3pt\vrule width 0.6pt}\hrule height 0.6pt}
  \kern1pt}}

\def\hatt{{\hat{\cal T}}}
\def\hatj{{\hat{\cal J}}}
\def\hatm{{\hat{\cal M}}}
\def\tilt{{\tilde{\cal T}}}
\def\tilj{{\tilde{\cal J}}}
\def\tilm{{\tilde{\cal M}}}
\def\itt{{\cal  T}}
\def\ijj{{\cal J}}
\def\imm{{\cal M}}

\def\dx{{\dot{X}}}

\def\bz{\Bbb Z}
\def\bc{\Bbb C}
\def\br{\Bbb R}

\begin{document}

\begin{flushright}
NSF-ITP-01-04\\
MIT-CTP-3078\\
hep-th/0101126
\end{flushright}
\vspace{0.9in}

\begin{center}

{\Large \bf M(atrix) Theory:\\
Matrix Quantum Mechanics as a
Fundamental Theory\\}

\end{center}

\vspace{7 mm}

\begin{center}

{\large
Washington Taylor}

\vspace{3mm}
{\small Permanent address:} \\
{\small \sl Center for Theoretical Physics} \\
{\small \sl MIT, Bldg. 6-306} \\
{\small \sl Cambridge, MA 02139, U.S.A.} \\
{\small \tt wati@mit.edu}\\

\vspace{3mm}
{\small Current address:} \\
{\small \sl Institute for Theoretical Physics} \\
{\small \sl University of California, Santa Barbara} \\
{\small \sl Santa Barbara, CA, U.S.A.} \\
{\small \tt wati@itp.ucsb.edu}

\vspace{8 mm}

\parbox{5in}{A self-contained review is given of the matrix model of
M-theory.  The 
introductory part of the review is intended to be accessible to the
general reader.  M-theory is an eleven-dimensional quantum theory of
gravity which is believed to underlie all superstring theories.  This
is the only candidate at present for a theory of fundamental physics
which reconciles gravity and quantum field theory in a potentially
realistic fashion.  Evidence for the existence of M-theory is still
only circumstantial---no complete background-independent formulation
of the theory yet exists.  Matrix theory was first developed as a
regularized theory of a supersymmetric quantum membrane.  More
recently, the theory appeared in a different guise as the discrete
light-cone quantization of M-theory in flat space.  These two
approaches to matrix theory are described in detail and compared.  It
is shown that matrix theory is a well-defined quantum theory which
reduces to a supersymmetric theory of gravity at low energies.
Although the fundamental degrees of freedom of matrix theory are
essentially pointlike, it is shown that higher-dimensional fluctuating
objects (branes) arise through the nonabelian structure of the matrix
degrees of freedom.  The problem of formulating matrix theory in a
general space-time background is discussed, and the connections
between matrix theory and other related models are reviewed.} 
\vspace{0.3in}

\noindent
To appear in Reviews of Modern Physics
\end{center}
\vspace{0.5in}
\begin{flushleft}
January 2001
\end{flushleft}

\newpage
\tableofcontents


\section{Introduction}
\label{sec:introduction}

In the last two decades, a remarkable structure has emerged as a
candidate for the fundamental theory of nature.  Until recently, this
structure was known primarily under the rubric ``string theory'', as
it was believed that the fundamental theory should be most effectively
described in terms of quantized fundamental stringlike degrees of
freedom.  Since 1995, however, several new developments have
drastically modified our perspective.  An increased understanding of
nonperturbative aspects of string theory has led to the realization
that all the known consistent string theories seem be special limiting
cases of a more fundamental underlying theory, which has been dubbed
``M-theory''.  While the consistent superstring theories give
microscopic models for quantum gravity in ten dimensions, M-theory
seems to be most naturally described in eleven dimensions.  We do not
yet have  a truly fundamental 
definition of M-theory.  It may be that in the most natural
formulation of the theory, the dimensionality of space-time emerges in
a smooth approximation to a non-geometrical mathematical system.

At the same time that string theory has been replaced by M-theory as
the most natural candidate for a fundamental description of the world,
the string itself has also lost its position as the main candidate for
a fundamental degree of freedom.  Both M-theory and string theory
contain dynamical objects of several different dimensionalities.  In
addition to one-dimensional string excitations (1-branes), string
theories contain pointlike objects (0-branes), membranes (2-branes),
three-dimensional extended objects (3-branes), and objects of all
dimensions up to eight or nine.  Eleven-dimensional M-theory, on the
other hand, seems to contain dynamical membranes and 5-branes.
Amongst all these degrees of freedom, there is no obvious reason 
why the ``string'' of string theory is any more fundamental
than, say, the pointlike or 3-brane excitations of string theory, or
the membrane of M-theory.  While the perturbative string expansion
makes sense in a regime of the theory where the string coupling is
small, there are also limits in which the theory is described by the
low-energy dynamics of a system of higher- or lower-dimensional
branes.  It seems that by considering the dynamics of any of these
sets of degrees of freedom, we can access at least some part of the
full physics of M-theory.

This review article concerns itself with a remarkably simple theory
which is believed to be equivalent to M-theory in a particular
reference frame.  The theory in question is a simple quantum mechanics
with matrix degrees of freedom.  The quantum mechanical degrees of
freedom are a finite set of bosonic $N \times N$ matrices and
fermionic partners, which combine to form a system with a high degree
of supersymmetry.  It is believed that this matrix quantum mechanics
theory provides a second-quantized description of M-theory
around a flat space-time background and in a light-front coordinate
system.  The finite integer $N$ serves as a regulator for the theory,
and the exact correspondence with M-theory in flat space-time emerges
only in the large $N$ limit.
Since this system has a finite number of degrees of freedom
for any value of $N$, it is manifestly a well-defined theory.  Since
it is a quantum mechanics theory rather than a quantum field theory,
it does not even exhibit the standard problems of renormalization and
other subtleties which afflict any but the simplest quantum field
theories.  

It may seem incredible that a simple matrix quantum mechanics model
can capture most of the physics of M-theory, and thus perhaps of the
real world.   This would imply that matrix theory provides a 
calculational framework in which, at least in principle, questions of
quantum effects in gravity and Planck scale corrections to the
standard model could be determined to an arbitrarily high degree of
accuracy by a large enough computer.  Unfortunately, however, although
it is only a quantum mechanics theory, matrix theory is a remarkably
tricky model in which to perform detailed calculations relevant to
understanding quantum corrections to general relativity, even at
very small values of $N$.  

Although it is technically difficult to study detailed aspects of
quantum gravity using the matrix theory approach, it is possible to
demonstrate analytically that classical 11-dimensional gravitational
interactions are produced by matrix quantum mechanics.  This has been
shown for all linearized gravitational interactions and a subset of
nonlinear interactions.  This is the
first time that it has been possible to explicitly show that a
well-defined microscopic quantum mechanical theory agrees with
classical gravity at long distances, including some nonlinear corrections
from general relativity.  Understanding the correspondence between
matrix quantum mechanics and classical supergravity in detail gives
some important new insights into the connections between quantum
mechanical systems with matrix degrees of freedom and gravity
theories.

One remarkable aspect of the matrix description of M-theory is the
fact that classical gravitational interactions are described in matrix
theory through quantum mechanical effects.  In classical matrix theory
separated objects experience no interactions.  Performing a one-loop
calculation in matrix quantum mechanics gives classical Newtonian
(linearized) gravitational interactions.  Higher-order general
relativistic corrections to the linearized gravity theory arise from
higher-loop calculations in matrix theory.  This connection between a
classical theory of gravity and a quantum system with matrix degrees
of freedom was the first example found of what now seems to be a very
general family of correspondences.  The celebrated AdS/CFT
correspondence, which relates classical ten-dimensional quantum
gravity on an anti-de Sitter background with a conformal quantum field
theory gives another wide class of examples of this type of
correspondence.  We discuss other examples of such connections in
the latter part of these notes.

Another remarkable aspect of matrix theory is the appearance of the
extended objects of M-theory (the supermembrane and M5-brane) in terms
of apparently pointlike fundamental degrees of freedom.  There is a
rich mathematical structure governing the way in which objects of
higher dimension can be encoded in
noncommuting matrices.  This structure may eventually lead us to
crucial new insights into the way in which all the many-dimensional
excitations of M-theory and string theory arise in terms of
fundamental degrees of freedom.

This review focuses primarily on some basic aspects of matrix theory:
the definitions of the theory through regularization of the
supermembrane and through light-front compactification of M-theory,
the appearance of classical supergravity interactions through quantum
effects in matrix theory, and the construction of the objects of
M-theory in terms of matrix degrees of freedom.  There are many other
interesting related directions in which progress has been made.
Reviews of matrix theory and related work which emphasize different
aspects of the subject are given in Bilal (1999), Banks (1998, 1999),
Bigatti and Susskind (1997), Taylor (1998, 2000), Nicolai and Helling
(1998), Obers and Pioline (1999), de Wit (1999), and Konechny and
Schwarz (2000).

In the remainder of this section, we give a brief overview of a number
of ideas which form the background for the discussion of matrix theory
and M-theory in the remainder of the review.  This section is intended
to be a useful introduction to these subjects for the non-specialist.
In subsection \ref{sec:ssm} we review some basic aspects of classical
supergravity theories and the appearance of strings and membranes in
these theories.  In subsection \ref{sec:dd} we discuss the two major
developments of the second superstring revolution: Duality and
D-branes.  We focus in particular on the duality relating M-theory to
a strongly coupled limit of string theory.  Subsection \ref{sec:mt}
gives a brief introduction to matrix theory in the context of the
developments summarized in \ref{sec:ssm}, \ref{sec:dd}.
The material in this subsection is essentially an overview of the
remainder of the review.

\subsection{Supergravity, strings, and membranes}
\label{sec:ssm}

The principal outstanding problem of theoretical physics at the close
of the 20th century is to find a theoretical framework which
combines the classical theory of general relativity at large distance
scales with the standard model of quantum particle physics at short
distance scales.  At the phenomenological and experimental level, the
next major challenge is to extend the standard model of particle
physics to describe physics at and above the TeV scale.  For both of
these endeavors, a potentially key structure is the idea of a
``supersymmetry'', which relates bosonic and fermionic fields through
a symmetry group with anticommuting (Grassmann)
generators $Q_{\alpha}$, where $\alpha$ is a spinor index.  For an
introduction to supersymmetry, see Wess and Bagger (1992).

In a supersymmetric theory in flat space, the anticommutator of a pair
of supersymmetry (SUSY) generators $Q_\alpha$ is a (linear combination
of)
translation generator(s):
$\{Q, Q\} \sim P_\mu$.  If supersymmetry plays any role in
describing physics in the real world, it must be necessary to
incorporate local supersymmetry into Einstein's theory of gravity.  The
supersymmetry generators cannot  simply describe a global symmetry of
the fundamental theory, since in general relativity the momentum
generator which appears as an anticommutator of two SUSY generators
becomes a local vector field generating a diffeomorphism of
space-time.  In a theory combining general relativity with
supersymmetry, supersymmetry generators  become spinor valued
fields on the space-time manifold.

It is possible to classify supersymmetric theories of gravity
(supergravity theories) by constructing supersymmetry algebras with
multiplets containing particles of spin 2 (gravitons).  In any
dimension greater than eleven, supersymmetry multiplets automatically
contain particles of spin higher than 2, so that the maximal dimension
for a supergravity theory is eleven.  Indeed, there is a unique such
classical theory in eleven dimensions with local supersymmetry
(Cremmer, Julia, and Scherk, 1978).  This theory has ${\cal N} = 1 $
supersymmetry, meaning that the supersymmetry generators live in a
single 32-component spinor representation of the 11D Lorentz group.
The generators $Q_\alpha$ extend the usual eleven-dimensional
Poincar\'e algebra into a super-Poincar\'e algebra.
Eleven-dimensional supergravity is in a natural sense the parent of
all other supergravity theories, since all supergravity theories in
lower dimensions can be derived from the eleven-dimensional theory by
compactifying some subset of the dimensions (or by considering a dual
limit of a compactification, as in the ten-dimensional type IIB
supergravity theory, which we will discuss momentarily).  We recall
here some basic features of eleven- and ten-dimensional supergravity
theories.  For more details the reader may consult Green, Schwarz, and
Witten (1987) or Townsend (1996b).

By examining the structure of the supersymmetry multiplet containing
the graviton, the set of classical fields which appear in any
supergravity theory may be determined.  In eleven-dimensional
supergravity, there are the following propagating fields\footnote{We
denote space-time indices in eleven dimensions by capital Roman
letters $I, J, K, \ldots \in \{0, 1, \ldots, 8, 9, 11\}$, and indices
in ten-dimensions by Greek letters $\mu, \nu, \ldots \in \{0, 1,
\ldots, 9\}$.}:
\vspace{0.03in}

\noindent
\hspace*{0.2in}
$e^a_I$: vielbein field (bosonic, with 44 components)\\
\noindent
\hspace*{0.2in}
$A_{I J K}$: 3-form potential (bosonic, with 84 components)\\
\noindent
\hspace*{0.2in}
$\psi_I$: Majorana fermion gravitino (fermionic, with 128
components).
\vspace{0.03in}

\noindent
The vielbein $e^a_I$ is an alternative description of the space-time
metric tensor $g_{IJ}$.  The 3-form field $A_{IJK}$ is antisymmetric
in its indices, and plays a role very similar to the vector potential
$A_{\mu}$ of classical electromagnetism.

In ten dimensions there are two supergravity theories with 32 SUSY
generators.  These are ${\cal N} = 2$ theories, since the
supersymmetry generators comprise two 16-component spinors.  In type
IIA supergravity these spinors have opposite chirality, while in type
IIB supergravity the spinors have the same chirality.  In addition to
the metric tensor/vielbein field, both type IIA and IIB supergravity
have several other propagating bosonic fields.  The IIA and IIB theory
both have a scalar field $\phi$ (the dilaton) and an antisymmetric
two-form field $B_{\mu \nu}$.  Each of the type II theories also has a
set of antisymmetric ``Ramond-Ramond'' $p$-form fields $C^{(p)}_{\mu_1
\cdots \mu_p}$.  For for the type IIA theory, $p \in\{1, 3\}$ is even,
and for the type IIB theory $p \in\{0, 2, 4\}$ is odd.

Like the 3-form field $A_{IJK}$ of 11D supergravity, the antisymmetric
2-form field $B_{\mu \nu}$ and the Ramond-Ramond $p$-form fields of
the type II supergravity theories are closely
analogous to the vector potential of electromagnetism.  In both the
type IIA and IIB supergravity theories there are classical stringlike extremal
black hole solutions of the field equations which are charged under
the 2-form field (Dabholkar, Gibbons, Harvey, and Ruiz Ruiz, 1990), as well as
higher-dimensional brane solutions which couple to the $p$-form fields
(for a review, see Duff, Khuri, and Lu, 1995).
The dynamics of these string- and brane-like solutions can be
described through an effective action living on the world-volume of  the
string or higher-dimensional brane.
Just as the
electromagnetic vector potential $A_{\mu}$ couples to an electrically
charged particle through a term of the form
\begin{equation}
\int_{\Lambda} A_{\mu} d X^\mu
\end{equation}
where $\Lambda$ is the trajectory of the particle, the 2-form field of
type II supergravity couples to the two-dimensional string world-sheet
through a term of the form
\begin{equation}
\int_\Sigma B_{\mu \nu} \epsilon^{ab} (\partial_a X^\mu) (\partial_b
X^\nu),
\label{eq:string-b}
\end{equation}
where $X^\mu$ are the embedding functions of the string world-sheet
$\Sigma$
in ten dimensions and $a, b \in\{0, 1\}$ are world-sheet indices.

The tension of the string is given by
$
T_s = 1/2 \pi  \alpha'
$, where $l_s = \sqrt{\alpha'}$ is the fundamental string length.  The
starting point for perturbative string theory is the quantization of
the world-sheet action on a  string, treating the
space-time coordinates $X^\mu$ as bosonic fields on the string
world-sheet.  The remarkable consequence of this quantization is that
quanta of all the fields in the supergravity multiplet
arise as massless excitations of the fundamental string.  It has been
shown that there are five consistent quantum superstring theories which
can be constructed by choosing different sets of fields on the string
world-sheet.  These are the type I, IIA, IIB, and heterotic $E_8 \times
E_8$ and $SO(32)$ theories.  In each of these cases, string theory
seems to give a consistent microscopic description of interactions
between gravitational quanta.  Besides the massless fields, there is
also an infinite tower of fields in each theory with masses on the
order of $1/l_s$.  In principle, any scattering process involving a
finite number of massless supergravity particles can be systematically
calculated as a perturbative expansion in string theory.  The strength
of string interactions is encoded in the dilaton field through the
string coupling $g = e^\phi$.  The perturbative string expansion makes
sense when $g$ is small.

We will not discuss string theory in any detail in this review; for a
comprehensive introduction to superstring theory, the reader should
consult the excellent textbooks by Green, Schwarz, and Witten (1987)
and by Polchinski (1998).  We would like, however, to emphasize the
following points:
\vspace{0.1in}

\noindent
{\it (i)} The world-sheet approach to superstring quantization yields
a theory which is a first-quantized theory of gravity from the point
of view of the target space---that is, a state in the string Hilbert
space corresponds to a single particle state in the target space
consisting of a single string.
\vspace{0.05in}

\noindent
{\it (ii)} The world-sheet approach to superstrings is perturbative in
the string coupling $g$.  As we will discuss in the following
subsection, there are many nonperturbative objects which should appear
in a consistent quantum theory of 10D supergravity.  
\vspace{0.1in}

In order to have a definition of string theory which corresponds to a
true quantum theory of gravity in space-time, it is necessary to
overcome these obstacles by developing a second-quantized theory of
strings.   Work has been done towards developing such a string
field theory (see for example Zwiebach, 1993; Gaberdiel and Zwiebach,
1997).  It is currently difficult to use this formalism to do practical
calculations or gain new insight into the theory, although the work of
Sen (1999) and others has recently generated a new wave of development
in this direction.

To summarize our discussion of string theory, it has been found that a natural
approach to finding a microscopic quantum theory of gravity whose
low-energy limit is ten-dimensional supergravity is to quantize
the stringlike degrees of freedom which couple to the antisymmetric
2-form field $B_{\mu \nu}$.   

Because eleven-dimensional supergravity seems to be in some sense more
fundamental than the ten-dimensional theory, it is natural to want to
find an analogous construction of a microscopic quantum theory of
gravity in eleven dimensions.  Unlike the ten-dimensional theories,
however, in eleven-dimensional supergravity there is no stringlike
black hole solution; indeed, there is no 2-form for it to couple to.
There is, however,  a ``black membrane'' solution in eleven dimensions,
which has a source extended infinitely in two
spatial dimensions.  Just as the black string couples to the 2-form
field through Eq.~(\ref{eq:string-b}), the black membrane solution of 11D
supergravity couples to the 3-form field through 
\begin{equation}
\int_\Sigma A_{IJK} \epsilon^{abc} (\partial_a X^I) (\partial_b
X^J)(\partial_c X^K).
\label{eq:membrane-a}
\end{equation}
where now $a, b, c\in \{0,1, 2\}$  are indices of coordinates on the
three-dimensional membrane world-volume.

It is tempting to imagine that a microscopic description of 11D
supergravity might be found by quantizing the supermembrane, just as a
microscopic description of 10D supergravity is found by quantizing the
superstring.  This idea was explored extensively in the 80's, when it
was first realized that a consistent classical theory of a
supermembrane could be realized in eleven dimensions.  At that time,
while no satisfactory covariant quantization of the membrane theory
was found, it was shown that the supermembrane could be quantized in
light-front coordinates.  As we will discuss in more detail in the
following sections, this construction leads to precisely the matrix
quantum mechanics theory which is the subject of this article.  Not
only does this matrix quantum mechanics theory provide a microscopic
description of quantum gravity in eleven dimensions, but, as more
recent work has demonstrated, it also bypasses the difficulties
mentioned above for string theory by directly providing a
nonperturbative definition of a theory which is second-quantized in
target space.

\subsection{Duality and D-branes}
\label{sec:dd}

Although eleven-dimensional supergravity and the quantum supermembrane
theory were originally discovered at around the same time as the five
consistent superstring theories, much more attention was given to
string theory in the decade from 1985-1995 than to the
eleven-dimensional theory.  There were several reasons for this lack
of attention to 11D supergravity and membrane theory by (much of) the
high energy community.  For one thing, heterotic string theory looked
like a much more promising framework in which to make contact with
standard model phenomenology.  In order to connect a ten- or
eleven-dimensional theory with 4-dimensional physics, it is necessary
to compactify all but 4 dimensions of space-time (or, as has been
suggested more recently by Randall and Sundrum (1999) and others, to
consider  our 4D space-time as a brane living in the higher dimensional
space-time).  There is no way to compactify eleven-dimensional
supergravity on a smooth 7-manifold in such a way as to give rise to
chiral fermions in the resulting 4-dimensional theory (Witten, 1981).
This fact made 11D supergravity for some time a very unattractive
possibility for a fundamental theory; more recently, however, singular
(orbifold) compactifications of eleven-dimensional M-theory have been
considered  (Ho\v{r}ava and Witten,
1996)
which lead to realistic models of phenomenology with chiral
fermions (see for example, Donagi, Ovrut, Pantev, and Waldron, 2000).
Another reason for which the quantum supermembrane was dropped from
the mainstream of research was the appearance of an apparent
instability in the membrane theory (de Wit, L\"uscher, and Nicolai,
1989).  As we will discuss in Section \ref{sec:BFSS}, rather than
being a problem this apparent instability is an indication of the
second-quantized nature of the membrane theory.

As was briefly discussed in the first two paragraphs of the introduction,
in 1995 two remarkable new ideas caused a substantial change in the
dominant picture of superstring theory.  The first of these ideas was
the realization that all five superstring theories, as well as
eleven-dimensional supergravity, seem to be related to one another by
duality transformations which exchange the degrees of freedom of one
theory for the degrees of freedom of another theory (Hull and
Townsend, 1995; Witten, 1995).  It is now generally believed that all
six of these theories are realized as particular limits of some more
fundamental underlying theory, which may be describable as a quantum
theory in eleven dimensions.  This eleven-dimensional theory quantum
theory of gravity, for which no rigorous definition has yet been
given, is often referred to as ``M-theory'' (Ho\v{r}ava and Witten,
1996)\footnote{Note: The term ``M-theory'' is usually used to refer to
an eleven-dimensional quantum theory of gravity which reduces to
${\cal N} = 1$ supergravity at low energies.  It is possible that a
more fundamental description of this eleven-dimensional theory and
string theory can be given by a model in terms of which the
dimensionality of space-time is either greater than 11 or is an
emergent aspect of the dynamics of the system.  Generally the term
M-theory does not refer to such models, but usage varies.  In this
article we mean by M-theory a consistent quantum theory of gravity in
eleven dimensions}.  

The second new idea in 1995 was the realization by Polchinski (1995)
that black $p$-brane solutions which are charged under the
Ramond-Ramond fields of string theory can be described in the language
of perturbative strings as ``Dirichlet-branes'', or ``D-branes'', that
is, as hypersurfaces on which open strings may have endpoints.  Type
IIA string theory contains D$p$-branes with $p = 0, 2, 4$, and 6, while
type IIB string theory contains D$p$-branes with $p = -1, 1, 3, 5$, and
7.  D$p$-branes with $p \geq 8$ also appear in certain situations;
they will not, however, be relevant to this article.  The D$p$-branes
with $p \leq 3$ couple to the Ramond-Ramond $(p + 1)$-form fields of
supergravity through expressions analogous to
Eqs.~(\ref{eq:string-b}), (\ref{eq:membrane-a}).  These branes are referred to
as being ``electrically coupled'' to the relevant Ramond-Ramond
fields.  The D$p$-branes with $p \geq 4$ have ``magnetic'' couplings
to the $(7-p)$-form Ramond-Ramond fields, which can be described in
terms of electric couplings to the dual fields $\tilde{C}^{(p + 1)}$
defined through $d\tilde{C}^{(p + 1)} ={}^*dC^{(7-p)}$.  D-branes are
nonperturbative structures in first-quantized string theory, but play
a fundamental role in many aspects of quantum gravity.  In recent
years D-branes have been used to construct stringy black holes and to
explore connections between string theory and quantum field theory.
Reviews of basic aspects of D-brane physics are given in Polchinski
(1996) and Taylor (1998); applications of D-branes to black holes are
reviewed in Skenderis (1999), Mohaupt (2000), and Peet (2000); a recent
comprehensive review of D-brane constructions of supersymmetric field
theories is given in Giveon and Kutasov (1999).

Combining the ideas of duality and D-branes, we have a new picture of
fundamental physics as being described by an as-yet unknown
microscopic structure, which reduces in certain limits to perturbative
string theory and to 11D supergravity.  In the ten- and
eleven-dimensional limits, there are a variety of dynamical extended
objects of various dimensions appearing as effective excitations.
There is no clear reason for strings to be any more fundamental in
this structure than the membrane in eleven dimensions, or even than
D0-branes or D3-branes in type IIA or IIB superstring theory.  At this
point, in fact, it seems likely that these objects should all be
thought of as equally important pieces of the theory.  On one hand,
the strings and branes can all be thought of as effective excitations
of some as-yet unknown set of degrees of freedom.  On the other hand,
by quantizing any of these objects to whatever extent is technically
possible for an object of the relevant dimension, it is possible to
study particular aspects of each of the theories in certain limits.
This equality between branes is often referred to as ``brane
democracy''.  

As we shall see in the remainder of this review, matrix theory can be
thought of alternatively as a quantum theory of membranes in eleven
dimensions, or as a quantum theory of pointlike D0-branes in ten
dimensions.  In order to relate these complementary approaches to
matrix theory, it will be helpful at this point to briefly review one
of the simplest links in the network of dualities connecting the
string theories with M-theory.  This is the duality which relates
M-theory to type IIA string theory (Townsend, 1995; Witten, 1995).
The connection between these theories essentially follows from the
fact that type IIA supergravity can be constructed from
eleven-dimensional supergravity by performing ``dimensional
reduction'' along a single dimension.  To implement this procedure we
assume that eleven-dimensional supergravity is defined on a space-time
with geometry $M^{10} \times S^1$ where $M^{10}$ is an arbitrary
ten-dimensional manifold and $S^1$ is a circle of radius $R$.  When
$R$ is small we can systematically neglect the dependence of all
fields in the 11D theory along the 11th (compact) direction, giving an
effective low-energy ten-dimensional theory, which turns out to be
type IIA supergravity.  In this dimensional reduction, the different
components of the fields of the eleven-dimensional theory decompose
into the various fields of the 10D theory.  The metric tensor $g_{IJ}$
in eleven dimensions has components $g_{\mu \nu}$, $0 \leq \mu, \mu
\leq 9$, which become the 10D metric tensor, components $g_{\mu\, 11}$
which become the Ramond-Ramond vector field $C^{(1)}_\mu$ in the
ten-dimensional theory, and a single component $g_{11\, 11}$ which
becomes the 10D dilaton field $\phi$.  Similarly, the 3-form field of
the eleven-dimensional theory decomposes into the 2-form field $B_{\mu
\nu}$ and the Ramond-Ramond 3-form field $C^{(3)}_{\mu \nu \lambda}$
in ten dimensions.  Just as the fields of eleven-dimensional
supergravity reduce to the fields of the type IIA theory under
dimensional reduction, the extended objects of M-theory reduce to
branes of various kinds in type IIA string theory.  The membrane of
M-theory can be ``wrapped'' around the compact direction of radius $R$
to become the fundamental string of the type IIA theory.  The
unwrapped M-theory membrane corresponds to the Dirichlet 2-brane
(D2-brane) in type IIA.  In addition to the membrane, M-theory has an
M5-brane (with 6-dimensional world-volume) which couples magnetically
to the 3-form field $A_{IJK}$.  Wrapped M5-branes become D4-branes in
type IIA, while unwrapped M5-branes become solitonic (NS) 5-branes in
type IIA, which are magnetically charged objects under the NS-NS
2-form field $B_{\mu \nu}$.

Through the dimensional reduction of 11D supergravity to type IIA
supergravity, the string coupling $g$
and string length $l_s$
in the
ten-dimensional theory can be related to the 11D Planck length
$l_{11}$ and the compactification radius $R$ through
\begin{eqnarray}
g  =   (\frac{R}{l_{11}} )^{3/2},& \;\;\;\;\; \;\;\;\;\;
\;\;\;\;\; \;\;\;\;\; &
l_s^2  =   \frac{l_{11}^3}{R}\, .  \label{eq:coupling-relations}
\end{eqnarray}
From these relations we see that in the strong coupling limit $g
\rightarrow \infty$, type IIA string theory ``grows'' an extra
dimension $R \rightarrow \infty$ and should be identified with
M-theory in flat space.  This motivates a definition of M-theory as
the strong coupling limit of the type IIA string theory (Witten,
1995); because there is no nonperturbative definition of type IIA
string theory, however, this definition is not completely
satisfactory.

When the compactification radius $R$ used to reduce M-theory to type
IIA is small, momentum modes in the 11th direction of the massless
fields associated with the eleven-dimensional graviton multiplet
become massive Kaluza-Klein particles in the ten-dimensional IIA
theory.  These particles couple to the components $g_{\mu\, 11}$ of
the eleven-dimensional metric, and therefore to $C^{(1)}_\mu$ in ten
dimensions.  Thus, these particles can be identified as the Dirichlet
0-branes of type IIA string theory.  This connection between momentum
in eleven dimensions and Dirichlet particles, first emphasized by
Townsend (1996a), is a crucial ingredient in understanding the
connection between the two perspectives on matrix theory which we
develop in this review.

\subsection{M(atrix) theory}
\label{sec:mt}

In this section we briefly summarize the development of matrix theory,
giving an overview of the material which we describe in detail in the
following sections.  As discussed above, it seems that a natural way
to try to construct a microscopic model for M-theory would be to
quantize the supermembrane which couples to the 3-form field of 11D
supergravity.  In general, quantizing any fluctuating geometrical
object of higher dimensionality than the string is a problematic
enterprise, and for some time it was believed that membranes and
higher-dimensional objects could not be described in a sensible
fashion by quantum field theory.  Almost two decades ago, however,
Goldstone (1982) and Hoppe (1982, 1987) found a very clever way of
regularizing the theory of the classical membrane.  They replaced the infinite
number of degrees of freedom representing the embedding of the
membrane in space-time by a finite number of degrees of
freedom contained in $N \times N$ matrices.  This approach, which
we describe in detail in the next section, was generalized by de Wit,
Hoppe, and Nicolai (1988) to the supermembrane.  The resulting theory
is a simple quantum mechanical theory with matrix degrees of freedom.
The Hamiltonian of this theory is given by
\begin{equation}
 H = \tr \left( \half \dot{{\bf X}}^i
\dot{{\bf X}}^i - {1 \over 4} [{\bf X}^i,{\bf X}^j] [{\bf X}^i,{\bf
X}^j] + \half {\bf \theta}^T \gamma_i[ {\bf X}^i, {\bf \theta}]
\label{eq:matrix-Hamiltonian} \right)\,.  
\end{equation}
In this expression,
${\bf X}^i$ are nine $N
\times N$ matrices, $\theta$ is a 16-component matrix-valued spinor of
$SO(9)$ and $\gamma_i$ are the $SO(9)$ gamma matrices in the
16-dimensional representation.  Even before its discovery as a
regularized version of the supermembrane theory, this quantum
mechanics theory had been studied as a particularly elegant example of
a quantum system with a high degree of supersymmetry
(Claudson and Halpern, 1985; Flume, 1985; Baake, Reinicke, and
Rittenberg, 1985).

Although the Hamiltonian Eq.~(\ref{eq:matrix-Hamiltonian}) describing
matrix theory and its connection with the supermembrane has been known
for some time, this theory was for some time believed to suffer from
insurmountable instability problems.  It was pointed out several years
ago by Townsend (1996a) and by Banks, Fischler, Shenker, and Susskind
(1997) that the Hamiltonian Eq.~(\ref{eq:matrix-Hamiltonian}) can
also be seen as arising from a system of $N$ Dirichlet particles in
type IIA string theory.  Using the duality relationship between
M-theory and type IIA string theory described above, Banks, Fischler,
Shenker and Susskind (henceforth ``BFSS'') made the bold conjecture
that in the large $N$ limit the system defined by
Eq.~(\ref{eq:matrix-Hamiltonian}) should give a complete description
of M-theory in the light-front (infinite momentum) coordinate frame.
This picture cleared up the apparent instability problems of the
theory in a very satisfactory fashion by making it clear that matrix
theory should describe a second-quantized theory in target space,
rather than a first-quantized theory as had previously been imagined.

Following the BFSS conjecture, there was a flurry of activity for
several years centered around the matrix model defined by
Eq.~(\ref{eq:matrix-Hamiltonian}).  In this period of time much progress
was made in understanding both the structure and the limits of this
approach to studying M-theory.  It has been shown that matrix theory can
indeed be constructed in a fairly rigorous way as a light-front
quantization of M-theory by taking a limit of spatial
compactifications.  A fairly complete picture has been formed of how the
objects of M-theory (the graviton, membrane, and M5-brane) can be
constructed from matrix degrees of freedom.  It has been shown that all
linearized supergravitational interactions between these objects and
some nonlinear general relativistic corrections can be derived from
quantum effects in matrix theory.  Some simple compactifications of
M-theory have been constructed in the matrix theory formalism, leading to
new insight into connections between certain quantum field theories
and quantum theories of gravity.    The goal of
this article is to review these developments in some detail and to
summarize our current understanding of both the successes and the
limitations of the matrix model approach to M-theory.

In the following sections, we develop the structure of matrix theory
in more detail.  Section \ref{sec:quantized-membrane} reviews the
original description of matrix theory in terms of a regularization of
the quantum supermembrane theory.  In Section \ref{sec:BFSS} we
describe the theory in the language of light-front quantized M-theory,
and discuss the second-quantized nature of the resulting space-time
theory.  The connection between classical supergravity interactions in
space-time and quantum loop effects in matrix theory is presented in
Section \ref{sec:matrix-interactions}.  In Section
\ref{sec:matrix-objects} we show how the extended objects of M-theory
can be described in terms of matrix degrees of freedom.  In Section
\ref{sec:matrix-general-background} we discuss extensions of the basic
matrix theory conjecture to other space-time backgrounds, and in
Section \ref{sec:related-models} we briefly review the connection
between matrix theory and several other related models.  Section
\ref{sec:matrix-summary} contains concluding remarks.

\section{Matrix theory from the quantized supermembrane}
\label{sec:quantized-membrane}

In this section we describe in some detail how matrix theory arises
from the quantization of the supermembrane.   In
\ref{sec:bosonic-membrane} we describe the theory of the relativistic
bosonic membrane in flat space.  The light-front description of this
theory is discussed in \ref{sec:membrane-light-front}, and the matrix
regularization of the theory is described in
\ref{sec:matrix-regularization}.  In \ref{sec:bosonic-general} we
briefly discuss the description of the bosonic membrane moving in a
general background geometry.  In \ref{sec:supermembrane} we extend the
discussion to the supermembrane. The problem of
finding a covariant membrane quantization is discussed in
\ref{sec:covariant-membrane}.

The material in this section roughly follows the original papers by
Hoppe (1982, 1987) and de Wit, Hoppe, and Nicolai (1988).  Note,
however, that the original derivation of the matrix quantum mechanics
theory was done in the Nambu-Goto-type membrane formalism, while we
use here the Polyakov-type approach.

\subsection{The bosonic membrane theory}
\label{sec:bosonic-membrane}

In this subsection we review the theory of a classical relativistic
bosonic membrane moving in flat $D$-dimensional Minkowski space.  This
analysis is very similar in flavor to the theory of a classical
relativistic bosonic string.  We do not assume familiarity with string
theory, and give a self-contained description of the membrane theory
here; readers unfamiliar with the somewhat
simpler classical bosonic string may wish to look at the texts by
Green, Schwarz, and Witten (1987) and by Polchinski (1998)  for
comparison with the discussion here.

Just as a particle sweeps out a trajectory described by a
one-dimensional world-line  
as it moves through space-time, a dynamical membrane moving in $D-1$
spatial dimensions
sweeps out a three-dimensional world-volume in
$D$-dimensional space-time.  We can think of the motion of the membrane in
space-time as being described by a map $X:{\cal V}
\rightarrow\br^{D-1, 1}$ taking a 3-dimensional manifold ${\cal V}$
(the membrane world-volume) into flat $D$-dimensional Minkowski space.
We can locally choose a set of 3 coordinates $\sigma^\alpha, \alpha
\in\{0, 1, 2\}$, on the world-volume of the membrane, analogous to the
coordinate $\tau$ used to parameterize the world-line of a
particle moving in space-time.  We will sometimes use the notation $\tau =
\sigma_0$ and we will use indices $a, b, \ldots$ to describe ``spatial''
coordinates $\sigma^a \in\{1, 2\}$ on the membrane world-volume.  In
such a coordinate system, the motion of the membrane through
space-time is described by a set of $D$ functions $X^\mu (\sigma^0,
\sigma^1, \sigma^2)$.

The natural classical action for a membrane moving in flat space-time
is given by the integrated proper volume swept out by the membrane.
This action takes the Nambu-Goto form
\begin{equation}
S = -T  \int d^3 \sigma \sqrt{-\det h_{\alpha \beta}}\, ,
\label{eq:Nambu-Goto-membrane}
\end{equation}
where  $T$ is a constant which can be interpreted as the membrane tension 
$
T = 1/ (2 \pi)^2 l_p^3
$, and
\begin{equation}
h_{ \alpha \beta} = \partial_\alpha X^\mu \partial_\beta X_\mu
\label{eq:induced-metric-membrane}
\end{equation}
is the pullback of the flat space-time metric (with signature $-+ +
\cdots +$) to the three-dimensional membrane world-volume.

Because of the square root, it is cumbersome to analyze the membrane
theory directly using this action.  There is a convenient
reformulation of the membrane theory which leads to the same classical
equations of motion using a polynomial action.  This is the analogue of
the Polyakov action for the bosonic string.  In order to describe the
membrane using this approach, we must introduce an auxiliary metric
$\gamma_{\alpha\beta}$ on the membrane world-volume.  We then take the
action to be
\begin{equation}
S = - \frac{T}{2}  \int d^3 \sigma \sqrt{-\gamma} \left(
\gamma^{\alpha \beta} \partial_\alpha X^\mu \partial_\beta X_\mu -1\right).
\label{eq:Polyakov-membrane}
\end{equation}
The final constant term $-1$ inside the parentheses does not appear in
the analogous string theory action.  This additional
``cosmological'' term is needed due to the absence of scale invariance in
the theory.  

Computing the equations of motion from Eq.~(\ref{eq:Polyakov-membrane}) by
varying
$\gamma_{\alpha \beta}$, we get
\begin{equation}
\gamma_{\alpha \beta} = h_{\alpha \beta} = \partial_\alpha X^\mu
\partial_\beta X_\mu.
 \label{eq:gamma-constraints}
\end{equation}
Replacing this in Eq.~(\ref{eq:Polyakov-membrane}) again gives
Eq.~(\ref{eq:Nambu-Goto-membrane}), so we see that the two forms of the
action are actually equivalent.  The equation of motion which arises
from varying $X^\mu$ in Eq.~(\ref{eq:Polyakov-membrane}) is
$
\partial_\alpha \left( \sqrt{-\gamma} \gamma^{\alpha \beta}
\partial_\beta X^\mu \right) = 0.
$

To simplify the analysis, we would now like to use the symmetries of
the theory to gauge-fix the metric $\gamma_{\alpha \beta}$.
Unfortunately, unlike in the case of the classical string, where there
are three components of the metric and three continuous symmetries
(two diffeomorphism symmetries and a scale symmetry), for the membrane
we have six independent metric components and only three
diffeomorphism symmetries.  We can use these symmetries to fix the
components $\gamma_{0\alpha}$ of the metric to be
\begin{eqnarray}
\gamma_{0a}  = 0,  & \;\;\;\;\; \;\;\;\;\; \;\;\;\;\; &
\gamma_{00}  =  -  \frac{4}{\nu^2}  \bar{h} \equiv
-\frac{4}{\nu^2}  \det h_{ab} \label{eq:3-gauge}
\end{eqnarray}
where $\nu$ is an arbitrary constant whose normalization has been
chosen to make the later matrix interpretation transparent.  Once we
have chosen this gauge, no further components of the metric
$\gamma_{ab}$ can be fixed.  This gauge can only be chosen when the
membrane world-volume is of the form $\Sigma \times\br$, where $\Sigma$
is a Riemann surface of fixed topology.  The membrane action becomes
in this gauge, using Eq.~(\ref{eq:gamma-constraints}) to eliminate $\gamma$,
\begin{equation}
S = \frac{T \nu}{4}  \int d^3 \sigma \left(
\dot{X}^\mu \dot{X}_\mu -\frac{4}{ \nu^2}  \bar{h} \right).
 \label{eq:membrane-action-h}
\end{equation}

It is natural to rewrite this action in terms of a canonical
Poisson bracket on the membrane where at constant $\tau$,
$\{f,g\} \equiv \epsilon^{ab} \partial_a f
\partial_b g$ with $\epsilon^{12} = 1$.
We will assume that the coordinates $\sigma$ are chosen so that with
respect to the symplectic form associated to
this canonical Poisson bracket the volume of the Riemann surface
$\Sigma$ is $\int d^2 \sigma  =4 \pi$.
In terms of the Poisson bracket, the membrane action becomes
\begin{equation}
S = \frac{T \nu}{4}  \int d^3 \sigma \left(
\dot{X}^\mu \dot{X}_\mu -\frac{2}{ \nu^2}  
\{X^\mu, X^\nu\}\{X_\mu, X_\nu\} \right).
\label{eq:covariant-action}
\end{equation}
The equations of motion for the fields $X^\mu$ are
\begin{equation}
\ddot{X}^\mu = {4 \over \nu^2} \partial_a \left(\bar{h} h^{ab}
\partial_b X^\mu\right)   
           = {4 \over \nu^2} \{\{X^\mu,X^\nu\},X_\nu\}\, .
\label{eq:membrane-eom} 
\end{equation}
The auxiliary constraints on the system arising from combining
Eqs.~(\ref{eq:gamma-constraints}) and (\ref{eq:3-gauge})
are
\begin{equation}
\dot{X}^\mu \dot{X}_\mu  =   -\frac{4}{ \nu^2}
\bar{h} 
  =   -\frac{2}{\nu^2} 
\{X^\mu, X^\nu\}\{X_\mu, X_\nu\} \label{eq:membrane-constraint-1}
\end{equation}
and
\begin{equation}
\dot{X}^\mu \partial_a X_\mu = 0.
\label{eq:membrane-constraint-2}
\end{equation}
It follows directly from Eq.~(\ref{eq:membrane-constraint-2}) that
\begin{equation}
\{\dot{X}^\mu, X_\mu\} = 0.
\end{equation}

We have thus expressed the classical bosonic membrane theory as a
constrained dynamical system.  The degrees of freedom of this system
are $D$ functions $X^\mu$ on the 3-dimensional world-volume of a
membrane with topology $\Sigma \times\br$, where $\Sigma$ is a
Riemann surface.  This theory is still completely covariant.  It is
difficult to quantize, however, because of the constraints and the
nonlinearity of the equations of motion.  The direct quantization of
this covariant theory will be discussed further in Section
\ref{sec:covariant-membrane}.

\subsection{The light-front bosonic membrane}
\label{sec:membrane-light-front}

We now consider the membrane theory in
light-front coordinates
\begin{equation}
X^{\pm} = (X^0 \pm X^{D-1})/\sqrt{2}.
\label{eq:xpm}
\end{equation}
The constraints
(\ref{eq:membrane-constraint-1},\ref{eq:membrane-constraint-2}) can be
explicitly solved in light-front gauge
\begin{equation}
 X^+ (\tau, \sigma_1, \sigma_2) =\tau.
\label{eq:light-cone-gauge-membrane}
\end{equation}
We have
\begin{eqnarray}
\dot{X}^- = \half \dot{X}^i \dot{X}^i + {1 \over \nu^2}
\{X^i,X^j\}\{X^i,X^j\},  & \;\;\;\;\; \;\;\;\;\; \;\;\;\;\; &
\partial_a X^- = \dot{X}^i \partial_a X^i\,. \label{eq:dx}
\end{eqnarray}
We can go to a Hamiltonian formalism by computing the canonically conjugate
momentum densities.
The total momentum in the direction $P^+$ is then
\begin{equation}
p^+ = \int d^2 \sigma P^+ = 2 \pi \nu T,
\end{equation}
and the Hamiltonian of the theory is given by
\begin{eqnarray}
H &=& {\nu T \over 4} \int d^2 \sigma \left( \dot{X}^i \dot{X}^i
+ {2 \over \nu^2} \{X^i,X^j\}\{X^i,X^j\} \right)\,. \label{eq:light-front-h}
\end{eqnarray}
The only remaining constraint that the transverse degrees of
freedom must satisfy is
\begin{equation}
\{\dot{X}^i,X^i\}=0\,.
\end{equation}
This theory has a residual invariance under time-independent
area-preserving diffeomorphisms.  Such diffeomorphisms do not change
the symplectic form and thus manifestly leave the Hamiltonian
(\ref{eq:light-front-h}) invariant.

We now have a Hamiltonian formalism for the light-front membrane
theory.  Unfortunately, this theory is still rather difficult to
quantize.  Unlike string theory, where the equations of motion are
linear in the analogous formalism, for the membrane the equations of
motion (\ref{eq:membrane-eom}) are nonlinear and difficult to solve.

\subsection{Matrix regularization}
\label{sec:matrix-regularization}

A remarkably clever regularization of the light-front membrane
theory was found by Goldstone (1982) and Hoppe (1982) in the case
where the membrane 
surface $\Sigma$ is a sphere $S^2$.  According
to this regularization procedure, functions on the membrane surface
are mapped to finite sized matrices.  Just as in the
quantization of a classical mechanical system defined in terms of a
Poisson brackets, the Poisson bracket appearing in the membrane theory
is replaced in the matrix regularization of the theory by a matrix
commutator.

It should be emphasized that this procedure of replacing functions by
matrices is a {\it completely classical} manipulation.  Although the
mathematical construction used is similar to those used in geometric
quantization of classical systems, after regularizing the continuous
classical membrane theory the resulting theory is a system
which has a finite number of degrees of freedom, but is still
classical.  After this regularization procedure has been carried out,
we can quantize the system just like any other classical system with a
finite number of degrees of freedom.

The matrix regularization of the theory can be generalized to
membranes of arbitrary topology, but is perhaps most easily understood
by considering the case originally discussed by Hoppe (1982), where the
membrane has the topology of a sphere $S^2$.
In this case the world-sheet of the membrane surface at fixed time can
be described by a unit sphere with an SO(3) invariant canonical
symplectic form.  Functions on this membrane can be described in terms
of functions of the three Cartesian coordinates $\xi_1, \xi_2, \xi_3$
on the unit sphere satisfying
$
\xi_1^2 + \xi_2^2 + \xi_3^2 = 1
$. The Poisson brackets of these functions are
given by
$
\{\xi_A, \xi_B\} = \epsilon_{ABC} \xi_C
$. This is the same algebraic structure as that defined by
the commutation relations of the generators of $SU(2)$.  It is
therefore natural to associate these coordinate functions on $S^2$
with the matrices generating $SU(2)$ in the $N$-dimensional
representation.  In terms of the conventions we are using here, when
the normalization constant $\nu$ is integral, the correct
correspondence is
\begin{equation}
\xi_A \rightarrow {2 \over N} {\bf J}_A
\label{eq:x-j-correspondence}
\end{equation}
where ${\bf J}_1, {\bf J}_2, {\bf J}_3$ are generators of the $N$-dimensional representation of
$SU(2)$ with $N = \nu$,
satisfying the commutation relations
$
-i [{\bf J}_A, {\bf J}_B] = \epsilon_{ABC} {\bf J}_C \,.
$

In general, any function on the membrane can be expanded as a sum of
spherical harmonics
\begin{equation}
f (\xi_1, \xi_2, \xi_3) = \sum_{l, m} c_{lm} Y_{lm} (\xi_1, \xi_2,
\xi_3)\, .
\label{eq:general-function}
\end{equation}
The spherical harmonics can in turn be written as  sums of
monomials in the coordinate functions:
\begin{equation}
Y_{lm} (\xi_1, \xi_2, \xi_3)=
 \sum_k t^{(lm)}_{A_1 \ldots A_l} \xi_{A_1} \cdots \xi_{A_l}\,
\end{equation}
where the coefficients $t^{(lm)}_{A_1 \ldots A_l}$ are symmetric and
traceless (because $\xi_A \xi_A = 1$).  Using the correspondence
(\ref{eq:x-j-correspondence}), matrix approximations ${\bf Y}_{lm}$ to
each of the spherical harmonics with $l < N$ can be constructed through
\begin{equation}
Y_{lm} (\xi_1, \xi_2, \xi_3)\rightarrow
{\bf Y}_{lm} = \left(2 \over N\right)^l
\sum t^{(lm)}_{A_1\ldots A_l}
{\bf J}_{A_1}
\cdots {\bf J}_{A_l}\, .
\label{eq:matrix-spherical}
\end{equation}
For a fixed value of $N$ only the spherical harmonics with $l < N$ can
be constructed
because higher order monomials in the generators ${\bf J}_A$ do not
generate linearly independent matrices.  Note that the number of
independent matrix entries is precisely equal to the number of
independent spherical harmonic coefficients which can be determined
for fixed $N$,
\begin{equation}
N^2 = \sum_{l = 0}^{N -1}  (2l + 1)\, .
\end{equation}
The matrix approximations (\ref{eq:matrix-spherical}) of the spherical
harmonics can be used to construct matrix approximations to an
arbitrary function of the form (\ref{eq:general-function})
\begin{equation}
\label{MMcorrespond}
f (\xi_1, \xi_2, \xi_3)\rightarrow
F =
\sum_{l < N, m}   c_{lm}
{\bf Y}_{lm}\, .
\end{equation}

The Poisson bracket in the membrane theory is replaced in the
matrix-regularized theory with the matrix commutator according to the
prescription
\begin{equation}
\{f, g\} \rightarrow \frac{-i N}{2} [F, G].
 \label{eq:Poisson-map}
\end{equation}
Similarly, an integral over the membrane at fixed $\tau$ is replaced
by a matrix trace through
\begin{equation}
\frac{1}{4 \pi} \int d^2 \sigma f \rightarrow \frac{1}{N}  {\rm
Tr}\; F\, .
 \label{eq:integral-map}
\end{equation}

The Poisson bracket of a pair of spherical harmonics takes the form
\begin{equation}
\{Y_{lm}, Y_{l' m'}\} = g_{lm, l' m'}^{l'' m''} Y_{l'' m''}.
\end{equation}
The commutator of a pair of matrix  spherical harmonics
(\ref{eq:matrix-spherical}) can be written
\begin{equation}
\left[{\bf Y}_{lm}, {\bf Y}_{l' m'}\right] = G_{lm, l' m'}^{l'' m''} {\bf Y}_{l'' m''}.
\end{equation}
It can be verified that in the large $N$ limit the structure constants
of these algebras agree
\begin{equation}
\lim_{N\rightarrow \infty} 
\frac{-i N}{ 2}  G_{lm, l' m'}^{l'' m''}
\rightarrow g_{lm, l' m'}^{l'' m''}\, .
\end{equation}
As a result,  it can be shown that
for any smooth functions $f, g$ on the membrane defined
in terms of convergent sums of spherical harmonics, with Poisson
bracket $\{f, g\} = h$ we have
\begin{eqnarray}
\lim_{N \rightarrow \infty}
\frac{1}{N} 
{\rm Tr}\; F & = &  \frac{1}{4 \pi}  \int d^2 \sigma f
\end{eqnarray}
and
\begin{eqnarray}
\lim_{N \rightarrow \infty} ((\frac{-i N}{2} )[F, G] -H) & = & 0\, .
\end{eqnarray}
This last relation is really shorthand for the statement that
\begin{equation}
\lim_{N \rightarrow \infty}\frac{1}{N} 
 {\rm Tr}\; \left(((\frac{-i N}{2} )[F, G] -H) J \right) = 0\,.
\end{equation}
where $J$ is the matrix approximation to any smooth function $j$ on
the sphere.

We now have a dictionary for transforming between continuum and
matrix-regularized quantities.  The correspondence is given by
\begin{equation}
\label{eq:mm-correspondence}
\xi_A \leftrightarrow {2 \over N} {\bf J}_A, \;\;\;\;\; \;\;\;\;\;
\{\cdot,\cdot\} \leftrightarrow  {-i N \over 2} [\cdot,\cdot],
\;\;\;\;\; \;\;\;\;\;
{1 \over 4 \pi} \int d^2 \sigma \leftrightarrow \frac{1}{N} 
\tr \;\, .
\end{equation}
The matrix regularized membrane Hamiltonian is therefore given by
\begin{eqnarray}
H  & = & (2 \pi l_p^3) \tr \left( \half {\bf P}^i {\bf P}^i
\right)
-  {1 \over  (2 \pi l_p^3)}\tr \left(
{1 \over 4} [{\bf X}^i,{\bf X}^j] [{\bf X}^i,{\bf X}^j]
\right) \nonumber\\
 & = &  {1 \over  (2 \pi l_p^3)} \tr \left( \half \dot{{\bf X}}^i \dot{{\bf
X}}^i - {1 \over 4} [{\bf X}^i,{\bf X}^j] [{\bf X}^i,{\bf X}^j]
\right)\,.\label{matrix-Hamiltonian}
\end{eqnarray}
This Hamiltonian gives rise to the matrix equations of
motion
\begin{equation}
\ddot{\bf X}^i + [[{\bf X}^i,{\bf X}^j],{\bf X}^j] = 0,
\end{equation}
which must be supplemented with the Gauss constraint
\begin{equation}
\label{MatrixGauss}
[\dot{\bf X}{}^i, {\bf X}^i] = 0 \,.
\end{equation}
This is a classical theory with a finite number of degrees of
freedom.  The quantization of such a system is straightforward,
although solving the quantum theory can in practice be quite tricky.

We have now described, following Goldstone and Hoppe, a well-defined
quantum theory arising from the matrix regularization of the
relativistic membrane theory in light-front coordinates.  This model
has $N \times N$ matrix degrees of freedom, and a symmetry group
$U(N)$ with respect to which the matrices $X^i$ are in the adjoint
representation.  The model just described arose from the
regularization of a membrane with world-volume topology $S^2 \times
\br$.  A similar regularization procedure can be followed for an
arbitrary genus Riemann surface.  Remarkably, the same $U(N)$ matrix
theory arises as the regularization of the theory describing a
membrane of any genus (Bordemann, Meinrenken, and Schlichenmaier,
1994).  While this result has only been demonstrated implicitly for
Riemann surfaces of genus greater than one, the toroidal case was
described explicitly by Fairlie, Fletcher, and Zachos (1989) and
Floratos (1989) (see also Fairlie and Zachos (1989)).  In this case a
natural basis of functions on the torus parameterized by $\eta_1,
\eta_2 \in\{[0, 2 \pi]\}$ is given by the Fourier modes
\begin{equation}
Y_{nm} (\eta_1, \eta_2) = e^{in\eta_1 + im\eta_2}\, .
 \label{eq:torus-Fourier}
\end{equation}
To describe the matrix approximations for these functions we use the
't Hooft matrices
\begin{equation}
U = 
\pmatrix{
1& & & & \cr
&q && & \cr
& &q^2 & &\cr
& & & \ddots & \cr
& & &&  q^{N-1}
}, \;\;\;\;\; \;\;\;\;\; \;\;\;\;\;
V =
\pmatrix{
 & 1 & && \cr
 &   & 1 && \cr
 &   &   & \ddots & \cr
& & & & 1\cr
1&   &   &  & 
} 
\end{equation}
where
\begin{equation}
q = e^{2\pi i\over N}.
\end{equation}
The matrices $U, V$ satisfy
\begin{equation}
UV = q^{-1}VU.
\end{equation}
In terms of these matrices we can define
\begin{equation}
{\bf Y}_{nm} = q^{nm/2}
U^nV^m = q^{-nm/2} V^mU^n\, .
 \label{eq:matrix-Fourier}
\end{equation}
The matrix approximation to an arbitrary function 
on the torus
is then given by
\begin{equation}
f(\eta_1, \eta_2) = \sum_{n, m}c_{nm} Y_{nm} (\eta_1, \eta_2)
\rightarrow
F = \sum_{n, m}c_{nm} {\bf Y}_{nm} .
 \label{eq:torus-function-map}
\end{equation}
Just as in the case of the sphere, the structure constants of the
Poisson bracket algebra of the Fourier modes (\ref{eq:torus-Fourier})
is reproduced by the commutators of the matrices
(\ref{eq:matrix-Fourier}) in the large $N$ limit, where the symplectic
form on the torus is taken to be proportional to $\epsilon_{ij}$.
Combining Eq.~(\ref{eq:torus-function-map}) with Eqs.~(\ref{eq:Poisson-map})
and (\ref{eq:integral-map}) then gives a consistent regularization of
the membrane theory on the torus, which again leads to the matrix
Hamiltonian (\ref{eq:matrix-Hamiltonian}).

The fact that the regularization of the membrane theory on a Riemann
surface of any genus gives rise to a family of theories with $U(N)$
symmetry can be related to the fact that the symmetry group of
area-preserving diffeomorphisms on the membrane can be approximated by
$U(N)$ for a surface of any genus.  This was emphasized in the case of the
sphere by Floratos, Iliopoulos, and Tiktopoulos (1989), and discussed
for arbitrary genus by Bordemann, Meinrenken, and
Schlichenmaier (1994).  How this connection should be understood in
the large $N$ limit is, however, a subtle issue.  It is possible to
construct, for example, sequences of matrices in $U(N)$ which
correspond in the large $N$ limit to singular area-preserving
diffeomorphisms of the membrane surface.  These singular maps may have
the effect of essentially changing the membrane topology by adding or
removing handles.  Thus, it probably does not make sense to think of
the matrix membrane theory as being associated with membranes of a
particular topology.  Indeed, as we will emphasize in Section
\ref{sec:BFSS}, matrix configurations with large values of $N$ can
approximate any system of multiple membranes with arbitrary
topologies.  Thus, in some sense the matrix regularization of the
membrane theory contains {\em more} structure than the smooth theory
it is supposed to be approximating.  This additional structure may be
precisely what is needed to make sense of M-theory as a quantized
theory of membranes.

Another way to mathematically describe the matrix regularization of a
theory on the membrane is in terms of the language of geometrical
quantization.  From this point of view the matrix membrane is like a
``fuzzy'' membrane which is in some sense discrete, and yet which may
preserve continuous symmetries such as the $SU(2)$
rotational symmetry of a spherical membrane.  This point of view
ties into recent developments in noncommutative geometry, and we will
discuss it again briefly in Section~\ref{sec:matrix-summary}.

In this section we have focused on the matrix regularization of closed
membranes (membranes without boundaries).  It is also possible to
consider a theory of open membranes with boundaries on an M5-brane
(Strominger, 1996; Townsend, 1996a).  The matrix regularization of the
open membrane theory has been constructed by Li (1996), Ezawa, Matsuo,
and Murakami (1998), and de Wit, Peeters, and Plefka (1998a).

\subsection{The bosonic membrane in a general background}
\label{sec:bosonic-general}

So far we have only considered the membrane in a flat background
Minkowski geometry.   It is natural to generalize
the discussion to a bosonic membrane moving in a general background
metric $g_{\mu \nu}$ and 3-form field $A_{\mu \nu \rho}$.  The
introduction of a general background metric modifies the Nambu-Goto
action by replacing $h_{\alpha \beta}$ in
Eq.~(\ref{eq:induced-metric-membrane}) with
\begin{equation}
h_{ \alpha \beta} = \partial_\alpha X^\mu \partial_\beta X^\nu
g_{\mu \nu} (X).
\label{eq:induced-metric-membrane-general}
\end{equation}
The membrane couples to the 3-form field as an electrically charged
object through Eq.~(\ref{eq:membrane-a}).
This gives a total action for the membrane in a
general background of the form
\begin{equation}
S = -T  \int d^3 \sigma \left( \sqrt{-\det h_{\alpha \beta}}
+6 \dot{X}^\mu \partial_1 X^\nu \partial_2 X^\rho
 A_{\mu \nu \rho}(X)\right).
\end{equation}
With an auxiliary world-volume metric, this action becomes
\begin{equation}
S  =   - \frac{T}{2}  \int d^3 \sigma 
\left[\sqrt{-\gamma} \left(
\gamma^{\alpha \beta} \partial_\alpha X^\mu \partial_\beta X^\nu
g_{\mu \nu} (X)
-1\right)
+12 \dot{X}^\mu \partial_1 X^\nu \partial_2 X^\rho
 A_{\mu \nu \rho} (X)\right]\, .
 \label{eq:bosonic-membrane-Polyakov-general}
\end{equation}

We can gauge fix the action
(\ref{eq:bosonic-membrane-Polyakov-general}) using the same gauge
(\ref{eq:3-gauge}) as in the flat space case.  We can then consider
carrying out a similar procedure for quantizing the membrane in a
general background as we described in the case of the flat background.
We will return to this possibility in section
\ref{sec:curved-background} when we discuss in more detail the
prospects for constructing matrix theory in a general background.

\subsection{The supermembrane}
\label{sec:supermembrane}

Now let us turn our attention to the supermembrane.  In order to make
contact with M-theory, and indeed to make the membrane theory
well-behaved it is necessary to add supersymmetry to the theory.
Supersymmetric membrane theories can be constructed classically in
dimensions 4, 5, 7, and 11.  These theories have different degrees of
supersymmetry, with 2, 4, 8, and 16 independent supersymmetric
generators respectively.  It is believed that all the supermembrane
theories other than the 11D maximally supersymmetric theory are
problematic quantum mechanically.  Thus, just as $D = 10$ is the
natural dimension for the superstring, $D = 11$ is the natural
dimension for the supermembrane.

The formalism for describing the supermembrane is rather technically
complicated.  We outline here very briefly the steps
involved in constructing the supermembrane theory and deriving the
associated supersymmetric matrix model.  For a more detailed
description of the supermembrane, the reader should refer to the
original paper of Bergshoeff, Sezgin, and Townsend (1988) or the
reviews of Duff (1996), Nicolai and Helling (1998), and de Wit (1999).

To understand how space-time supersymmetry can be incorporated into
membrane theory, it is useful to consider the analogous situation in
string theory.  There are two very different approaches to
incorporating space-time supersymmetry in string theory.  One approach
is the Neveu-Schwarz-Ramond (NSR) approach (see for example, Green,
Schwarz, and Witten, 1987), in which the world-volume string theory
itself is extended to have supersymmetry.  This formalism gives a
theory which is easy to quantize, and which can be used in a
straightforward fashion to describe the spectra of the five
superstring theories.  One disadvantage of this formalism, however, is
that the target space supersymmetry of the theory is difficult to show
explicitly.  The second approach to incorporating space-time
supersymmetry into string theory is the Green-Schwarz formalism (Green
and Schwarz, 1984a, 1984b), in which the target space supersymmetry of
the theory is manifest.  In the Green-Schwarz formalism 
Grassmann (anticommuting)
degrees of freedom are introduced which transform as
space-time spinors but as world-sheet vectors.  These correspond to
space-time superspace coordinates for the string.  The Green-Schwarz
superstring action does not have a standard world-sheet supersymmetry
(it cannot, since there are no world-sheet fermions).  The theory does,
however, have a novel type of supersymmetry known as a
$\kappa$-symmetry.  The existence of the $\kappa$-symmetry in the
classical Green-Schwarz string theory restricts the theory to
space-time dimension $D = 3, 4, 6$ or 10.  No such restriction occurs
for the classical superstring with
world-sheet supersymmetry.

Unlike the superstring, there is no known way of formulating the
supermembrane in a world-volume supersymmetric fashion (although see
Duff (1996) for references to some recent progress in this direction).
A $\kappa$-symmetric formulation of the supermembrane in a general
background was first found by Bergshoeff, Sezgin, and Townsend (1988).
An interesting feature of the Green-Schwarz actions for the string and
membrane is that $\kappa$-symmetry on the string/membrane world-volume
is only possible when the background fields satisfy the supergravity
equations of motion.  Thus, 11D supergravity emerges from the membrane
theory even at the classical level.  The $\kappa$-symmetry of the
membrane can be gauge-fixed, reducing the number of propagating
fermionic degrees of freedom to 8.  This is also the number of
propagating bosonic degrees of freedom, as can be seen by going to a
static gauge the membrane theory where $X^{0, 1, 2}$ are identified
with $\tau, \sigma_{1, 2}$ so that only the 8 transverse directions
appear as propagating degrees of freedom.
In general, gauge-fixing the $\kappa$-symmetry in any particular way
will break the Lorentz invariance of the theory.  This makes it quite
difficult to find any way of quantizing the theory without breaking
Lorentz symmetry.  This situation is again analogous to the
Green-Schwarz superstring theory, where fixing of $\kappa$-symmetry
also breaks Lorentz invariance and no covariant quantization scheme is
known.

Beginning with the general supermembrane action, specializing to flat
space-time, fixing light-cone coordinates $X^+ = \tau$, and gauge
fixing $\kappa$-symmetry through $\Gamma^+ \theta = 0$,
the light-front supermembrane Hamiltonian
becomes
\begin{equation}
H = {\nu T \over 4} \int d^2 \sigma \left( \dot{X}^i \dot{X}^i
+ {2 \over \nu^2} \{X^i,X^j\}\{X^i,X^j\} -\frac{2}{\nu}  
\theta^T \gamma_i\{X^i, \theta\}
\right)
\end{equation}
where $\theta$ is a 16-component Majorana spinor of $SO(9)$
de Wit, Hoppe, and Nicolai (1988).
It is straightforward to apply the matrix regularization procedure
discussed in Section \ref{sec:matrix-regularization} to this
Hamiltonian.  This gives the supersymmetric form of matrix theory
\begin{equation}
H = {1 \over  (2 \pi l_p^3)} \tr \left( \half \dot{{\bf X}}^i \dot{{\bf
X}}^i - {1 \over 4} [{\bf X}^i,{\bf X}^j] [{\bf X}^i,{\bf X}^j] +
\half
{\bf \theta}^T \gamma_i[ {\bf X}^i, {\bf \theta}]
\label{eq:matrix-Hamiltonian-2} 
\right)\,.  \end{equation}

\subsection{Covariant membrane quantization}
\label{sec:covariant-membrane}

It is natural to think of generalizing the matrix regularization
approach to the covariant formulation of the bosonic and
supersymmetric membrane theories.  For the
bosonic membrane it is straightforward to implement the matrix
regularization procedure.  The only difficulty is that the BRST charge
needed to implement the gauge-fixing procedure cannot be simply
expressed in terms of the Poisson bracket on the membrane (Fujikawa
and Okuyama,  1997).  For the
supermembrane, there is a more serious complication related to the
$\kappa$-symmetry of the theory.  As mentioned above, any
gauge-fixing of the $\kappa$-symmetry will break the eleven-dimensional
Lorentz invariance of the theory.  This is the same difficulty as one
encounters when trying to construct a covariant quantization of the
Green-Schwarz superstring.  Fujikawa and Okuyama (1998) considered
the possibility of fixing the $\kappa$-symmetry in a way which
breaks the 32 of SO(10, 1) into $16_R + 16_L$ of SO(9, 1).  Thus, they
found a matrix formulation of a theory with explicit SO(9, 1)
Lorentz symmetry.  Although this theory does not have the desired
complete SO(10, 1) Lorentz symmetry of M-theory, there are 
questions which might be addressed using this theory with limited Lorentz
invariance.  

Another approach to finding a covariant version of the
matrix membrane involves the quantization of the Nambu bracket.  The
Poisson bracket used to transform Eq.~(\ref{eq:membrane-action-h}) to
Eq.~(\ref{eq:covariant-action}) can be generalized to a higher-dimensional
algebraic structure known as the classical Nambu bracket (Nambu,
1973).  On a 3-manifold, the Nambu bracket is given by
\begin{equation}
\{f, g, h\} =  \epsilon^{\alpha \beta \gamma}
(\partial_\alpha f)  (\partial_\beta g)  (\partial_\gamma h)\,.
\end{equation}
The Nambu-Goto form of the membrane action
(\ref{eq:Nambu-Goto-membrane}) can be rewritten in terms of the
classical Nambu bracket as
\begin{equation}
S = -T  \int d^3 \sigma \sqrt{-\det h_{\alpha \beta}}
=-T  \int d^3 \sigma \sqrt{- \frac{1}{6} \{X^\mu, X^\nu, X^\lambda\}
\{X_\mu, X_\nu, X_\lambda\}}\,.
\end{equation}
If a finite matrix regularization of the Nambu bracket could be
constructed analogous to the usual quantization of the Poisson
bracket, it would lead to a matrix regularization of the covariant
membrane theory analogous to the light-cone theory we have been
discussing.  Some progress in this direction was made by Awata, Li,
Minic, and Yoneya (1999) and Minic (1999); the reader is referred to
these papers for further references on this interesting subject.
An alternative approach to a covariant matrix membrane theory was
described by Smolin (1998).

\section{The BFSS conjecture}
\label{sec:BFSS}


As we have already discussed, it has been known for over a decade that
the light-front supermembrane theory can be regularized and described
as a supersymmetric quantum mechanics theory.  At the time that this
theory was first developed, however, it was believed that the quantum
supermembrane theory suffered from instabilities which would make the
low-energy interpretation as a theory of quantized gravity impossible.
In 1996 supersymmetric matrix quantum mechanics was brought back into
currency as a candidate for a microscopic description of an
eleven-dimensional quantum mechanical theory containing gravity by
Banks, Fischler, Shenker, and Susskind (1997, henceforth ``BFSS'').
This suggestion, which quickly became known as the ``Matrix Theory
Conjecture'', was primarily motivated not by the quantum supermembrane
theory, but by considering the low-energy theory of a system of many
D0-branes as a partonic description of light-front M-theory.

In this section we discuss the apparent instability of the quantized
membrane theory and the BFSS conjecture.  We describe the membrane
instability in subsection \ref{sec:instability}.  We describe the BFSS
conjecture in subsection \ref{sec:BFSS-s}.  In subsection
\ref{sec:second-quantized} we describe the resolution of the apparent
instability of the membrane theory by an interpretation of matrix
theory in terms of a second-quantized theory of gravity.  Finally, in
subsection \ref{sec:proof} we review an argument due to Seiberg and
Sen which shows that matrix theory should be equivalent to a discrete
light-front quantization of M-theory, even at finite $N$, assuming
that M-theory and its compactification to type IIA string theory can
be defined in a consistent fashion.

\subsection{Membrane ``instability''}
\label{sec:instability}

When de Wit, Hoppe, and Nicolai (1988) showed that the regularized
supermembrane theory could be described in terms of supersymmetric
matrix quantum mechanics, the general hope of the community seems to
have been that the quantized supermembrane theory would have a
discrete spectrum of states.  In string theory the spectrum of states
in the Hilbert space of the string can be put into one-to-one
correspondence with elementary particle-like states in the target
space.  It is crucial for this interpretation that the massless
particle spectrum contains a graviton and that there is a mass gap
separating the massless states from massive excitations.  For the
supermembrane theory, however, the spectrum does not seem to have
these properties.  This can be seen in both the classical and quantum
membrane theories.

\begin{figure}[ht]
\begin{center}
\epsfig{figure=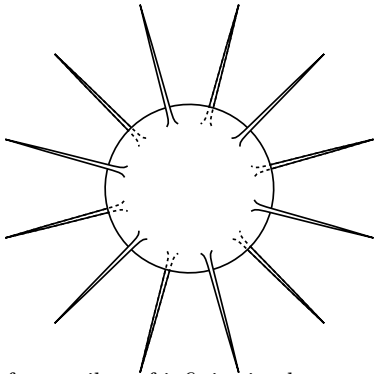, width=0.6\textwidth}
\caption{Classical membrane instability arises from spikes of
infinitesimal area\label{fig:spikes}}
\end{center}
\end{figure}

The simplest way to see the instability of the membrane theory at the
classical level is to consider a bosonic membrane whose energy is
given by the area of the membrane times a constant tension $T$.  Such
a membrane can have long narrow spikes at very low cost in energy (See
Fig.~\ref{fig:spikes}).  
If the spike is roughly cylindrical and has
a radius $r$ and length $L$ then the energy is $2 \pi rL T$.  For a
spike with very large $L$ but a small radius $r\ll 1/TL$ the energy
cost is small but the spike is very long.  This heuristic picture
indicates that a quantum membrane will tend to have many fluctuations
of this type, making it difficult to conceive of the membrane as an
object which is well localized in space-time.  Note that the quantum
string theory does not have this problem since a long spike in a
string always has energy proportional to the length of the string.
In the matrix regularized version of the membrane theory, this
instability appears as a set of flat directions in the classical
theory.  For example,
if we have a pair of $N = 2$ matrices with nonzero entries of the form
\begin{equation}
X^1 = \left(\begin{array}{cc}
x & 0\\0 & 0
\end{array}\right),
\;\;\;\;\;
X^2 = \left(\begin{array}{cc}
0 &  y\\ y & 0
\end{array}\right)\label{eq:}
\end{equation}
then a potential term ${\rm Tr}\;[X^1, X^2]^2$ corresponds to a term
proportional to $x^2 y^2$.  If either $x = 0$ or $y = 0$ then the
other (nonzero) variable is unconstrained, giving flat directions in
the moduli space of solutions to the classical equations of motion.
This corresponds classically to a marginal instability in the matrix
theory with $N > 1$.  (Note that in the previous section we
distinguished matrices ${\bf X}^i$ from related functions $X^i$ by
using bold font for matrices.  We will henceforth drop this font
distinction as long as the difference can easily be distinguished from
context.)

In the quantum bosonic membrane theory, the apparent instability from
the flat directions is cured because of the zero-modes of off-diagonal
degrees of freedom.  In the above example, for instance, if $x$ takes
a large value then $y$ corresponds to a harmonic oscillator degree of
freedom with a large mass.  The zero point energy of this oscillator
becomes larger as $x$ increases, giving an effective confining
potential which removes the flat directions of the classical theory.
This would seem to resolve the instability problem.  Indeed, in the
matrix regularized quantum bosonic membrane theory, there is a
discrete spectrum of energy levels for the system of  $N \times N$
matrices (Simon, 1983).

When we consider the supersymmetric theory, on the other hand, the
problem returns.  The zero point energies of the fermionic degrees of
freedom conspire to precisely cancel the zero point energies of the
bosonic oscillators.  This cancellation gives rise to a continuous
spectrum in the supersymmetric matrix theory.  This result was proven
by de Wit, L\"uscher, and Nicolai (1989).  They showed that for any
$\epsilon > 0$ and any energy $E\in [0, \infty)$ there exists a state
$\psi$ in the $N = 2$ maximally supersymmetric matrix model which is
normalizable ($\int | \psi |^2 = | | \psi | |^2 =1$) and which has
\begin{equation}
|| (H-E) \psi | |^2 < \epsilon.
\end{equation}
This implies that the spectrum of the supersymmetric matrix quantum
mechanics theory is continuous\footnote{Note that de Wit, L\"uscher
and Nicolai did not resolve the question of whether a state existed
with identically vanishing energy $H = 0$ (see section
\ref{sec:gravitons}).}.  This result
indicated that it would not be possible to have a simple
interpretation of the states of the theory in terms of a discrete
particle spectrum.  After this work there was little further
development on the supersymmetric matrix quantum mechanics theory as a
theory of membranes or gravity until almost a decade later.

\subsection{The BFSS conjecture}
\label{sec:BFSS-s}

Motivated by recent work on D-branes and string dualities, Banks,
Fischler, Shenker, and Susskind (1997) proposed that the large $N$
limit of the supersymmetric matrix quantum mechanics model described by
Eq.~(\ref{eq:matrix-Hamiltonian}) should describe all of M-theory in a
light-front coordinate system.  Although this conjecture fits neatly
into the framework of the quantized membrane theory, the starting
point of BFSS was to consider M-theory compactified on a circle $S^1$,
with a large momentum in the compact direction.  As discussed in
Section \ref{sec:dd}, when M-theory is compactified on $S^1$ the
resulting ten-dimensional theory is type IIA string theory.  The
quanta corresponding to momentum in the compact direction $x^{11}$ are
the D0-branes of the IIA theory.  In the ``infinite momentum frame''
of M-theory, where the momentum $p_{11}$ is taken to be very large,
the dynamics of the theory becomes nonrelativistic (Weinberg, 1966;
Kogut and Susskind, 1973).  BFSS argued that this dynamics should be
described by the large $N$ limit of a nonrelativistic system of
D0-branes.  

The low-energy Lagrangian for a system of $N$ type IIA
D0-branes is the matrix quantum mechanics Lagrangian arising from the
dimensional reduction to 0 + 1 dimensions of the 10D super Yang-Mills
Lagrangian (Witten, 1996; see Polchinski, 1996 or Taylor, 1998 for a
review)
\begin{equation}
{\cal L} = \frac{1}{2gl_s}  {\rm Tr}\; \left[
\dot{X}^a \dot{X}^a+\frac{1}{2}[X^a, X^b]^2 + \theta^T (i \dot{\theta}
-\Gamma_a[X^a, \theta]) \right]\,.
\label{eq:D0-Lagrangian}
\end{equation}
In this action the gauge has been fixed to $A_0 = 0$.  Just as in
Eq.~(\ref{eq:matrix-Hamiltonian-2}), $X^a$ are 9 $N \times N$ bosonic
matrices and $\theta$ are 16 Grassmann $N \times N$ matrices.  Using
the relations $R = g^{2/3}l_{11} = gl_s$ from
Eq.~(\ref{eq:coupling-relations}), we see that in string units ($2 \pi
l_s^2 = 1$) we can replace $gl_s = R = 2 \pi l_{11}^3$.  Thus, the
Hamiltonian associated with Eq.~(\ref{eq:D0-Lagrangian}) is in fact
precisely equivalent to the matrix membrane Hamiltonian
(\ref{eq:matrix-Hamiltonian-2}).  This connection and its possible
significance was first pointed out by Townsend (1996a).  The fact that
$l_{11}$ arises as the basic length scale in D0-brane quantum
mechanics was discussed by Kabat and Pouliot (1996) and Douglas,
Kabat, Pouliot, and Shenker (1997); this was an early indication that
D0-branes might play a fundamental role as constituents of M-theory
(see also Shenker, 1995, for a discussion of substring distance
scales).  The matrix theory Hamiltonian is often written, following
BFSS, in the form
\begin{eqnarray}
H & = &   \frac{R}{2}    \tr \left( P^i P^i
 - {1 \over  2} [X^i,X^j] [X^i,X^j] +
 \theta^T \gamma_i[X^i, \theta]\right)
\end{eqnarray}
where we have rescaled $X/g^{1/3} \rightarrow X$ and written the
Hamiltonian in Planck units $l_{11} = 1$.
It is this Hamiltonian which BFSS conjectured should correspond with
the infinite momentum limit of M-theory when $N \rightarrow \infty$.

The original BFSS conjecture was made in the context of the large $N$
theory.  It was later argued by Susskind (1997a) that the finite $N$
matrix quantum mechanics theory should be equivalent to the discrete
light-front quantized (DLCQ; see for example Pauli and Brodsky, 1985)
sector of M-theory with $N$ units of compact momentum.  We describe in
section (\ref{sec:proof}) below an argument due to Seiberg and Sen
which makes this connection more precise and which justifies the use
of the low-energy D0-brane action in the BFSS conjecture.

While the BFSS conjecture was based on a different philosophy from
that underlying the matrix quantization of the supermembrane theory we
have discussed above, the fact that the M-theory membrane can be
described as a classical configuration in the matrix quantum mechanics
theory was a substantial piece of additional evidence given by BFSS
for the validity of their conjecture.  Two additional pieces of
evidence were given by BFSS which extended their conjecture beyond the
previous work on the matrix membrane theory.

One important point made by BFSS is that the Hilbert space of the
matrix quantum mechanics theory naturally contains multiple particle
states.  This observation, which we discuss in more detail in the
following subsection, resolves the problem of the continuous spectrum
discussed above.  Another piece of evidence given by BFSS for their
conjecture is the fact that quantum effects in matrix theory give rise
to long-range interactions between a pair of gravitational quanta
(D0-branes).  These interactions have precisely the structure expected
from light-front supergravity.  This result was first shown for
D0-branes by a calculation of Douglas, Kabat, Pouliot, and Shenker
(1997); we will discuss this result and its generalization to
more general matrix theory interactions in Section
\ref{sec:matrix-interactions}.

\subsection{Matrix theory as a second-quantized theory}
\label{sec:second-quantized}

\begin{figure}[ht]
\begin{center}
\epsfig{figure=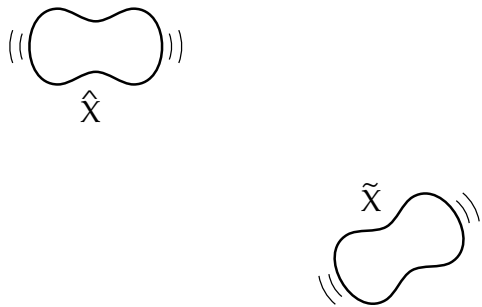, width=0.6\textwidth}
\caption{Two matrix theory objects described by block-diagonal
matrices\label{fig:multiple-objects}}
\end{center}
\end{figure}

The classical equations of motion for a bosonic matrix configuration
with the Hamiltonian (\ref{eq:matrix-Hamiltonian}) are (up to an
overall constant)
\begin{equation}
\ddot{X}^i = -[[X^i, X^j], X^j].
\label{eq:classical-eom}
\end{equation}
If we consider a block-diagonal set of matrices
\begin{equation}
X^i = \left(\begin{array}{cc}
\hat{X}^i & 0\\
0 & \tilde{X}^i
\end{array} \right)
\end{equation}
with first time derivatives $\dot{X}^i$ which are also of
block-diagonal form, then the classical equations of motion for the
blocks are separable
\begin{equation}
\ddot{\hat{X}}^i  =  -[[\hat{X}^i, \hat{X}^j], \hat{X}^j], \;\;\;\;\; \;\;\;\;\;
\ddot{\tilde{X}}^i  =  -[[\tilde{X}^i, \tilde{X}^j], \tilde{X}^j]\, .
\end{equation}
If we think of these blocks as describing two matrix theory
objects with centers of mass
\begin{equation}
 \hat{x}^i = \frac{1}{ \hat{N}}  {\rm Tr}\; \hat{X}^i,
\;\;\;\;\; \;\;\;\;\;
 \tilde{x}^i = \frac{1}{ \tilde{N}}  {\rm Tr}\; \tilde{X}^i,
\end{equation}
then we have two objects obeying classically independent equations of
motion (See Fig.~\ref{fig:multiple-objects}).  
It is straightforward
to generalize this construction to a block-diagonal matrix
configuration describing $k$ classically independent objects.  This
gives a simple indication of how matrix theory can encode, even in
finite $N$ matrices, a configuration of multiple objects.  In this
sense it is natural to think of matrix theory as a second-quantized
theory from the point of view of the target space.

Given the realization that matrix theory should describe a second
quantized theory, the puzzle discussed above regarding the continuous
spectrum of the theory is easily resolved.  Assume that there is a
state in matrix theory corresponding to a single graviton of M-theory
with $H = 0$, which is roughly a localized state (we will discuss such
states in more detail in section \ref{sec:gravitons}).  By taking two
of these gravitons to have a large separation and a small relative
velocity $v$ it should be possible to construct a two-body state with
an arbitrarily small total energy using block diagonal matrices.
Since the D0-branes of the IIA theory correspond to gravitons in
M-theory with a single unit of longitudinal momentum, we therefore
naturally expect to find a continuous spectrum of energies even in the
theory with $N = 2$.  This resolves the puzzle found by de Wit,
L\"uscher, and Nicolai in a very pleasing fashion, and suggests that
matrix theory is perhaps even more powerful than perturbative string
theory, which only gives a first-quantized theory in the target space.

\begin{figure}[ht]
\begin{center}
\epsfig{figure=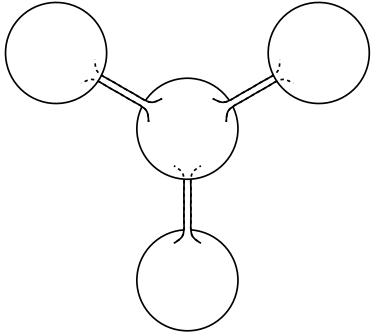, width=0.6\textwidth}
\caption{Membrane of fixed (spherical) topology mapped to multiple
membranes  connected by
tubes in the target space\label{fig:multiple-membranes}}
\end{center}
\end{figure}

The second-quantized nature of matrix theory can also be seen
heuristically in the continuous membrane theory.  Recall that the
instability of membrane theory appears in the classical theory of a
continuous membrane when we consider the possibility of long thin
spikes of negligible energy, as discussed in section
\ref{sec:instability}.  In a similar fashion, it is possible for a
classical smooth membrane of fixed topology to be mapped to a
configuration in the target space which looks like a system of
multiple distinct macroscopic membranes connected by infinitesimal
tubes of negligible energy (See Fig.~\ref{fig:multiple-membranes}).
In the limit where the tubes become very small, their effect on the
classical dynamics of the multiple membrane configuration becomes
negligible, and we effectively have a system of multiple independent
membranes moving in the target space.  At the classical level, the sum
of the genera of the membranes in the target space must be equal to or
smaller than the genus of the single world-sheet membrane, but when
quantum effects are included handles can be added to the membrane as
well as removed.  These considerations seem to indicate that any
consistent quantum theory which contains a continuous membrane in its
effective low-energy theory must contain configurations with arbitrary
membrane topology and must therefore be a ``second-quantized'' theory
from the point of view of the target space.

\subsection{Matrix theory and DLCQ M-theory}
\label{sec:proof}

A theory which has been compactified on a lightlike circle can be
viewed as a limit of a theory compactified on a spacelike circle where
the size of the spacelike circle becomes vanishingly small in the
limit.  This point of view was used by Seiberg (1997b) and Sen (1998)
to argue that light-front compactified M-theory is described through
such a limiting process by the low-energy Lagrangian for many
D0-branes, and hence by matrix theory.  In this section we review
this argument in detail.  Other perspectives on the DLCQ limit are
given in Balasubramanian, Gopakumar, and Larsen (1998) and de Alwis
(1999).  A nice synthesis of the various approaches to the matrix theory
limit is given in Polchinski (1999).

Consider a space-time which has been compactified on a lightlike
circle by identifying
\begin{equation}
\left(\begin{array}{c}
x\\
t
\end{array} \right) \sim
\left(\begin{array}{c}
x-R/\sqrt{2}\\
t+ R/\sqrt{2}
\end{array} \right) \, .
\label{eq:light-front-compact}
\end{equation}
This theory has a quantized momentum in the compact direction
$
P^+ = N/R
$. The compactification (\ref{eq:light-front-compact}) can be described
as a limit of a family of spacelike compactifications
\begin{equation}
\left(\begin{array}{c}
x\\
t
\end{array} \right) \sim
\left(\begin{array}{c}
x-\sqrt{R^2/2 + R_s^2}\\
t+ R/\sqrt{2}
\end{array} \right) 
\label{eq:spacelike}
\end{equation}
parameterized by the size $R_s \rightarrow 0$ of the spacelike
circle, which is taken to vanish in the limit.

The system satisfying Eq.~(\ref{eq:spacelike}) is related 
to a system with the identification
\begin{equation}
\left(\begin{array}{c}
x'\\
t'
\end{array} \right) \sim
\left(\begin{array}{c}
x'- R_s\\
t'
\end{array} \right) 
\label{eq:space-compact}
\end{equation}
through a boost
with boost parameter $\beta$ given by
\begin{equation}
\beta = \frac{1}{ \sqrt{1 + \frac{2 R_s^2}{R^2} }}  \equiv
1-\frac{R_s^2}{ R^2} .
\end{equation}

We are interested in compactifying M-theory on a lightlike circle.
This is related through the above limiting process to a family of
spacelike compactifications of M-theory, which we know can be
identified with the IIA string theory.  At first glance, it may seem
that the limit we are considering here is difficult to analyze from
the IIA point of view.  The IIA string coupling and string length are
related to the compactification radius and 11D Planck length as in
Eq.~(\ref{eq:coupling-relations}) by
\begin{eqnarray}
g=  (\frac{R_s}{l_{11}} )^{3/2}, & \;\;\;\;\; \;\;\;\;\; \;\;\;\;\; &
l_s^2  \frac{l_{11}^3}{R_s} \, . \nonumber
\end{eqnarray}
Thus, in the limit $R_s \rightarrow 0$ the string coupling $g$ becomes small as
desired; the string length $l_s$, however, goes to $\infty$.  Since
$l_s^2 = \alpha'$, this corresponds to a limit of vanishing string
tension.  Such a limiting theory is very complicated and would not
seem to provide a useful alternative description of the theory.

Let us consider, however, how the energy of the states we are
interested in behaves in the class of limiting theories with
spacelike compactification.  If we want to describe the behavior of a
state which has light-front energy $P^-$ and compact momentum 
$P^+ = N/R$ then the spatial momentum in the theory with spatial $R_s$
compactification is $P' = N/R_s$.  The energy in the spatially
compactified theory is 
\begin{equation}
E' = N/R_s + \Delta E,
\label{eq:compact-energy}
\end{equation}
where $\Delta E$  is at the energy scale we are interested in
understanding.  The term $N/R_s$ in the energy is simply the
mass-energy of the $N$ D0-branes which correspond to the momentum in
the compactified M-theory direction.  Relating back to the near
lightlike compactified theory we have
\begin{equation}
P^-=\frac{1}{ \sqrt{2}}  (E-P) =
 \frac{1}{ \sqrt{2}}  \frac{1+\beta}{ \sqrt{1-\beta^2}}  \Delta E
 \approx \frac{R}{R_s}  \Delta E\, .
\end{equation}
As a result we see that the energy $ \Delta E$ of the IIA
configuration needed to approximate the light-front energy $P^-$
is given by
$
 \Delta E \approx P^-R_s/R
$. We know that the string mass scale $1/l_s$ becomes small as $R_s
\rightarrow 0$.  We can compare the energy scale of interest to this
string mass scale, however, and find
\begin{equation}
\frac{\Delta E}{(1/l_s)}  = \frac{P^-}{R}  R_sl_s
=\frac{P^-}{R}  \sqrt{R_sl_{11}^3}\, .
\end{equation}
This ratio vanishes in the limit $R_s \rightarrow 0$, which implies
that although the string scale vanishes, the energy scale of interest
is smaller still.  Thus, it is reasonable to study the lightlike
compactification through a limit of spatial compactifications in this
fashion.

To make the correspondence between the light-front compactified theory
and the spatially compactified limiting theories more transparent,
we perform a change of units to a new Planck length
$\tilde{l}_{11}$ in the spatially compactified theories in such a way
that the energy of the states of interest is independent of $R_s$.
For this condition to hold we must have
\begin{equation}
\Delta E\, \tilde{l}_{11}  = P^-\frac{R_s {l}_{11}^2}{ R \,\tilde{l}_{11}} 
\end{equation}
where $E, R$ and $P^-$ are independent of $R_s$ and all units have
been explicitly included.  This requires us to keep the quantity
$
R_s/\tilde{l}_{11}^2
$ fixed in the limiting process.  Thus, in the limit $\tilde{l}_{11}
\rightarrow 0$.

We can summarize the preceding discussion as follows: to
describe the sector of M-theory corresponding to light-front
compactification on a circle of radius $R$ with light-front momentum
$P^+ = N/R$ we may consider the limit $R_s \rightarrow 0$
of a family of IIA
configurations with $N$ D0-branes where the string coupling and string
length
\begin{eqnarray}
\tilde{g}  =  (R_s/\tilde{l}_{11})^{3/2}\rightarrow 0 ,
& \;\;\;\;\; \;\;\;\;\; &
\tilde{l}_s  =   \sqrt{ \tilde{l}_{11}^3/R_s} \rightarrow 0
\end{eqnarray}
are defined in terms of a Planck length $\tilde{l}_{11}$ and
compactification length $R_s$ which satisfy
$
R_s / \tilde{l}_{11}^2 = R/l_{11}^2
$. All transverse directions scale normally through
$
\tilde{x}^i/\tilde{l}_{11} = x^i/l_{11}
$.

To give a very concrete example of how this limiting process works,
let us consider a system with a single unit of longitudinal momentum 
$
P^+ =  1/R
$. We know that in the corresponding IIA theory, we
have a single D0-brane whose Lagrangian has the relativistic Born-Infeld form
\begin{equation}
{\cal L} = -\frac{1}{\tilde{g} \tilde{l}_s} 
\sqrt{1-\dot{\tilde{x}}^i \dot{\tilde{x}}^i}\, .
\end{equation}
Expanding the square root we have
\begin{equation}
{\cal L} = -\frac{1}{\tilde{g} \tilde{l}_s} 
\left( 1
-\frac{1}{2}\dot{\tilde{x}}^i \dot{\tilde{x}}^i +{\cal O}
(\dot{\tilde{x}}^4)
\right).
\end{equation}
Replacing $\tilde{g} \tilde{l}_s \rightarrow R_s$ and
$\tilde{x}\rightarrow x \tilde{l}_{11}/l_{11}$ gives
\begin{equation}
{\cal L} = -\frac{1}{R_s}  + \frac{1}{2 R}  \dot{x}^i \dot{x}^i +{\cal
O} (R_s/R).
\end{equation}
Thus, we see that all the higher order terms in the Born-Infeld action
vanish in the $R_s \rightarrow 0$ limit.  The leading term is the
D0-brane energy $1/R_s$ which we subtract to compare to the M-theory
light-front energy $P^-$.  Although we do not know the full form of
the nonabelian Born-Infeld action describing $N$ D0-branes in IIA, it
is clear that an analogous argument shows that all terms in this
action other than those in the nonrelativistic supersymmetric matrix
theory action (\ref{eq:D0-Lagrangian}) will vanish in the limit $R_s
\rightarrow 0$.

This argument apparently demonstrates that matrix theory gives a
complete description of the dynamics of DLCQ M-theory.  There are
several caveats which should be taken into account, however, with
respect to this discussion.  First, in order for this argument to be
correct, it is necessary that there exists a well-defined theory with
the properties expected of M-theory, and that there exist a
well-defined IIA string theory which arises as the compactification of
M-theory.  Neither of these statements is at this point definitely
established.  Thus, this argument must be taken as contingent upon the
definition of these theories.  Second, although we know that 11D
supergravity arises as the low-energy limit of M-theory, this argument
does not necessarily indicate that matrix theory describes DLCQ
supergravity in the low-energy limit.  It may be that to make the
connection to supergravity it is necessary to deal with subtleties of
the large $N$ limit.  

While there are reasons to suspect that care must be taken with this
argument when $N$ becomes large, there are also some aspects of the
story which become much clearer at large $N$.  For large $N$, the rest
energy $N/R_s$ from Eq.~(\ref{eq:compact-energy}) becomes very large,
and modifies the flat space-time geometry we have assumed in this
discussion.  As discussed in Hyun, Kiem and Shin (1998),
Balasubramanian, Gopakumar, and Larsen (1998), Itzhaki, Maldacena,
Sonnenschein, and Yankielowicz (1998), and Polchinski (1999), this
back-reaction produces a ``bubble'' of eleven-dimensional space-time
in the vicinity of the 0-branes, which in the large $N$ limit
decompactifies the space-time.  From the ten-dimensional point of
view, this provides an explanation of how eleven-dimensional physics
can be described by weakly coupled string theory in ten dimensions.
The local gravitational effects of
the D0-branes also have the effect of making the periodic dimension
spacelike rather than lightlike, so that the problems of zero-modes
usually associated with light-front field theories are avoided.

In the following sections we will discuss some more explicit
approaches to connecting matrix theory with supergravity.  In
particular, we will see how far it is possible to go in demonstrating
that 11D supergravity arises from calculations in the finite $N$
version of matrix theory, which is a completely well-defined theory.

\vspace{0.2in}

\section{Interactions in matrix theory}
\label{sec:matrix-interactions}

As we discussed in section \ref{sec:second-quantized}, a many-body
system is described in matrix theory by a set of block-diagonal
matrices.  Classically, the blocks describing each object evolve
independently, so that there are no classical interactions in matrix
theory between separated objects.  How, then, can matrix theory be
said to describe even classical gravitational interactions?

The answer to this question is quite remarkable, and is one of the
most important features of this theory.  It turns out that {\it
classical gravitational interactions arise through quantum loop
effects in matrix theory}.  The first example of this
classical-quantum correspondence was used as a piece of evidence for
the validity of the 1996 matrix theory conjecture by BFSS.  These
authors pointed out that earlier work on D0-brane scattering in type
IIA string theory by Douglas, Kabat, Pouliot, and Shenker (1997) gave
a leading-order interaction between a pair of individual D0-branes
which agrees precisely with the interaction between a pair of
gravitons in eleven-dimensional supergravity according to the
conjectured matrix theory correspondence.  The interaction between the
D0-branes in this calculation arises from a one-loop quantum
mechanical calculation, while the leading interaction between
gravitons in eleven dimensions is purely a classical effect arising
from the linearized gravity theory.  It was shown by Paban, Sethi, and
Stern (1998a) that the leading one-loop interaction term in matrix
theory is exact and is protected by a supersymmetric
nonrenormalization theorem.  A power counting argument (Becker,
Becker, Polchinski, and Tseytlin, 1997) indicates that $n$th order
nonlinear gravitational effects require $n$-loop interactions in the
matrix quantum mechanics.  This leads to the hypothesis that perhaps
all classical gravitational interactions can be reproduced by
perturbative calculations in matrix theory.  At the linearized level
this statement seems to be correct.  It was shown by Kabat and Taylor
(1998b) that all linearized gravitational interactions between a pair
of bosonic sources can be reproduced by a general one-loop matrix
theory calculation.  Taylor and Van Raamsdonk (1999a) generalized this
result to quadratic order in fermions.  Beyond the linearized theory,
however, this hypothesis is less strongly supported.  There is
evidence that some simple nonlinear gravitational interactions are
correctly reproduced by perturbative calculations in matrix theory,
including an impressive demonstration of agreement between
three-graviton interactions and a two-loop calculation in $N = 3$
matrix theory by Okawa and Yoneya (1999a, 1999b).  It has also been
argued by Dine, Echols, and Gray (2000), however, that terms arise in
a three-loop matrix theory calculation which cannot correspond to
third order gravitational effects.  While some of the one-loop and
two-loop interaction terms are protected by supersymmetric
nonrenormalization theorems (Paban, Sethi, and Stern, 1998a, 1998b),
there is no evidence that higher-loop effects are similarly
constrained.  Thus, even if the matrix theory conjecture is correct,
it may not be possible to directly demonstrate the correspondence with
supergravity by perturbative finite $N$ calculations in matrix quantum
mechanics.  In this section we describe the perturbative matrix theory
calculations just discussed and the correspondence with supergravity
interactions in some detail, as well as discussing the known
supersymmetric nonrenormalization theorems and their consequences.

In subsection \ref{sec:two-body} we discuss perturbative calculations of
two-body interactions in matrix theory.  We begin by reviewing the
perturbative Yang-Mills formalism in background field gauge, which can
be used to carry out loop calculations in matrix theory.  We describe
in detail the one-loop calculation for a pair of D0-branes with
relative velocity $v$.  We then summarize the results of the one-loop
calculation for a general two-body interaction and show that these
interaction terms can be described by a sum of linearized supergravity
interactions arising from the exchange of a single graviton, 3-form
quantum or gravitino.  We review results on spin effects and higher
order terms for interactions between a pair of matrix theory
gravitons, and we discuss nonrenormalization theorems and their
implications for two-graviton interactions.  In subsection
\ref{sec:N-body} we discuss interactions between more than two
objects.  We discuss the N-body problem in general, and we review
positive results for 3-graviton scattering and negative results for
scattering of 4 or more gravitons.  Section \ref{sec:longitudinal}
contains a brief discussion of interactions involving longitudinal
momentum transfer, which correspond to nonperturbative processes in
matrix theory.  In Subsection \ref{sec:interaction-summary} we review
the status of the correspondence between matrix theory and
supergravity, and discuss possible subtleties in the large $N$ limit.
We also briefly discuss work on reproducing quantum corrections to
supergravity interactions from the matrix theory point of view.
All the analysis in this section deals with matrix theory in a flat
space-time background.  Some progress towards understanding matrix
theory in a curved space-time background is reviewed in Section
\ref{sec:curved-background}.

\subsection{Two-body interactions}
\label{sec:two-body}

\subsubsection{The background field formalism}
\label{sec:background-field}

In this subsection we review the background formalism for
Yang-Mills theory in the context of the (0+1)-dimensional matrix
quantum mechanics theory.  For a more complete introduction to the
background field method, see for example Abbott (1981, 1982).
Starting from the dimensionally reduced Yang-Mills action describing a
system of  $N$ D0-branes (without gauge-fixing $A$),
the
matrix theory Lagrangian is
\begin{equation}
{\cal L}= \frac{1}{2 R}  {\rm Tr}\; \left[
D_0 X^i\;D_0 X^i
+\frac{1}{2} [X^i, X^j]^2+  \theta^T (i\dot{\theta}
- \gamma_i[X^i, \theta]) \right]
\label{eq:matrix-Lagrangian}
\end{equation}
where
\begin{equation}
D_0 X^i = \partial_t X^i-i[A, X^i].
\end{equation}

We wish to expand each of the matrix theory fields around a classical
background.  We will assume here for simplicity that the background
has a vanishing gauge field and vanishing fermionic fields.  The
general situation with background fermionic fields as well as bosonic
fields is described in Taylor and Van Raamsdonk (1999a).  We expand the
bosonic fields in terms of a background plus a fluctuation
\begin{eqnarray*}
X^i & = &  B^i + Y^i.
\end{eqnarray*}
We choose the background field gauge
\begin{equation}
D^{{\rm bg}}_{\mu} A^\mu =
\partial_tA-i[B^i, X^i] = 0.
\end{equation}
This gauge can be implemented by adding a term $-(D^{{\rm bg}}_{\mu}
A^\mu )^2$ to the action and including the appropriate ghosts.  The
nice feature of this gauge is that the terms quadratic in the bosonic
fluctuations simplify considerably.

The complete gauge-fixed action including ghosts is written in
Euclidean time $\tau = it$ as
\begin{equation}
S = S_0 + S_2 + S_3 + S_4
\end{equation}
where
\begin{eqnarray}
S_0 & = &\frac{1}{2 R}   \int d \tau {\rm Tr}\;
 \left[ \partial_\tau B^i \partial_\tau B^i+\frac{1}{2}[B^i, B^j]^2
\right]\nonumber
\\
S_2 & = &\frac{1}{2 R}   \int d \tau {\rm Tr}\;
 \left[ \partial_\tau Y^i \partial_\tau Y^i
-[B^i, Y^j][B^i, Y^j]  -[B^i, B^j][Y^i, Y^j]\right.
\nonumber\\
& &\left.
\hspace{0.8in}
+ \partial_\tau A \partial_\tau A
 -[B^i, A][B^i, A]-2i \dot{B}^i[A, Y^i]\right.
\label{eq:s-2}\\
& &\left.
\hspace{0.8in}
 + \partial_\tau \bar{C} \partial_\tau C
  -[B^i,\bar{C}][B^i, C] +
 \theta^T \dot{\theta} -\theta^T \gamma_i[B^i, \theta]
\right] \nonumber
\end{eqnarray}
and where $S_3$ and $S_4$ contain terms cubic and quartic in the
fluctuations $Y^i, A, C, \theta$.  These interaction terms are given
explicitly in Becker and Becker (1997).  Note that we have taken $A
\rightarrow -iA$ as appropriate for the Euclidean formulation.

This gauge-fixed action can be used to perturbatively compute the
effective action governing the interaction between any set of matrix
theory objects.  This effective action in turn determines the
scattering phase shift of the objects in the eikonal approximation.
In general, to calculate the effective interaction potential to
arbitrary order it is necessary to include the terms $S_3$ and $S_4$
in the action.  The propagators for each of the fields can be computed
from the quadratic term $S_2$.  A systematic diagrammatic expansion
then yields the effective potential to arbitrarily high order.
The only calculations which we describe in detail here are one-loop
terms in the effective potential, for which the quadratic action $S =
S_0 + S_2$ is sufficient.

\subsubsection{Two-graviton interactions at leading order}
\label{sec:2-graviton}

According to the BFSS conjecture, a $1 \times 1$ matrix describing a
single D0-brane in type IIA string theory corresponds to a graviton of
M-theory with longitudinal momentum $p^+ = 1/N$.  As we will discuss
in further detail in section \ref{sec:gravitons}, at distances large
compared to the size of the wavefunction describing the single
graviton, it is sufficient to use a classical approximation $X^i = a^i
+ v^i t$ for the bosonic fields describing a single D0-brane moving
along a linear trajectory in transverse 9-dimensional space with
velocity $v^i$ and position $a^i$ at time $t = 0$.  Given this
interpretation of a classical $1 \times 1$ matrix, and the many-body
interpretation of block-diagonal matrices described in
\ref{sec:second-quantized}, a classical background describing a pair
of gravitons with relative velocity $v$ and impact parameter $b$ is
given in the center of mass frame in the Euclidean theory by
\begin{eqnarray}
B^1  =  \frac{-i}{2}   \left(\begin{array}{cc}
v\tau & 0\\
0 & -v\tau
\end{array} \right), \;\;\;\;\;
B^2  =  \frac{1}{2}   \left(\begin{array}{cc}
b & 0\\
0 & - b
\end{array} \right), \;\;\;\;\;
B^i  =   0, i > 2\, .
\label{eq:3-background}
\end{eqnarray}
In assuming this classical background we have ignored polarization
effects, which are discussed in subsections
\ref{sec:2-graviton-general} and \ref{sec:quantum-gravitons}.
Following Douglas, Kabat, Pouliot, and Shenker (1997), we can use the
matrices (\ref{eq:3-background}) as a background and perform a
one-loop calculation to find the leading long-range interaction
between these two matrix theory gravitons.  Related earlier
calculations were performed by Bachas (1996) and Lifschytz (1996).

Inserting the background (\ref{eq:3-background}) into
Eq.~(\ref{eq:s-2}) we see that at a fixed value of time the Lagrangian
at quadratic order for the 10 complex bosonic off-diagonal components
of $A$ and $Y^i$ is that of a system of 10 harmonic oscillators with
mass matrix
\begin{equation}
(\Omega_b)^2 = \left(\begin{array}{ccccc}
r^2 & -2 iv& 0 & \cdots & 0\\
2iv & r^2 & 0 & \ddots & 0\\
0 & 0 & r^2 & \ddots & \vdots\\
\vdots & \ddots &  \ddots & \ddots & 0\\
0 & 0 & \cdots & 0 & r^2
\end{array} \right)
\end{equation}
where $r^2 = b^2 + (vt)^2$ is the instantaneous separation between the
gravitons.
There are two complex off-diagonal ghosts with $\Omega^2 = r^2$.
There are 16 fermionic oscillators with a mass-squared matrix 
\begin{equation}
(\Omega_f)^2 = r^2 \identity_{16 \times 16} + v \gamma_1\, .
\end{equation}

To perform a completely general calculation of the two-body
effective interaction potential to all orders in $1/r$ it is necessary
to perform a diagrammatic expansion using the exact propagators for the
bosonic and fermionic fields.  For example, the bosonic propagator
satisfying
\begin{equation}
(-\partial^2 + b^2 + v^2 \tau^2) \Delta_B
(\tau, \tau' | b^2 + v^2 \tau^2) = \delta (\tau -\tau')
\end{equation}
is given by the expression (Becker and Becker, 1997)
\begin{eqnarray}
\Delta_B
(\tau, \tau' | b^2 + v^2 \tau^2)  & = & 
\int_0^\infty ds \; e^{-b^2 s} \sqrt{\frac{v}{2 \pi \sinh 2sv} }\,
\exp \left( -\frac{v}{2 \sinh 2sv}((\tau^2 + \tau'^2) \cosh 2sv-2 \tau
\tau') \right).\label{eq:exact-propagator}
\end{eqnarray}
In general, even for a simple 2-graviton calculation there is a fair
amount of algebra involved in extracting the effective potential using
propagators of the form (\ref{eq:exact-propagator}).  If,
however, we are only interested in calculating the leading term in the
long-range interaction potential we can simplify the calculation by
making the quasi-static assumption\footnote{The
validity of this approximation is discussed in Tafjord and Periwal
(1998).} that all the oscillator frequencies
$\omega$ of interest are large compared to the ratio $v/r$.
In this approximation, all the
oscillators stay in their ground state over the time of the
interaction, so that the effective potential between the two objects
is simply given by the sum of the ground-state energies of the boson,
ghost and fermion oscillators
\begin{equation}
V_{ {\rm qs}} = \sum_{b}\omega_b-\sum_{g}  \omega_g -\frac{1}{2}
\sum_{f} \omega_f.
 \label{eq:oscillator-sum}
\end{equation}
(Note that the bosonic and ghost oscillators are complex so that no
factor of 1/2 is included.)

In the situation of two-graviton scattering we can therefore calculate
the effective potential by diagonalizing the frequency matrices
$\Omega_b, \Omega_g$, and $\Omega_f$.  We find that 
the bosonic oscillators have frequencies
\begin{eqnarray*}
\omega_b & = &   r \;\;\;\;\; {\rm with\ multiplicity\ 8}\\
\omega_b & = &  \sqrt{r^2 \pm 2v} \;\;\;\;\;
{\rm with\ multiplicity\ 1\ each}.
\end{eqnarray*}
The 2 ghosts have frequencies 
$
\omega_g = r,
$ and the 16 fermions have frequencies
\begin{equation}
\omega_f = \sqrt{r^2 \pm v} \;\;\;\;\;
{\rm with\ multiplicity\ 8\ each}.
\end{equation}

The effective potential for a two-graviton system with instantaneous
relative velocity $v$ and separation $r$ is thus given by the leading
term in a $1/r$ expansion of the expression
\begin{equation}
V  = \sqrt{r^2 + 2v} + \sqrt{r^2 -2v} + 6r
-4 \sqrt{r^2 + v} + 4 \sqrt{r^2 -v}.
\end{equation}
Expanding in $v/r^2$ we see that the terms of order $r, v/r, v^2/r^3$
and $v^3/r^5$ all cancel.  The leading term is
\begin{equation}
V = -\frac{15}{16}  \frac{v^4}{r^7}  +{\cal O} (\frac{v^6}{r^{11}} )\, .
\end{equation}

As mentioned above, this result agrees with the leading term in the
effective potential between two gravitons with $P^+ = 1/R$ in
light-front 11D supergravity.  We will discuss the supergravity side
of this calculation in more detail in the following section.

\subsubsection{General two-body systems and linearized supergravity
at leading order}
\label{sec:two-body-linear}

We now generalize the background to include an arbitrary pair of bosonic
matrix theory objects,  described by block-diagonal matrices
\begin{equation}
B^i = \left(\begin{array}{cc}
\hat{X}^i & 0\\
0 & \tilde{X}^i
\end{array}\right)
 \label{eq:general-two-body-background}
\end{equation}
where $\hat{X}^i$ and $\tilde{X}^i$ are $\hat{N} \times \hat{N}$ and
$\tilde{N} \times \tilde{N}$ matrices.  The separation distance between the objects, which we
will use as an expansion parameter, is given by
\begin{equation}
r^i = \frac{1}{ \hat{N}}  {\rm Tr}\; \hat{X}^i
-\frac{1}{ \tilde{N}}  {\rm Tr}\; \tilde{X}^i\, .
\end{equation}

To compute the leading term in the interaction potential, following
Kabat and Taylor (1998b), we insert
Eq.~(\ref{eq:general-two-body-background}) into Eq.~(\ref{eq:s-2}) and,
as in the simpler two-graviton example, compute the frequency matrices
for the bosons, ghosts, and fermions.  We summarize here the results
of this calculation.  Expanding the frequency matrices as before in
powers of $1/r$, and using Eq.~(\ref{eq:oscillator-sum}), for a
completely arbitrary pair of objects the potential again vanishes to
order $1/r^7$.  At this order the potential is
\begin{eqnarray}
V_{\rm leading}& = & \tr \left(\Omega_b\right) - \half \tr
\left(\Omega_f\right) - 2 \tr \left(\Omega_g\right)
=-\frac{5}{128r^7} {\rm STr}\;{\cal F}
\end{eqnarray}
where
\begin{equation}
{\cal F} = 
    24 F^\mu{}_\nu F^\nu{}_\lambda F^\lambda{}_\sigma F^\sigma{}_\mu
-  6  F_{\mu\nu} F^{\mu \nu} F_{\lambda \sigma} F^{\lambda \sigma}
\end{equation}
and ${\rm STr}$ indicates that the trace is symmetrized over all
possible orderings of  the $F$'s.  The field strength
$F_{\mu \nu}$ is a linear combination of contributions from each of
the two objects
\begin{equation}
F_{\mu \nu} = \hat{F}_{\mu \nu} - \tilde{F}_{\mu \nu},
 \label{eq:fff}
\end{equation}
where $\hat{F}_{\mu \nu}$ and $\tilde{F}_{\mu  \nu}$ are defined
through
\begin{eqnarray}
F_{0i} =  \dot{X}^i, \;\;\;\;\; \;\;\;\;\;
F_{ij} =  i[X^i, X^j] \label{eq:f-definition}
\end{eqnarray}
in terms of $\hat{X}$ and
$\tilde{X}$ respectively.  

Because of the linear structure of Eq.~(\ref{eq:fff}), it is possible to
decompose the potential $V_{{\rm leading}}$ into a sum of terms which
are written as products of a function of $\hat{X}$ and a function of
$\tilde{X}$, where the terms can be grouped according to the number of
Lorentz indices contracted between the two objects.  With some
algebra,  this potential can be rewritten in the suggestive form
\bea
\label{eq:matrix-potential}  
V_{\rm leading} & = & V_{\rm gravity} + V_{\rm electric} + V_{\rm magnetic} \\
V_{\rm gravity} & = & - {15 R^2 \over 4 r^7} \left( \hatt^{IJ} \tilt_{IJ}
- {1 \over 9} \hatt^I{}_I \tilt^J{}_J\right) 
\label{eq:v-gravity} \\
V_{\rm electric} & = & - {45 R^2 \over r^7} \hatj^{IJK} \tilj_{IJK} 
\label{eq:v-electric} \\
V_{\rm magnetic} & = & - {45 R^2\over r^7} \hatm^{+-ijkl}
\tilm^{-+ijkl} \, .
\label{eq:v-magnetic}
\eea 
This is, as we shall discuss shortly, precisely the form of the
interactions we expect to see from 11D supergravity in light-front
coordinates, where ${\cal T},\ijj$, and $\imm$ play the role of the
(integrated) stress tensor, membrane current, and  M5-brane current of
the two objects.  The quantities appearing in this decomposition are
defined as follows.

The matrix stress tensor
$\itt^{IJ}$ is a
symmetric tensor with components
\bea
\label{eq:matrix-t}
\itt^{--} & = & {1 \over R} \; \str \frac{{\cal F}}{96}  \\
\itt^{-i} & = & {1 \over R} \;\str \left(\half \dx^i \dx^j \dx^j +
{1 \over 4} \dx^i F^{jk} F^{jk} 
                  + F^{ij} F^{jk} \dx^k \right) \nonumber \\
\itt^{+-} & = & {1 \over R} \;\str \left(\half \dx^i \dx^i + {1
\over 4} F^{ij} F^{ij} \right) \nonumber \\ 
\itt^{ij} & = & {1 \over R}  \;\str \left( \dx^i \dx^j + F^{ik}
F^{kj} \right) \nonumber \\ 
\itt^{+i} & = & {1 \over R} \;\str \dx^i \nonumber \\
\itt^{++} & = & {N \over R} \nonumber\, .
\eea

The matrix membrane current
$\ijj^{IJK}$ is a totally antisymmetric tensor with components
\bea
\ijj^{-ij} & = & {1 \over 6 R} \str \left( \dx^i \dx^k F^{kj} -
\dx^j \dx^k F^{ki} - \half \dx^k \dx^k F^{ij} 
 + {1 \over 4} F^{ij} F^{kl} F^{kl} +F^{ik}
F^{kl} F^{lj} \right) \nonumber \\ 
\label{eq:matrix-j}
\ijj^{+-i} & = & {1 \over 6 R} \str \left( F^{ij} \dx^j \right)  \\
\ijj^{ijk} & = & - {1 \over 6 R} \str \left( \dx^i F^{jk} + \dx^j F^{ki} + \dx^k F^{ij} \right) \nonumber \\
\ijj^{+ij} & = & - {1 \over 6 R} \str F^{ij} \nonumber\, .
\eea
Note that we retain some quantities --- in particular $\ijj^{+-i}$ and
$\ijj^{+ij}$ --- which vanish at finite $N$ (by the Gauss constraint and
antisymmetry of $F^{ij}$, respectively).  These terms
represent membrane charges
which are only present in the large $N$ limit.  We also define higher
moments of these terms below which can be
non-vanishing at finite $N$; the existence of these higher moments
makes it useful to include these formally vanishing terms even at
finite $N$.

The matrix M5-brane current
$\imm^{IJKLMN}$ is a totally antisymmetric tensor with
\begin{equation}
\label{eq:matrix-m}
 \imm^{+-ijkl} = {1 \over 12 R} \str \left(F^{ij} F^{kl} + F^{ik}
F^{lj} + F^{il} F^{jk}\right)\,.
\end{equation}
At finite $N$ this vanishes by the Jacobi identity, but we shall
retain it for the reasons noted above.  This term represents the
charge of an M5-brane wrapped in the longitudinal ($X^-$) direction.
The other components of ${\cal M}^{IJKLMN}$ do not appear in the
Matrix potential.  In principle, we expect another component of the
M5-brane current, ${\cal M}^{-ijklm}$, to be well-defined.  This term
arises from a moving longitudinal M5-brane.  This term does not appear
in the two-body interaction formula because it would couple to the
charge ${\cal M}^{+ijklm}$ of a transverse (unwrapped)
M5-brane.  As we discuss in \ref{sec:5-branes}, this charge is
expected to vanish classically in matrix theory.  The component ${\cal
M}^{-ijklm}$ can, however, be determined from the conservation of the
M5-brane current, and was shown by Van Raamsdonk (1999) to be given by
\begin{equation}
{\cal M}^{-ijklm} = \frac{5}{4 R}  {\rm STr}\;
\left( \dot{X}^{[i} F^{jk} F^{lm]} \right).
\end{equation}

Let us now compare the interaction potential
Eq.~(\ref{eq:matrix-potential}) with the leading long-range interaction
between two objects in eleven-dimensional light-front compactified
supergravity.  The scalar propagator in eleven dimensions is
\begin{equation}
\laplace{}^{-1}(x) = {1 \over 2 \pi R} \sum_n \int{dk^- d^{\,9}k_\perp \over (2 \pi)^{10}}
{e^{-i{n \over R} x^- - i k^- x^+ + i k_\perp \cdot x_\perp} \over
2 {n \over R} k^- - k_\perp^2}
\end{equation}
where $n$ counts the number of units of longitudinal momentum $k^+$.
To compare the leading term in the long-distance potential with matrix
theory we  extract the  $n = 0$ term,
corresponding to interactions mediated by exchange of a supergraviton
with no longitudinal momentum,
\begin{equation}
\label{eq:propagator}
\laplace{}^{-1}(x-y) = {1 \over 2 \pi R} \delta(x^+ - y^+) { - 15
\over 32 \pi^4 \vert x_\perp-y_\perp\vert^7 } \, .
\end{equation}
Note that the exchange of quanta with zero longitudinal momentum gives
rise to interactions that are instantaneous in light-front time, as
emphasized in Hellerman and Polchinski (1999).  This is
precisely the type of instantaneous interaction that arises at one
loop in Matrix theory.  Such action-at-a-distance potentials are
allowed by the Galilean invariance manifest in the light-front
formalism.

The graviton propagator can be written in terms of this scalar
propagator as
\begin{equation}
D_{\rm graviton}^{IJ,KL}  =  2 \kappa^2 \left(\eta^{IK} \eta^{JL} +
\eta^{IL} \eta^{JK} - {2 \over 9} \eta^{IJ} \eta^{KL} \right)
\laplace{}^{-1}(x-y),
\end{equation}
where $ 2 \kappa^2 =(2 \pi)^5 R^3$ in string units.  The effective
supergravity interaction between two objects having stress tensors
$\hat{T}_{IJ}$ and $\tilde{T}_{KL}$ can then be expressed as
\begin{equation}
S = - {1 \over 4} \int d^{11} x d^{11} y \, \, \hat{T}_{IJ}(x) D_{\rm
graviton}^{IJ,KL}(x-y) \tilde{T}_{KL}(y)\, .
 \label{eq:gravity-interaction}
\end{equation}
This interaction has a leading term of precisely the form
Eq.~(\ref{eq:v-gravity}) if we define $\itt^{IJ}_g$ to be the integrated
component of the stress tensor
\begin{equation}
\itt^{IJ}_g \equiv \int dx^- d^{\, 9}
x_\perp T^{IJ}(x).
\end{equation}
It is straightforward to show in a similar fashion that
Eqs.~(\ref{eq:v-electric}) and (\ref{eq:v-magnetic}) are precisely
the forms of the leading supergravity interaction mediated by 3-form
exchange between membrane currents and M5-brane currents of a pair of
objects.

In this section we have summarized the analysis of the leading two-body
interaction potential between an arbitrary pair of bosonic matrix
theory objects.  This analysis was generalized to include all
quadratic and some quartic terms in the fermionic matrices by Taylor
and Van Raamsdonk (1999a).  With the inclusion of fermions and the
added assumption that the background fields satisfy the classical
equations of motion, the general form of the leading matrix theory
potential in Eq.~(\ref{eq:matrix-potential}) remains essentially
unchanged, although the integrated matrix theory currents given by
Eq.~(\ref{eq:matrix-t})-(\ref{eq:matrix-m}) acquire additional terms at
quadratic and higher order in the fermions.  Furthermore, with the
inclusion of fermionic backgrounds, new interaction terms
between fermionic sources appear which correspond to linearized
gravitational interactions mediated by the gravitino.  These
interaction terms allow for the identification of the fermionic
components of the matrix theory supercurrent.

The fact that Eq.~(\ref{eq:matrix-potential}) (and its generalization
to include fermions) is exactly the form of the leading long-range
supergravity interaction between a general pair of supercurrent
sources implies that matrix theory correctly reproduces all leading
order linearized supergravity interactions\footnote{Prior to and
following the proof
of this general result,
the agreement
between one-loop matrix calculations and leading long-distance
interactions due to linearized supergravity was verified in 
specific examples of two-body backgrounds by 
Aharony and Berkooz (1997), 
Lifschytz and Mathur (1997),
Berenstein and Corrado (1997),
Balasubramanian and Larsen (1997),
Lifschytz (1997, 1998a),
Chepelev and Tseytlin (1997, 1998a),
Maldacena (1998a, 1998b),
Keski-Vakkuri and Kraus (1998a, 1998b),
Pierre (1997, 1998),
Gopakumar and Ramgoolam (1998),
Brandhuber, Itzhaki, Sonnenschein, and Yankielowicz (1998),
Kabat and Taylor (1998a),
Fatollahi, Kaviani, Parvizi (1998),
Hari Dass and Sathiapalan (1998),
Bill\'o, Di Vecchia, Frau, Lerda, Pesando, Russo, and Sciuto (1998)
Hyun, Kiem, and Shin (1999b), and
Massar and Troost (2000).}, and that the integrated
stress tensor, membrane current, and M5-brane current of M-theory
objects are encoded in the $N \times N$ matrix degrees of freedom of
matrix theory through Eqs.~(\ref{eq:matrix-t})-(\ref{eq:matrix-m}).  
An alternative approach to finding the matrix theory stress tensor and
membrane current is to compute the stress tensor and membrane current
of the membrane from the continuous theory defined by the action
(\ref{eq:bosonic-membrane-Polyakov-general}) for the bosonic
membrane in a general background.  The matrix theory stress tensor and
membrane current should then follow from the matrix-membrane
correspondence given in Eq.~(\ref{eq:mm-correspondence}).
This calculation was performed in Kabat and Taylor (1998b) and
Dasgupta, Nicolai, and Plefka (2000).
It turns out that indeed the matrix definitions given above for the
stress tensor and membrane current are compatible with the expressions
for the analogous expressions for continuum membrane, including
higher
moment terms which we discuss in the next subsection. 
The matrix expressions are not uniquely determined by this
correspondence, however.  Additional terms appear in the matrix theory
currents which depend upon the higher degree of sensitivity to
operator ordering afforded by the matrix description.  These terms,
like the appearance of longitudinal M5-branes,
seem to be examples of new physical properties which are mysteriously
added to the system in the matrix regularization process, making the
regularized theory in many ways richer than the initial continuous
membrane theory would suggest.

\subsubsection{General two-body interactions}
\label{sec:two-body-general}

In the previous subsections we have considered only the leading
$1/r^7$ terms in the two-body interaction potential.  In this subsection
we discuss higher order terms in the interaction potential between a
general pair of sources.

Let us begin by considering the series of subleading terms in the
linearized supergravity potential between a general pair of sources
arising from higher multipole moments of the supergravity currents for
the two sources.
Performing a Taylor series expansion around the origin for each of the
two stress tensor sources in Eq.~(\ref{eq:gravity-interaction}), for
example, we find an infinite series of terms in the effective
potential arising from linearized graviton exchange
\begin{eqnarray}
V_{{\rm  gravity}}  &= &
\sum_{m  \leq n = 0}^{\infty}
-\frac{15 R^2}{4 \; r^7}  \left[
\frac{(-1)^{n-m}}{(n-m) ! m !}
\hatt_g^{I J(i_1i_2 \cdots i_{n-m})} \left(
 \eta_{I K} \eta_{J L}
-\frac{1}{9}  \eta_{I J} \eta_{K L} \right)
\tilt_g^{K L(j_1j_2 \cdots j_m)} \right. \nonumber\\
& &\hspace{1.2in} \left.\times
\partial_{i_1} \partial_{i_2} \cdots
\partial_{i_{n-m}}
\partial_{j_1} \partial_{j_2} \cdots
\partial_{j_m}
 (\frac{1}{r^7}) \right].
\label{eq:graviton-many}
\end{eqnarray}
where the moments of the stress tensor in the supergravity theory are
defined through
\[
\itt_g^{IJ (i_1 i_2 \cdots i_n)}  \equiv \int dx^- d^{\, 9}
x_\perp \left( T^{IJ}(x) x^{i_1} x^{i_2} \cdots x^{i_n} \right).
\]
Similar multipole interactions arise from the exchange of 3-form field
quanta, generalizing the leading interaction terms given in
Eqs.~(\ref{eq:v-electric}) and (\ref{eq:v-magnetic}).

Let us now consider how higher-order terms of the form
(\ref{eq:graviton-many}) can be reproduced by loop calculations
in matrix theory.  If we consider all possible Feynman diagrams which
might contribute to higher-order terms, it is straightforward to
demonstrate by power counting that the complete two-body potential can
be written as a sum of terms of the form
\begin{equation}
V = \sum_{n, k, l, m, p, \alpha} V_{n, k, l, m, p, \alpha} R^{n-1}
\frac{X^l D^p F^k \psi^{2m}}{ r^{3n + 2k + l + 3m+ p-4}}.
\label{eq:potential-general}
\end{equation}
where $n$ counts the number of loops in the relevant diagrams and
$\psi$ describes the fermionic background fields.
Each $D$ either indicates a time derivative or a commutator with an
$X$, as in $\psi[X, \psi]$.
The summation over the index $\alpha$ indicates a sum over many
possible index contractions for every combination of $F$'s, $X$'s, and
$D$'s and
$\Gamma$ matrices between the $\psi$'s.

For a completely general pair of objects, only terms in the one-loop
effective action have been understood in terms of supergravity.  At
one-loop order, when the fields are taken on-shell by imposing the
matrix theory equations of motion, all terms with $k + m + p < 4$
which have been calculated vanish.  All terms with $k + m + p = 4$
which have been calculated have $m \geq p$ and can be written in the
form
\begin{equation}
 V_{1,  4-m-p, l, m, p, \alpha} 
\frac{X^l  F^{(4-m-p)} \psi^{2(m-p)} (\psi D \psi)^p}{ r^{7 + m-p+ l}}.
\label{eq:potential-one-loop}
\end{equation}
In this expression, the grouping of $\psi$ terms indicates the
contraction of spinor indices.  The terms can be ordered
in an arbitrary fashion when considered as $U(N)$ matrices, and each
ordering is associated with a different index $\alpha$ and overall
coefficient.  The terms in Eq.~(\ref{eq:potential-one-loop}) have been
explicitly determined for $m < 2$ in Taylor and Van Raamsdonk (1998a),
Kabat and Taylor (1998b), and Taylor and Van Raamsdonk (1999a).  These
terms precisely correspond to linearized supergravitational
interactions of the form of Eq.~(\ref{eq:graviton-many}).  We now
briefly summarize some of the details of this correspondence.

\noindent {\bf Matrix multipole moments:}  Associated with each of the
components of the integrated matrix theory supercurrents given in
Eqs.~(\ref{eq:matrix-t})-(\ref{eq:matrix-m}) there is an infinite
sequence of higher multipole moments.  The bosonic parts of these
multipole moments are formed by simply including $l$ extra matrices
$X^{i}$ into the formula for a given supercurrent component, and
symmetrizing over all possible orderings of the matrices
$\dot{X}^i,F^{ij}$, and $X^i$ inside the trace.  For example, the
higher multipole moments of
the component $\itt^{+-}$ of the matrix theory stress tensor are given
by
\begin{equation}
\itt^{+-(i_1 i_2 \cdots i_n)} = {1 \over R} {\rm STr}\; \left[ \left(\half \dx^i \dx^i + {1
\over 4} F^{ij} F^{ij}  \right) X^{i_1} X^{i_2}\cdots
X^{i_n} \right]\ ,
\label{eq:tpm-moments}
\end{equation}
where ${\rm STr}$ denotes a symmetrized trace.
In the one-loop matrix theory potential between a general pair of
objects, these higher multipole moments appear in the long-range
potential in precisely the form of Eq.~(\ref{eq:graviton-many}) and
its generalization for interactions mediated by 3-form and gravitino
exchange.
The simplest
example of such an interaction is a term in the interaction potential
Eq.~(\ref{eq:potential-one-loop}) of the form 
$\hat{{\cal J}}^{ij}\tilde{{\cal T}}^{-i}r^j/r^9 \sim F^4 X/r^8$
which appears in the case of a graviton moving in the long-range
gravitational field of a matrix theory object with angular momentum
\begin{equation}
{\cal J}^{ij} = {\cal T}^{+ i (j)} -{\cal T}^{+ j (i)}
 \label{eq:matrix-angular-momentum}
\end{equation}
where the first moment of the matrix theory stress tensor component
${\cal T}^{+ i}$ is defined through
\begin{equation}
{\cal T}^{+ i (j)} = \frac{1}{R}  {\rm Tr}\; \left( \dot{X}^i X^j
 \right)\, .
 \label{eq:bosonic-angular-momentum}
\end{equation}

\noindent {\bf Matrix 6-brane current:}
At order $1/r^8$ new ``dyonic'' interaction terms describing
higher-moment membrane-M5-brane and D0-brane-D6-brane interactions
appear in addition to the interactions mentioned above (Dhar and
Mandel 1998;
Bill\'o, Di
Vecchia, Frau, Lerda, Russo, and Sciuto, 1998; Taylor and Van
Raamsdonk 1999a).
These interactions again are exactly in agreement with those of
linearized supergravity, providing that we define (bosonic) components
of a 6-brane current through
\begin{eqnarray}
{\cal S}^{+ijklmn} = \frac{1}{R}  \str\left(F_{[ij} F_{kl}
F_{mn]}\right),
\;\;\;\;\; \;\;\;\;\;
{\cal S}^{ijklmnp} = 
\frac{7}{R}  \str\left(F_{[ij} F_{kl} F_{mn} \dot{X}_{p]}\right)\,.
\end{eqnarray}
It is interesting that this current appears in the matrix theory
interaction potential, since the D6-brane of type IIA string theory
corresponds to a Kaluza-Klein monopole descending from eleven
dimensions, rather than an electrically or magnetically charged brane
like the membrane or M5-brane (Townsend, 1995).

\noindent {\bf Fermion multipole moments:}
In addition to the purely bosonic components of the higher multipole
moments, there are fermionic contributions. There are
fermionic contributions to the integrated supercurrent components,
as well as fundamentally fermionic contributions to higher moments of
the supercurrent, where no derivatives act on the fermions.
The simplest example of a term of the latter type is the spin
contribution to the matrix theory angular momentum,
first noted in the context of spinning gravitons
by Kraus (1998).
\begin{equation}
{\cal J}^{ij}_{{\rm fermion}} = \frac{1}{4 R}  {\rm Tr}\;
\left( \psi \gamma^{ij} \psi \right)
\label{eq:spin-angular}
\end{equation}
Like the term in Eq.~(\ref{eq:bosonic-angular-momentum}) above,
this angular momentum term couples to the component ${\cal T}^{-i}\sim F^3$
of the matrix theory stress-energy tensor through terms of the form
$\hat{{\cal J}}^{ij}\tilde{{\cal T}}^{-i}r^j/r^9$.
\vspace{0.08in}

This summarizes all that is known about the two-body interaction for a
completely general (and not necessarily supersymmetric) pair of matrix
theory objects.  All linearized supergravity interactions between an
arbitrary pair of sources are reproduced by a one-loop matrix theory
calculation up to quadratic order in fermions.  No higher-loop
calculations have been done for general backgrounds.  It seems likely
that the agreement between one-loop matrix theory calculations and
linearized supergravity persists to higher order in the fermions, but
the relevant contributions to the multipole moments of the
supergravity currents have not yet been calculated for a general
matrix theory object.  It is quite plausible that the one-loop matrix
theory interactions corresponding with linearized supergravity are all
protected by supersymmetric nonrenormalization theorems, although this
has not yet been demonstrated.

\subsubsection{General two-graviton interactions}
\label{sec:2-graviton-general}

Aside from the general one-loop results described in the previous
subsection, almost all other perturbative results on two-body
interactions in matrix theory are for a pair of gravitons, which we
discuss in this subsection.  In the case of a pair of gravitons, the
general interaction potential (\ref{eq:potential-general}) simplifies
to
\begin{equation}
V =\sum_{n, k, m} V_{n, k, m} R^{n -1}
\frac{v^k \psi^{2m}}{r^{3n + 2k + 3m-4}} \, .
\label{eq:graviton-potential-general}
\end{equation}

The leading terms for each value of $m \leq 4$ have been computed in
the eikonal approximation using the one-loop approach, and are in
agreement with the spin-spin interaction terms between gravitons in
supergravity.  The sum of these terms is summarized in Plefka, Serone,
and Waldron (1998b), and is given by\footnote{Aspects of these
fermionic contributions to the two-graviton interaction potential were
studied by Harvey (1998), Morales, Scrucca, and Serone (1998a, 1998b),
Kraus (1998), Barrio, Helling, and Polhemus (1998), Plefka, Serone,
and Waldron (1998a), McArthur (1998), Hyun, Kiem, and Shin (1999b,
1999c), and Nicolai and Plefka (2000).}
\begin{eqnarray}
V_{(1)} & = & -\frac{15}{16}  \left[
v^4 + 2v^2 v_i D^{ij} \partial_j
+2v_i v_j D^{ik} D^{jl} \partial_k 
\partial_l \right. \label{eq:spin-terms} \\ & &\left.\hspace{0.3in}
+ \frac{4}{9}  v_iD^{ij} D^{km} D^{lm} \partial_j \partial_k\partial_l
+ \frac{2}{63}  D^{in} D^{jn} D^{km} D^{lm}
\partial_i\partial_j \partial_k
\partial_l \right] \frac{1}{r^7}  \nonumber
\end{eqnarray}
where
$
D^{ij} = \psi \gamma^{ij} \psi
$. The term with a single $D$ proportional to $1/r^8$ arises from the
spin angular momentum term described in Eq.~(\ref{eq:spin-angular}).

No further checks have been made on the matrix theory/supergravity
correspondence for terms with nontrivial fermion backgrounds.
Simplifying to the spin-independent terms, the complete effective potential
(\ref{eq:graviton-potential-general}) simplifies still further to
\begin{equation}
V =\sum_{n, k}
 V_{n, k} R^{n -1}
\frac{v^k}{r^{3n + 2k-4}} .
\end{equation}
Following Becker, Becker, Polchinski, and Tseytlin (1997), we write
these terms in matrix form 
\begin{equation}
 \begin{array}{cccccccccc}
V & = &\frac{1}{R}V_{0,2} \;v^2 & &  & & & & & \\
 & & + & V_{1,4} \frac{v^4}{r^7}  & + &
 V_{1,6} \frac{v^6}{r^{11}}  & + &
 V_{1,8} \frac{v^8}{r^{15}}  & + &\cdots\\
 & & + & R \;V_{2,4}  \;\frac{v^4}{r^{10}}  & + &
R\; V_{2,6} \; \frac{v^6}{r^{14}}  & + &
R\; V_{2,8} \; \frac{v^8}{r^{18}}  & + &\cdots\\
 & & + &R^2 \;V_{3,4} \; \frac{v^4}{r^{13}}  & + &
R^2\; V_{3,6} \; \frac{v^6}{r^{17}}  & + &
R^2\; V_{3,8} \; \frac{v^8}{r^{21}}  & + &\cdots\\
 && + & \vdots & + & \vdots & + & \vdots & + &\ddots
\end{array}
\label{eq:matrix-terms}
\end{equation}
where each row gives the contribution at fixed loop order.
We will now give a brief review of what is known about these
coefficients.  First, let us note that in Planck units this potential
is (restoring factors of $\alpha' = l_{11}^3/R$ by dimensional analysis)
\begin{equation}
V =\sum_{n, k}
 V_{n, k} \frac{l_{11}^{3n+ 3k - 6} }{ R^{k-1}} 
\frac{v^k}{r^{3n + 2k-4}} .
 \label{eq:2-graviton-terms}
\end{equation}
Since the gravitational coupling constant is $\kappa^2 = 2^7
\pi^8l_{11}^9,$ we only expect terms with
$
n + k \equiv 2 \; ({\rm mod} \; 3)
$ to correspond with classical supergravity interactions, since all
terms in the classical theory have integral powers of $\kappa$.  Of
the terms explicitly shown in Eq.~(\ref{eq:matrix-terms}) only the
diagonal terms satisfy this criterion.  By including factors of
$\hat{N}$ and $ \tilde{N}$ for semi-classical graviton states with
finite momentum $P^+$ and comparing to supergravity, one finds that
the terms on the diagonal are precisely those which should correspond
to classical supergravity.  The terms beneath the diagonal have extra
powers of $N$ for a fixed power of $v$ and would therefore dominate
the diagonal terms in a fixed-$r$, large $N$ limit.  It has been
suggested that the terms above the diagonal correspond to quantum
gravity corrections.  It was shown by Becker, Becker, Polchinski, and
Tseytlin (1997) that the sum of diagonal terms corresponding to the
effective classical supergravity potential between two gravitons
should be given by an expansion in $v^2$ of the potential
\begin{equation}
V_{{\rm classical}} = \frac{2 r^7}{15 R^2} \left( 1-\sqrt{1-\frac{15 R}{
2}\frac{v^2}{r^7}  }\right).
\label{eq:two-classical}
\end{equation}

Now let us discuss the individual terms in Eq.~(\ref{eq:matrix-terms}).
As we have discussed, the one-loop analysis gives a term
$
V_{1, 4} = -15/16
$, which agrees with linearized supergravity.  The analysis of Section
\ref{sec:2-graviton} can be extended to the remaining one-loop terms.  
The next one-loop term vanishes
$
V_{1,  6} = 0
$. Some
efforts have been made to relate the higher order terms $V_{1, 8},
\ldots$ to quantum effects in 11D supergravity, but so far this
interpretation is not clear.  We briefly return to this question
in Section \ref{sec:interaction-summary}.  The term
$
V_{2, 4} = 0
$ was computed  in Becker and Becker (1997).  As expected,
this term vanishes.  The term
$
V_{2, 6} = 225/32
$ was computed in Becker, Becker, Polchinski, and Tseytlin (1997).  This
term agrees with the expansion of Eq.~(\ref{eq:two-classical}).  A
general expression for the two-loop effective potential given by the
second line of Eq.~(\ref{eq:matrix-terms}) was given in
Becker and Becker (1998b), although there is no known connection
between these terms and quantum corrections to supergravity.

It was argued by Paban, Sethi, and Stern (1998a, 1998b) that there can
be no (below-diagonal) higher-loop corrections to the $v^4$ and $v^6$
terms on the diagonal.  These authors considered the terms with the
maximal number of fermions which are related to the $v^4$ and $v^6$
terms by supersymmetry.  (For example, the $\psi^8/r^{11}$ term in
Eq.~(\ref{eq:spin-terms}) in the case $v^4$).  They showed that these
fermionic terms are uniquely determined by supersymmetry, and
suggested that this in turn should uniquely fix the form of the
bosonic terms proportional to $v^4$ and $v^6$.  Explicit arguments
along these lines for the nonrenormalization of the terms in
Eq.~(\ref{eq:spin-terms}) were given by Hyun, Kiem, and Shin (1999d),
Okawa (1999), Nicolai and Plefka (2000), and Kazama and Muramatsu
(2000).  The results of these authors support the conclusion that all
these terms are protected by supersymmetry, although the full
supersymmetric off-shell action has not yet been constructed.  The
connection between the $v^6$ terms and the related terms with 12
fermions appears to be more subtle than in the $v^4$ case (Okawa,
1999), particularly when the action is taken off-shell.  The
nonrenormalization results for the $v^4$ and $v^6$ terms indicate that
$V_{(n > 1), 4} = V_{(n > 2), 6} = 0$. The existence of such
nonrenormalization theorems in matrix theory was originally
conjectured by BFSS in analogy to similar known theorems for
higher-dimensional theories.

This completes our summary of what is known about  interactions
between two unpolarized gravitons
in matrix theory.  The complete set of known terms is given by
\begin{equation}
 \begin{array}{ccccccccccc}
V & = &\frac{1}{2R} v^2 & &  & & & & & &\\
& & & + & -\frac{15}{16}  \frac{v^4}{r^7}  & + &
0& + &({\rm known})  & \rightarrow &\\
& & & + & 0 & + &
 \frac{225}{32} R  \frac{v^6}{r^{14}}   & + &
({\rm known})  & \rightarrow &\\
& & & + &0 & + &
0 & + &
?  & + &\cdots\\
& &&  & \downarrow &  & \downarrow & + & \vdots & + &\ddots
\end{array}
\label{eq:matrix-terms-2}
\end{equation}

It has been proposed that for arbitrary $N$ the analogues of the
higher-loop diagonal terms should naturally take the form of a
supersymmetric Born-Infeld type action
(Chepelev and Tseytlin, 1998a, 1998b; Keski-Vakkuri and Kraus, 1998b;
Balasubramanian, Gopakumar and Larsen, 1998).
This would give rise in the case $N = 2$ to a sum of the form
Eq.~(\ref{eq:two-classical}).  There is as yet, however, no proof of this
statement beyond two loops.  One particular obstacle to calculating
the higher-loop terms in this series is that it is necessary to
integrate over loops containing propagators of massless fields.  These
propagators can give rise to subtle infrared problems with the
calculation.  Some of these difficulties can be avoided by trying to
reproduce higher-order supergravity interactions from interactions of
more than two objects in matrix theory, the subject to which we
turn in the next subsection.  One interesting example of a two body
interaction which has been considered at higher loop order involves
the scattering of a  D0-brane from a bound state of D0-branes and
D6-branes.  It was shown by Branco (1998) that the form of the $F^6$
term in the supersymmetric Born-Infeld action proposed by Chepelev and
Tseytlin correctly reproduces the supergravity interaction in this
situation.  Dhar (1999) found, however, that the two-loop matrix
theory calculation in this background suffers from divergences.  It
would be very interesting to understand whether indeed the higher
terms in the nonabelian Born-Infeld action can be organized in such a
way  as to reproduce nonlinear gravitational effects between general
sources.

\subsection{The N-body problem}
\label{sec:N-body}

So far we have seen that the linearized theory of supergravity is
correctly reproduced by an infinite series of terms arising from
one-loop calculations in matrix theory.  We have also discussed 2-loop
calculations of two-graviton interactions which  agree with
supergravity.  If matrix theory is truly to reproduce all of classical
supergravity, however, it must reproduce all the nonlinear effects of
the fully covariant gravitational theory.  The easiest way to study
these nonlinearities is to consider N-body interaction processes.  The
first nonlinear gravitational effects appear at order $\kappa^4$ in
the gravitational coupling.  An example of such a nonlinear effect is
the effect on a third object of the nonlinear contribution to the
long-range gravitational field produced by the interaction of the
fields from two distinct sources.  This effect can be seen in
classical gravity from a ``Y''-shaped tree diagram connecting three
separate objects.  From the same dimensional analysis leading to
Eq.~(\ref{eq:2-graviton-terms}), we expect these nonlinear effects to
arise in a two-loop matrix theory calculation, and to have a leading
term of the general form $v^6/r^{14}$, where the $v$'s are the
velocities of the three bodies and the $r$'s are their relative
positions.  The simplest three-body interaction is that of three
unpolarized gravitons, which can be described by the classical
background
\begin{equation}
B^i =\left(\begin{array}{ccc}
 r_1^i + v_1^i & 0 & 0\\
0 & r_2^i + v_2^i & 0\\
0 &0 & r_3^i + v_3^i
\end{array} \right)\,.
\end{equation}
Finding the leading terms in the two-loop effective action for
an $N = 3$ matrix configuration such as this is technically quite
complicated.  In an impressive pair of papers, Okawa and Yoneya
(1999a, 1999b) carried out a complete perturbative calculation of all
terms of order $v^6/r^{14}$ in the three-graviton effective action.
(There are many such terms, which can be expressed as different
functions of the relative velocities $v_{ij} = v_i-v_j$ and relative
positions $r_{ij}= r_i-r_j$  of the three gravitons.)
They found that there is an exact agreement between the two-loop
matrix theory calculation and nonlinear corrections to supergravity at
this order\footnote{Previous partial results on the three-graviton
problem had been found by Dine and Rajaraman (1999), 
Fabbrichesi, Ferretti, and Iengo (1998), Echols and Gray (1998),
and Taylor and Van
Raamsdonk (1998b).
Further work on this problem is described in McCarthy, Susskind, and
Wilkins (1998), Helling, Plefka, Serone, and Waldron (1999), Dine,
Echols, and Gray (2000), and Refolli, Terzi, and Zanon (2000).}.
Sethi and  Stern (1999) have argued that, like the $v^4$ and $v^6$
terms in the $N = 2$ theory, all these $v^6/r^{14}$ terms in the $N =
3$ theory are protected from higher loop corrections by a
supersymmetric nonrenormalization theorem.

One would naturally like to extend these results both by considering a
general 3-body system, and by going beyond the 3-body problem to the
general N-body problem.  To date, however there has been very little
progress on the problem of understanding higher order nonlinearities
in the theory beyond those involved in the 3-graviton system.  One
foray into the general N-body calculation was made in Dine, Echols, and
Gray (2000).  These authors considered a subset of the terms in the
general N-graviton interaction potential for arbitrary $N$.  They
found some terms at higher loop orders which agree with supergravity.
They also identified terms, however, which appear in the 3-loop
calculation of the 4-graviton effective action and which scale as
$v^6/r^{17}$.  These terms have improper scaling to correspond to
supergravity terms, and are in fact ``below the diagonal'' as seen in
Eq.~(\ref{eq:matrix-terms}).  The appearance of such terms in the
matrix theory perturbation series indicates a breakdown of the
correspondence between perturbative matrix theory calculations and
classical supergravity.  This is the first concrete calculation where
the two perturbative expansions have been shown to disagree.  There
are several subtleties in this calculation which may require further
consideration before the case is completely closed.  There are
infrared divergences in this calculation which must be handled
carefully, and which may lead to unexpected cancellations in some
situations.  There is also an issue of gauge choices; while the
scattering S-matrix is gauge-independent, the effective action derived
from Eq.~(\ref{eq:s-2}) is gauge dependent.  The manifest agreement
discussed above between the one-loop matrix theory effective action
and linearized supergravity seems to rely upon a fortuitous choice of
gauge on both sides.  If other gauges had been chosen, it might have
been necessary to perform a complicated field redefinition to see the
correspondence explicitly.  It may be that for $N > 2$ the background
field gauge is not suitable for direct comparison to $N$-body
interaction terms in the supergravity effective action.  The issue of
gauge-dependence was discussed in Hata and Moriyama (1999).  As one
goes to higher loop order it may also be necessary to understand
recoil effects and the off-shell effective action; these issues are
discussed in Periwal and von Unge (1998), Okawa and Yoneya (1998b), 
Okawa (1999), and Kazama and Muramatsu (2000).  Despite these
concerns, however, it seems most likely that this result is correct as
stated, and that the correspondence between perturbative calculations
in matrix theory and classical supergravity breaks down once
high-order nonlinear effects are taken into consideration.  As we will
discuss in slightly more detail below, the nonrenormalization theorems
which protect the one-loop and two-loop terms for $N \leq 3$ do not
seem to extend to higher loops and larger values of $N$, so there is
no contradiction between this breakdown of perturbative matrix theory
and supersymmetry.  It does, however, mean that we must work 
harder if we wish to demonstrate that classical
eleven-dimensional supergravity is reproduced by matrix theory in the
large $N$ limit.

\subsection{Longitudinal momentum transfer}
\label{sec:longitudinal}

In this section we have concentrated on interactions in matrix
theory and supergravity where no longitudinal momentum is transferred
from one object to another.  A supergravity process in which
longitudinal momentum is transferred from one object to another is
described in the IIA theory by a process where one or more D0-branes
are exchanged between coherent states consisting of clumps of
D0-branes.  Such processes are exponentially suppressed since the
D0-branes are massive, and thus are not relevant for the expansion of
the effective potential in terms of $1/r$ which we have been
discussing.  In the matrix theory picture, this type of exponentially
suppressed process can only appear from nonperturbative effects.
Clearly, however, for a full understanding of interactions in Matrix
theory it will be necessary to study processes with longitudinal
momentum transfer in detail and to show that they also correspond
correctly with processes in supergravity and M-theory.  Some progress
has been made in this direction.  Polchinski and Pouliot (1997)  have
calculated the scattering amplitude for two D2-branes for processes in
which a D0-brane is transferred from one D2-brane to the other.  In the
Yang-Mills picture on the world-volume of the D2-branes, the incoming
and outgoing configurations in this calculation are described in terms
of an $U(2)$ gauge theory with a scalar field taking a VEV which
separates the branes.  The transfer of a D0-brane corresponds to an
instanton-like process where a unit of flux is transferred from one
brane to the other.  The amplitude for this process was computed by
Polchinski and Pouliot and shown to be in agreement with expectations
from supergravity.  This result suggests that processes involving
longitudinal momentum transfer may be correctly described in Matrix
theory.  It should be noted, however, that the Polchinski-Pouliot
calculation is not precisely a calculation of membrane scattering with
longitudinal momentum transfer in Matrix theory since it is carried
out in the D2-brane gauge theory language.  In the T-dual Matrix theory
picture the process in question corresponds to a scattering of
D0-branes in a toroidally compactified space-time with the transfer of
membrane charge.  Processes with D0-brane transfer and the relationship
between these processes and graviton scattering in matrix theory have
been studied further in Dorey, Khoze, and Mattis (1997), Banks,
Fischler, Seiberg, and Susskind (1997), Keski-Vakkuri and Kraus
(1998d), Paban, Sethi, and Stern (1998c),
Hyun, Kiem, and Shin (1999a), and de Boer, Hori, and Ooguri (1998).

\subsection{Summary and outlook for the matrix theory-supergravity correspondence}
\label{sec:interaction-summary}

In this section we have described a variety of perturbative matrix
theory calculations describing interactions between two or more
``objects'' described by blocks in a matrix theory background.  For
most of these calculations the perturbative results of matrix quantum
mechanics precisely reproduce classical supergravity interactions
between appropriate sources.  It seems that all linearized
supergravity interactions between arbitrary sources can be reproduced
by a one-loop calculation in matrix theory.  Some more specific
nonlinear effects in supergravity, namely the second-order
interactions in systems of two and three unpolarized gravitons, are
also reproduced by two-loop matrix theory calculations.  We have
discussed supersymmetric nonrenormalization theorems which guarantee
that the one-loop and two-loop graviton interaction calculations are
protected by supersymmetry and cannot be corrected by higher-loop
effects in matrix theory.  While it has not been explicitly proven, it
is tempting to believe that similar supersymmetric nonrenormalization
theorems protect all the terms in the one-loop matrix theory effective
action for any $N$, with backgrounds describing an arbitrary pair of
interacting supergravity sources.

While it may be that all one-loop interactions and two-loop
interactions for $N \leq 3$ are protected by supersymmetric
nonrenormalization theorems as we have discussed, there is little
evidence that higher-loop terms or two-loop terms for $N > 3$ are
protected by supersymmetry.  Indeed, it was argued by Dine, Echols, and
Gray (1998) that even in the $N = 2$ theory terms of order $v^8$ and
higher should experience higher loop corrections.  Similarly, it was
argued by Sethi and Stern (1999) that when $N > 3$ even two-loop terms
cannot be shown to be nonrenormalized using the same arguments as for
one-loop terms and two-loop terms for small values of $N$.  Given that
the supersymmetric nonrenormalization theorems start to break down at
this point, it is perhaps not surprising that Dine, Echols, and Gray (2000)
found discrepancies between  3-loop calculations for $N = 4$
and classical supergravity. If the correspondence does indeed break down at
higher-loop order, then either we must accept that matrix theory does
not successfully model M-theory, or there must be a more complicated
way of understanding the correspondence between these theories. At
this time there is not universal agreement as to how this question
will eventually be resolved, but the only possible alternatives seem
to be the following:

\vspace{0.05in}

\noindent {\bf i)} Matrix theory is correct in the large $N$ limit,
and noncompact supergravity is reproduced by a naive large $N$ limit
of the standard perturbative matrix theory calculations.

\noindent {\bf ii)} Matrix theory is correct in the large $N$ limit,
but to connect it with classical supergravity it
is necessary to deal with subtleties in the large $N$ limit.  (i.e.,
there are problems with the standard perturbative analysis at higher
order)

\noindent {\bf  iii)} Matrix theory is simply wrong, and further terms
need to be added to the dimensionally reduced super Yang-Mills action
to find agreement with M-theory even in the large $N$ limit.

\vspace{0.05in}

Let us review the evidence:

\noindent $\bullet$
Assuming that the result of Dine, Echols, and Gray is correct, and
has been correctly interpreted, clearly (i) is not possible.
The fact that the methods of Paban, Sethi, and Stern for proving
nonrenormalization theorems  in the $SU(2)$ theory break down for
$SU(3)$ at two loops and at higher loop order
also hints that (i) may not be correct.

\noindent $\bullet$
The analysis of Seiberg and Sen seems to indicate that
either possibility (i) or (ii) should hold.
\vspace{0.04in}

It seems that (ii) is the most likely possibility, given this limited
evidence.  There are several issues which are extremely important in
understanding how this problem will be resolved.  The first is the
issue of Lorentz invariance.  If a theory contains linearized gravity
and is Lorentz invariant, then it is well known that it must be either
the complete generally covariant gravity theory or just the pure
linearized theory.  Since we know that matrix theory has some
nontrivial nonlinear structure which reproduces part of the
nonlinearity of supergravity, it would seem that the conjecture must
be valid if and only if the theory is Lorentz invariant.
Unfortunately, so far there is no complete understanding of whether
the quantum theory is Lorentz invariant (classical Lorentz invariance
was demonstrated in de Wit, Marquard, and Nicolai, 1990).  It was
suggested in Lowe (1998) that the problems found in Dine, Echols, and
Gray (2000) might be related to a breakdown of Lorentz invariance and
that in fact extra terms must be added to the theory to restore this
invariance; this would lead to possibility (iii) above.

Another critical issue in understanding how the perturbative matrix
theory calculations should be interpreted is the issue of the order of
limits.  In the perturbative calculations discussed here we have
assumed that the longitudinal momentum parameter $N$ is fixed for each
of the objects we are taking as a background, and we have then taken
the limit of large separations between each of the objects.  Since the
size of the wavefunction describing a given matrix theory object will
depend on $N$ but not on the separation from a distant object, this
gives a systematic approximation scheme in which the bound state and
wavefunction effects for each of the bodies can be ignored in the
perturbative analysis.  If we really are interested in the large $N$
theory, however, the correct order of limits to take is the opposite.
We should fix a separation distance $r$ and then take the large $N$
limit.  Unfortunately, in this limit we have no systematic
approximation scheme.  The wavefunctions for the separate objects
overlap significantly as the size of the objects grows.  Indeed, it
was argued recently by Polchinski (1999) that the size of the bound
state wavefunction of $N$ D0-branes will grow at least as fast as
$N^{1/3}$.  As emphasized by Susskind (1999), this overlap of
wavefunctions makes the theory very difficult to analyze.  Indeed, if
possibility (ii) above is correct, it may be very difficult to use
matrix theory to reproduce all the nonlinear structure of classical
supergravity, let alone to derive new results about quantum
supergravity.  On the other hand, it may be that whatever mechanism
allows the one-loop and two-loop matrix theory results to correctly
reproduce the first few terms in supergravity and to evade the problem
of wavefunction overlap may persist at higher orders through a more
subtle mechanism than those currently understood.  Indeed, one of
the must important outstanding questions regarding matrix theory is to
understand precisely which terms in the naive perturbative expansion
of the quantum mechanics will agree with classical supergravity, and
more importantly, {\it why} these terms agree.

In this section we have focused on the problem of deriving classical
eleven-dimensional supergravity from matrix theory.  A very
interesting, but more difficult, question is whether matrix theory can
also successfully reproduce string/M-theory corrections to classical
supergravity.  The first such corrections would be ${\cal R}^4$
corrections to the Einstein-Hilbert action (Fradkin and Tseytlin,
1983).  It was argued by Susskind (1997b) and Berglund and Minic (1997)
that such terms should be reproduced by the $v^8/r^{18}$ terms which
arise in the two-loop effective potential\footnote{An alternative
suggestion was made by Serone (1998)} (\ref{eq:matrix-terms}).  It was
shown by Keski-Vakkuri and Kraus (1998c) and Becker and Becker (1998a)
that in a two-body interaction between a pair of gravitons with
longitudinal momentum $N/R$ this term has the wrong scaling
in $N$.  This discrepancy was sharpened by Helling, Plefka, Serone, and
Waldron (1999), who performed a two-loop calculation in a
three-graviton background, and showed that the tensor structure of the
$v^8/r^{18}$ terms disagrees with that expected from a ${\cal R}^4$
correction to gravity.  While more work needs to be done in this
direction, the results of these authors indicate that the perturbative
loop expansion in matrix theory probably does not correctly reproduce
quantum effects in M-theory.  The most likely explanation for this
discrepancy is that, like the higher-loop diagonal terms discussed
above, such terms are not subject to nonrenormalization theorems, and
are only reproduced in the large $N$ limit if the matrix theory
conjecture is correct.

\section{M-theory objects from matrix theory}
\label{sec:matrix-objects}

In this section we discuss how the matrix theory degrees of freedom
can be used to construct the various objects of M-theory: the
supergraviton, supermembrane, and M5-brane.  We discuss classical and
quantum supergravitons in matrix theory in subsection
\ref{sec:gravitons}.  We present a general discussion of the structure
of extended objects and their charges in subsection
\ref{sec:extended}, following which we discuss the matrix
constructions of membranes and M5-branes in subsections
\ref{sec:matrix-membranes} and \ref{sec:5-branes}, respectively.

\subsection{Supergravitons}
\label{sec:gravitons}

In DLCQ M-theory, for every integer $N$ there should be a localized
state corresponding to a longitudinal graviton with $p^+ = N/R$ and
arbitrary transverse momentum $p^i$.  We expect from the massless
condition $m^2 = -p^Ip_I = 0$ that such an object will have matrix
theory energy
$
E = p_i^2/2p^+
$. We discuss such states first classically and then in the quantum theory.

\subsubsection{Classical supergravitons}
\label{sec:classical-gravitons}

The classical matrix theory potential is $-[X^i, X^j]^2$, from which
we have the classical equations of motion 
\begin{equation}
\ddot{X}^i = -[[X^i, X^j], X^j].
\end{equation}
One simple class of solutions to these equations of motion can be
found when the matrices minimize the potential at all times and
therefore all commute.  Such solutions are of the form
\begin{equation}
X^i = \left(\begin{array}{cccc}
x^i_1 + v^i_1 t & 0 &0 &  \ddots\\
0 & x^i_2 +v^i_2 t &\ddots & 0\\
0 & \ddots & \ddots & 0\\
 \ddots & 0 & 0 & x^i_N +v^i_N t
\end{array}\right)\, .
\label{eq:matrix-gravitons}
\end{equation}
This corresponds to a classical $N$-graviton solution, where each
graviton has
\begin{equation}
p_a^+ = 1/R, \;\;\;\;\; p^i_a = v^i_a/R, \;\;\;\;\; E_a = v_a^2/(2 R) =
(p_a^i)^2/2p^+\,.
\end{equation}
A single classical graviton with $p^+ = N/R$ can be formed by setting
\begin{equation}
x_1^i = \cdots = x_N^i, \;\;\;\;\; v_1^i = \cdots = v_N^i
\label{eq:all-same}
\end{equation}
so that the trajectories of all the components are identical.
This simple model for gravitons was used in all the spin-independent
matrix theory calculations described in the previous section.

The classical graviton gives a simple example with which to understand
the matrix theory stress tensor (\ref{eq:matrix-t}).  The
integrated stress tensor of a graviton can be written in the form
\begin{equation}
\itt^{IJ} = \frac{p^Ip^J}{p^+} ,
\end{equation}
where
\begin{equation}
p^+ = N/R,\;\;\;\;\; p^i = p^+ \dot{x}^i,\;\;\;\;\; p^- = p_\perp^2 / 2 p^+\,.
\end{equation}
These expressions agree precisely with the matrix expressions for the
stress tensor (\ref{eq:matrix-t}) using the matrices
(\ref{eq:matrix-gravitons}) with Eq.~(\ref{eq:all-same}).

\subsubsection{Quantum supergravitons}
\label{sec:quantum-gravitons}

The picture of a supergraviton in quantum matrix theory is somewhat
more subtle than the simple classical picture just discussed.  Let us
first consider the case of a single supergraviton with $p^+ = 1/R$.
This corresponds to the U(1) case of the super Yang-Mills quantum
mechanics theory.  The Hamiltonian is simply 
\begin{equation}
H = \frac{1}{2 R}  \dot{X}^2,
\end{equation}
since all commutators vanish in this theory.  The bosonic part of the
theory is simply a free nonrelativistic particle.  In the fermionic
sector there are 16 spinor variables with anticommutation relations
$
\{\theta_\alpha, \theta_\beta\} = \delta_{\alpha \beta}
$. By using the standard trick of writing these as 8 fermion creation
and annihilation operators
\begin{equation}
\theta_i^{\pm} = \frac{1}{ \sqrt{2}}  (\theta_i \pm \theta_{i + 8}), \;
\;\;\;\;\; 1 \leq i \leq 8,
\end{equation}
we see that the Hilbert space for the fermions is a standard fermion
Fock space of dimension $2^8 = 256$.  Indeed, this is precisely the
number of states needed to represent all the polarization states of
the graviton (44), the antisymmetric 3-tensor field (84), and the
gravitino (128).  For details of how the polarization states are
represented in terms of the fermionic Fock space, see de Wit, Hoppe
and Nicolai (1988), Plefka and Waldron (1998), Morales, Scrucca, and
Serone (1998b), and Millar, Taylor, and Van Raamsdonk (2000).

The case when $N > 1$ is much more subtle.  We can factor out the
overall $U(1)$ so that every state in the $SU(N)$ quantum mechanics
theory has 256 corresponding states in the full theory.  For the
matrix theory conjecture to be correct, as BFSS pointed out, it should
then be the case that for every $N$ there exists a unique threshold
bound state in the $SU(N)$ theory with $H = 0$.  As mentioned before,
no definitive answer as to the existence of such a state was given in
the early work on matrix theory.  The existence of a unique ground
state for the matrix quantum mechanics can be demonstrated by showing both
that the Witten index of the system is equal to one, and that there are
no fermionic ground states.  The first of these statements was finally
proven for $N = 2$ in Sethi and Stern (1998), demonstrating that at
least one threshold bound state exists.  These authors showed that the
Witten index breaks up into a bulk and a boundary contribution, each
of which is separately fractional.  
The uniqueness of the bound state for SU(2) was shown in Sethi and
Stern (2000a).
The existence of a bound
state for $N > 2$ was demonstrated when Moore, Nekrasov, and
Shatashvili (2000) computed the bulk contribution to the index for
general $N$ and Green and Gutperle (1998) computed the corresponding
boundary contribution.  (This boundary contribution has also been
checked by numerical methods in Krauth and Staudacher, 1998.)  Related
work was done by Yi (1997), Porrati and Rozenberg (1998), and Konechny
(1998).  The Witten index for groups other than SU(N) was
determined in Kac and Smilga (2000), Hanany, Kol, and Rajaraman
(1999), and Staudacher (2000), where a puzzling discrepancy between
the predictions of different methods was noted for the exceptional
group $G_2$.  

The exact determination of the bound state wavefunction, even for $N =
2$, is a difficult problem on which little progress has been made.  A
more tractable and still very interesting problem is the determination
of the asymptotic form of the ground state.  It was shown by Sethi and
Stern (2000a) that for the SU(2) theory the asymptotic form of the
wavefunction is invariant under the SO(9) R-symmetry group of the
quantum mechanics theory.  Combining this with the conditions of
integrability and SU(2) invariance are sufficient to uniquely fix the
asymptotic form of the SU(2) wavefunction.  Halpern and Schwartz
(1998) used a second order Born-Oppenheimer approach to determine the
form of the asymptotic wavefunction in the $SU(2)$ theory.  This
asymptotic form was reproduced using a first-order approach (based on
the supersymmetry generators rather than the Hamiltonian) by Graf and
Hoppe (1998) and Fr\"ohlich, Graf, Hasler, Hoppe, and Yau
(2000)\footnote{Related work appeared in Hoppe (1997b).}.  It was
shown by Bordemann, Hoppe, and Suter (1999) that the analogous
condition of R-symmetry invariance is not sufficient to uniquely
determine the asymptotics of the ground state for $N > 2$.  The extra
information needed to determine the asymptotic form of the $SU(3)$
ground state was, however, described in Hoppe (1999) and the
asymptotic form was found in Hoppe and Plefka (2000).

While little progress has been made so far towards an exact analytic
description of the bound state wavefunction of two D0-branes, Sethi
and Stern (2000b) considered the related problem of the bound state
wavefunction of a D0-brane and a D4-brane.  They found a system of
equations describing this bound state and arrived at the surprising
conclusion that the unique normalizable bound state could be described
by a single equation in terms of a single unknown function.  This
extraordinary simplification hints that perhaps there is some hidden
structure even in the case of two D0-branes which might eventually
allow for an analytic description of the graviton bound state.

As discussed in \ref{sec:interaction-summary}, to understand
interactions between matrix theory gravitons in the large $N$ limit, 
it is crucial to understand how the size of the bound state wavefunction grows
with $N$.  With the limited information we have at this time about
the wavefunctions for small values of $N$, it is difficult to
rigorously determine the asymptotics as $N$ becomes large.  It was
argued by Nekrasov (1999) that the ground state wavefunction has a
size which scales as $N^{1/3}$, saturating the lower bound found by
Polchinski (1999).  How such large wavefunctions interact when
scattered at an impact parameter which is fixed as $N$ becomes large
is a puzzle which must be better understood if we are
to develop a deeper understanding of matrix theory as a model of
quantum gravity.

\subsection{Extended objects from matrices}
\label{sec:extended}

We have discussed the construction of localized graviton states as
classical and quantum matrix theory configurations.  In addition to
these pointlike objects, we would like to construct M-theory membranes
and M5-branes from the fundamental matrix degrees of freedom.  These
will be objects extended in one, two, and four spatial dimensions.  In
this subsection we make some general comments about the structure of
these extended objects in matrix theory.

Let us begin by discussing the charges associated with the extended
objects in matrix theory.  In \ref{sec:two-body-linear} we used the
correspondence between one-loop matrix theory interactions and
linearized supergravity to construct an integrated stress tensor,
membrane current and M5-brane current for a general matrix theory
configuration.  The components of these tensors with a $+$ index
correspond to conserved charges in the theory.  The components
$\itt^{++}$ and $\itt^{+i}$ of the matrix stress tensor correspond to
longitudinal and transverse momentum $N/R$ and $p^i$ respectively.
The components $\ijj^{+ -i}$ and $\ijj^{+ ij}$ of the membrane current
correspond to charges for membranes which are wrapped and unwrapped
in the longitudinal direction, and the component $\imm^{+ -ijkl}$ of
the M5-brane current corresponds to a charge for wrapped
(longitudinal) M5-branes.  No charge associated with unwrapped
(transverse) M5-branes appears in the one-loop matrix theory
interaction potential.  The charges associated with extended objects
all vanish at finite $N$; this corresponds physically to the fact that
any finite-size configuration of membranes and M5-branes must have net
charges which vanish, as all the branes must be compact.

An alternative understanding of the conserved charges associated
with extended objects in matrix theory follows from the
supersymmetry algebra of the theory.  The eleven-dimensional
supersymmetry algebra takes the form
\begin{equation}
\{Q_\alpha, Q_\beta\} \sim P^I (\gamma_I)_{\alpha \beta} + Z^{I_1 I_2}
(\gamma_{I_1 I_2})_{\alpha \beta}+ Z^{I_1 \ldots I_5}
(\gamma_{I_1 \ldots I_5})_{\alpha \beta}
\label{eq:11D-SUSY}
\end{equation}
where the central terms $Z^{I_1 I_2}, Z^{I_1 \ldots I_5}$ correspond
to 2-brane and M5-brane charges.  The supersymmetry algebra of Matrix
theory was explicitly computed\footnote{Similar calculations were
performed previously by Claudson and Halpern (1985) and by de Wit,
Hoppe, and Nicolai (1988); in these earlier analyses, however, terms
such as ${\rm Tr}\;[X^i, X^j]$ and ${\rm Tr}\;X^{[i} X^{j} X^k X^{l]}$
were dropped since they vanish for finite $N$.} by Banks, Seiberg, and
Shenker (1997).  The full supersymmetry algebra of the theory takes
the schematic form
\begin{equation}
\{Q, Q\} \sim P^I + z^i + z^{ij} + z^{ijkl},
\label{eq:matrix-SUSY}
\end{equation}
as we would expect for the light-front supersymmetry algebra
corresponding to Eq.~(\ref{eq:11D-SUSY}).
The charge
\begin{equation}
z^i \sim\ijj^{+ -i} \sim i {\rm Tr}\; \left(\{P^j,[X^i, X^j]\} +
{}[{}[X^i, \theta^{\alpha}], \theta^{\alpha}] \right)
\end{equation}
corresponds to longitudinal membranes (strings),
the charge
\begin{equation}
z^{ij} \sim\ijj^{+ ij} \sim -i {\rm Tr}\;[X^i, X^j]
\end{equation}
corresponds to transverse membranes and
\begin{equation}
z^{ijkl} \sim\imm^{+ -ijkl} \sim {\rm Tr}\;X^{[i} X^{j} X^k X^{l]}
\end{equation}
corresponds to longitudinal M5-brane charge.

Yet another way to motivate these charge identifications is through
T-duality in the type IIA picture.  This approach is described in
Taylor (1998, 2000), Taylor and Van Raamsdonk (1999a), and Myers (1999).

The perturbative matrix theory calculations described in
\ref{sec:two-body} determine not only the conserved charges of the
theory, but also the higher multipole moments of these charges.  For
example the multipole moments of the membrane charge $z^{ij}= -2 \pi i
{\rm Tr}\;[X^i, X^j]$ can be written in terms of the matrix moments
\begin{equation}
z^{ij (k_1 \cdots k_n)} = -2 \pi i \;{\rm STr}\; \left([X^i, X^j] X^{k_1}
\cdots X^{k_n} \right),
\end{equation}
which are the matrix analogues of the moments
\begin{equation}
\int d^2 \sigma \; \{X^i, X^j\} X^{k_1} \cdots X^{k_n}
\label{eq:membrane-moments}
\end{equation}
for the continuous membrane.  The symbol ${\rm STr}$ as usual
indicates a symmetrized trace, wherein the trace is averaged over all
possible orderings of the terms $[X^i, X^j]$ and $X^{k_\nu}$ appearing
inside the trace.  This corresponds to a particular ordering
prescription in applying the matrix-membrane correspondence to
Eq.~(\ref{eq:membrane-moments}).  There is no {\it a priori}
justification for this ordering prescription, but it is a consequence
of the explicit calculations of interactions between general matrix
theory objects described above.  The same prescription can be used to
define the multipole moments of the longitudinal membrane and M5-brane
charges.  These multipole moments can also be derived from T-duality
arguments in type IIA, as in Taylor and Van Raamsdonk (1999b) and Myers
(1999), but the ordering information implied by the symmetrized trace
cannot be determined in this fashion.

Although as we have mentioned, the conserved charges in matrix theory
corresponding to extended objects all vanish at finite $N$, the same
is not true of the higher moments of these charges.  As we will
discuss in the following sections, it is possible to construct compact
membrane and M5-brane configurations in matrix theory whose multipole
moments of membrane and M5-brane charge are nonvanishing and agree to
within terms of order $ 1/N^2$ with the continuous versions of these
multipole moments.  Conversely, by calculating the multipole moments
of a fixed matrix configuration we can essentially reproduce the
complete spatial dependence of the matter configuration to which the
matrices correspond.  

It is interesting to note that while the superalgebra in
Eq.~(\ref{eq:matrix-SUSY}) does not contain a 6-brane charge, such a
charge does appear in the one-loop matrix theory effective action,
associated with dyonic interactions between D6-branes and D0-branes.
The construction of  6-branes from multiple D0-branes was discussed in
Taylor (1997b).

\subsection{Membranes}
\label{sec:matrix-membranes}

In this section we discuss the construction of M-theory membranes in
terms of the matrix quantum mechanics degrees of freedom. It is clear
from the derivation of matrix theory as a regularized supermembrane
theory that there must be matrix configurations which in the large $N$
limit give arbitrarily good descriptions of any membrane
configuration.  It is somewhat instructive, however, to study some
aspects of the  geometry of  simple matrix membranes at finite $N$.
In subsection \ref{sec:compact-membranes} we describe
some explicit examples of compact membrane configurations and discuss
how membrane geometry is encoded in a system of finite-size matrices.
In subsection \ref{sec:infinite-membranes} we discuss noncompact
matrix membranes, and in subsection \ref{sec:matrix-strings} we
discuss wrapped membranes, which appear as string excitations in
matrix theory.
\subsubsection{Compact membranes}
\label{sec:compact-membranes}

One extremely simple example of a membrane configuration, which makes
it clear that a smooth membrane geometry can be approximated quite
well even at finite $N$ by simple matrix configurations, is the
symmetric spherical membrane (Kabat and Taylor, 1997a).  Imagine that
we wish to construct a membrane embedded in an isotropic sphere
$
x_1^2 + x_2^2 + x_3^2 = r^2
$ in the first three dimensions of ${\br}^{11}$.  The embedding functions
for such a continuous membrane can be written as linear functions
$
X^i = r \xi^i, 1 \leq i \leq 3,
$ of the three Euclidean coordinates $\xi^i$ on the spherical
world-volume.  Using the matrix-membrane correspondence
(\ref{eq:mm-correspondence}) we see that the matrix approximation to
this membrane will be given by the $N \times N$ matrices
\begin{equation}
X^i = \frac{2r}{ N}  J^i \; \;\;\;\;\; 1 \leq i \leq 3
\label{eq:matrix-sphere}
\end{equation}
where $J^i$ are the generators of $SU(2)$ in the
$N$-dimensional representation.

It is interesting to see how many of the geometrical and
physical properties of the sphere can be extracted from the algebraic
structure of these matrices, even for small values of $N$.  We list
here some of these properties.

\noindent {\bf i) Spherical locus:} The matrices
Eq.~(\ref{eq:matrix-sphere}) satisfy
\begin{equation}
X_1^2 + X_2^2 + X_3^2 = \frac{4r^2}{ N^2}  C_2 (N)\identity
= r^2 (1-1/N^2)\identity,
\end{equation}
where $C_2 (N) = (N^2 -1)/4$ is the quadratic Casimir of $SU(2)$ in
the $N$-dimensional representation.  This shows that the D0-branes are
in a noncommutative sense ``localized'' on a sphere of radius $r
+{\cal O} (1/N^2)$.

\noindent {\bf ii)  Rotational invariance:} The matrices
Eq.~(\ref{eq:matrix-sphere}) satisfy
\begin{equation}
R_{ij} X_j = U (R) \cdot X_i \cdot U (R^{-1}),
\end{equation}
where $R \in SO(3)$ and $U (R)$ is the $N$-dimensional representation
of $R$.  Thus, the spherical matrix configuration is rotationally
invariant up to a gauge transformation, even though the smooth
membrane sphere has been ``discretized'' to a finite number of degrees
of freedom.

\noindent {\bf iii)   Spectrum:} The matrix $X^3 = 2r{\bf J}_3/N$ (as well
as the other matrices) has a spectrum of eigenvalues which are
uniformly distributed in the interval $[-r, r]$.  This is precisely
the correct distribution if we imagine a perfectly symmetric sphere
with D0-branes distributed uniformly on its surface and project this
distribution onto a single axis.

\noindent {\bf iv)  Membrane dipole moment:} 
The spherical matrix membrane
has nonvanishing
membrane dipole moments
\begin{eqnarray}
z^{12 (3)} = z^{23 (1)} = z^{31 (2)}  & = & -2 \pi i {\rm Tr}\;
\left([X^1, X^2] X^3 \right) \nonumber\\
& = & \frac{4 \pi r^3}{3}  (1-1/N^2),
\end{eqnarray}
which agrees with the membrane dipole moment $4 \pi r^3/3$ of the
smooth spherical membrane up to terms of order $1/N^2$.

\noindent {\bf v)  Energy:}
In M-theory we expect the tension energy of a (momentarily) stationary
membrane sphere to be
\begin{equation}
e = \frac{4 \pi r^2}{(2 \pi)^2 l_{11}^3}  = \frac{r^2}{\pi l_{11}^3}\, .
\end{equation}
Using $p^I p_I = -e^2$ we see that the light-front energy should be
\begin{equation}
E = \frac{e^2}{ 2p^+} 
\label{eq:membrane-energy}
\end{equation}
in 11D Planck units.  The matrix membrane energy
is given by
\begin{equation}
E = -\frac{1}{4 R} [X^i, X^j]^2 = \frac{2r^4}{ N R}  +{\cal O} (N^{-3})
\end{equation}
in string units, which is easily seen to agree with
Eq.~(\ref{eq:membrane-energy}).

It is also straightforward to verify that the equations of motion for
the membrane are correctly reproduced in matrix theory.  Like the
smooth membrane, the matrix membrane oscillates periodically according
to an equation of motion of the form $\ddot{r} = -ar^3$ for a constant
$a$ (Collins and Tucker, 1976; Kabat and Taylor, 1997a).  Related
ellipsoidal oscillating membrane solutions were considered in Rey
(1997b),  Harmark and Savvidy (2000), and Savvidy (2000).

Thus, we see that many of the geometrical and physical properties of
the membrane can be extracted from algebraic information about the
structure of the appropriate membrane configuration.  The discussion
we have carried out here has only applied to the simple case of the
rotationally invariant spherically embedded membrane.  It is
straightforward to extend the discussion to a membrane of spherical
topology and arbitrary  shape, however, simply by using the
matrix-membrane correspondence (\ref{eq:mm-correspondence}) to
construct matrices approximating an arbitrary smooth spherical
membrane.  

The proceeding discussion can similarly be extended to the torus by
using the genus one matrix membrane regularization described in
Section (\ref{sec:matrix-regularization}).
As a concrete example let us consider embedding a torus into ${\br}^4
\subset{\br}^9$ so that the membrane fills the locus of points
satisfying
\begin{equation}
X_1^2 + X_2^2 = r^2, \;\;\;\;\; \;\;\;\;\;
X_3^2 + X_4^2 = s^2.
\label{eq:torus-locus}
\end{equation}
Such a membrane configuration can be realized through the following
matrices
\begin{eqnarray}
 X_1=  \frac{ r}{2} (U + U^{\dagger}),  & \;\;\;\;\; \;\;\;\;\;
 \;\;\;\;\; &
 X_2 =   \frac{-ir}{2} (U - U^{\dagger}),\label{eq:symmetric-torus}\\
 X_3 =  \frac{s}{2} (V + V^{\dagger}),& \;\;\;\;\; \;\;\;\;\;
 \;\;\;\;\;  &
 X_4 =  \frac{-is}{2} (V - V^{\dagger})\,.\nonumber
\end{eqnarray}
It is straightforward to check that this matrix configuration has
geometrical properties analogous to those of the matrix membrane
sphere discussed in the previous subsection.  In particular,
Eq.~(\ref{eq:torus-locus}) is satisfied identically as a matrix
equation.  Note, however that this configuration is not gauge
invariant under $U(1)$ rotations in the 12 and 34 planes---only under
a $\bz_N$ subgroup of each of these $U(1)$'s.

Since the matrix regularization procedure for higher genus Riemann
surfaces has not yet been described as explicitly as for the sphere
and torus, it is more difficult to construct explicit matrices
approximating smooth higher-genus surfaces.  Some
progress in this direction has been made by Bars (1991) and Hoppe
(1997a).  Despite the increased technical complications presented by
giving explicit matrix-regularized representations of general
higher-genus surfaces, in principle there is
no obstacle to constructing systems of matrices which describe an
arbitrary configuration of multiple membranes of any genera to an
arbitrary degree of accuracy.  As mentioned in
\ref{sec:two-body-general}, the linearized coupling of matrix membranes
to background supergravity fields is precisely in accord with the
matrix-regularized expressions for the coupling of smooth membranes,
so it is clear that compact membranes of arbitrary genus will interact
with one another and with gravitons in a way consistent with
eleven-dimensional supergravity coupled to membrane sources, at least
at the level of linearized supergravity.

\subsubsection{Infinite membranes}
\label{sec:infinite-membranes}

So far we have discussed compact membranes, which can be described in
terms of finite-size $N \times N$ matrices.  In the large $N$ limit it
is also possible to construct membranes with infinite spatial extent.
The matrices $X^i$ describing such configurations are
infinite-dimensional matrices which correspond to operators on a
Hilbert space.  Infinite membranes are of particular interest because
they can be BPS (supersymmetric) states which solve the classical
equations of motion of matrix theory.  Extended compact membranes
cannot be static solutions of the equations of motion since their
membrane tension always causes them to contract and/or oscillate, as in
the case of the spherical membrane.

The simplest infinite membrane is the flat planar membrane
corresponding in IIA theory to an infinite D2-brane (Banks, Fischler,
Shenker, and Susskind, 1997).  This solution
can be found by looking at the limit of the spherical membrane at
large radius.  It is simpler, however, to  directly construct
the solution by regularizing the flat membrane of M-theory.  As in the
compact case, we wish to quantize the Poisson bracket
algebra of functions on the brane.  Functions on the infinite membrane
can be described in terms of two coordinates $x_1, x_2$ with a
symplectic form $\omega_{ij} = \epsilon_{ij}$
giving a Poisson bracket
\begin{equation}
\{f (x_1, x_2),g (x_1, x_2)\} = \partial_1f \partial_2g-\partial_1g
\partial_2f.
\end{equation}
This algebra of functions can be ``quantized'' in the standard fashion
to the algebra of
operators generated by $Q, P$ satisfying
$
[Q, P] = i \epsilon^2 \identity/2 \pi,
$ where $\epsilon$ is a constant parameter.  
This gives a map from functions on ${\br}^2$ to operators, which allows
us to describe fluctuations around a flat membrane geometry with a
single unit of $P^+ = 1/R$ in each region of area $\epsilon^2$ on the
membrane.
(As usual in the
quantization process there are operator-ordering ambiguities which
must be resolved in determining a general map from functions expressed
as polynomials in $x_1, x_2$
to operators expressed as polynomials of $Q, P$.)

In addition to the flat membrane solution there are other infinite
membranes which are static solutions of M-theory in flat space.  In
particular, there are BPS solutions corresponding to membranes which
are holomorphically embedded in ${\bc}^4 ={\br}^8 \subset{\br}^9$.  These
are static solutions of the membrane equations of motion.  Finding a
matrix theory description of such membranes is possible but requires
choosing a regularization which
preserves the complex structure of the brane.  The details of this
construction for a general holomorphic membrane are discussed in
Cornalba and Taylor (1998).

\subsubsection{Wrapped membranes as matrix strings}
\label{sec:matrix-strings}

So far we have discussed M-theory membranes which are unwrapped in the
longitudinal direction and which therefore appear as D2-branes in the
IIA language of matrix theory.  It is also possible to describe
wrapped M-theory membranes which correspond to strings in the IIA
picture.  The charge in matrix theory which measures the number of
strings present is proportional to
\begin{equation}
 \frac{i}{ R}  {\rm Tr}\;
\left([X^i, X^j] \dot{X}^j +[[X^i, \theta^{\dot{\alpha}}],
\theta^{\dot{\alpha}}] \right)\, .
\label{eq:string-charge}
\end{equation}
Configurations with nonzero values of this charge were considered by
Imamura (1997).

To realize a classical configuration in matrix theory which contains
fundamental strings extended in some direction $X^i$ it is clear from
the form of the charge that we need to construct a configuration with
local membrane charge extended in a pair of directions $X^i, X^j$ and
then to give the D0-branes velocity in the $X^j$ direction.  For
example, we could consider an infinite planar membrane (as discussed
in the previous subsection) sliding along itself according to 
\begin{eqnarray}
X^1 =  Q+ t \identity, \;\;\;\;\; \;\;\;\;\;
X^2 =  P \,.
\end{eqnarray}
This corresponds to an M-theory membrane which has a projection onto
the $X^1, X^2$ plane and which wraps around the compact direction as a
periodic function of $X^1$ so that the IIA system contains a D2-brane
with infinite strings extended in the $X^2$ direction.
The dependence of the compact coordinate $X^-$ on $X^1$ in this configuration
can be seen easily in the corresponding smooth membrane
configuration, where $\partial_a X^-= \dot{X}^i \partial_a X^i$ as in
Eq.~(\ref{eq:dx}).

It is interesting to note that there is no classical matrix theory
solution corresponding to a string which is truly
one-dimensional and has no local membrane charge.  This follows from the
appearance of the commutator $[X^i, X^j]$ in the string charge, which
vanishes unless the matrices describe a configuration with at least
two dimensions of spatial extent.  We can come very close to a
one-dimensional classical string configuration by considering a
one-dimensional array of D0-branes at equal intervals on the $X^1$
axis with small off-diagonal matrix elements connecting adjacent
D0-branes.  In the classical theory, this configuration can have
arbitrary string charge.  If the off-diagonal modes are quantized,
then the string charge is quantized in the correct units.  This string
configuration is almost  one-dimensional but has a small additional
extent in the $X^2$ direction corresponding to the extra dimension of
the M-theory membrane.  From the M-theory point of view this extra
dimension must appear because the membrane cannot have momentum in a
direction parallel to its direction of extension (since it has no
internal degrees of freedom).  Thus, the momentum in the compact
direction represented by the D0-branes must appear on the membrane as
a fluctuation in some transverse direction.

\subsection{5-branes}
\label{sec:5-branes}

The M-theory 5-brane can appear in two possible
guises in type IIA string theory.  If the M5-brane is wrapped around
the compact direction it becomes a D4-brane in the IIA theory,
corresponding to a longitudinal M5-brane in matrix theory,
while if
it is unwrapped it appears as an NS 5-brane in IIA, corresponding to a
transverse M5-brane in matrix theory.  
{\it A priori}, one
might think that it should be possible to see both types of M5-branes in
matrix theory.    
Several calculations, however, indicate that the transverse M5-brane
does not carry a classical conserved charge which can be described in terms of
the matrix degrees of freedom.
As we have discussed, no transverse M5-brane charge appears either in
the matrix theory supersymmetry algebra discussed in
\ref{sec:extended} or the linearized supergravity interactions
described in \ref{sec:two-body-linear}.

One way of understanding this apparent puzzle is by comparing to the
situation for D-branes in light-front string theory (Banks, Seiberg
and Shenker, 1997).
Due to the Virasoro
constraints, strings in the light-front
formalism must have Neumann boundary conditions in both the
light-front directions $X^+,X^-$.  Thus, in light-front string theory
there are no transverse D-branes which can be used as boundary
conditions for the string.  A similar situation holds for
membranes in M-theory, which can end on M5-branes.  The
boundary conditions on the bosonic membrane fields which can be
derived from the action Eq.~(\ref{eq:membrane-action-h})
state that
\begin{equation}
(\bar{h} h^{ab} \partial_b X^i) \delta X^i = 0\, .
\end{equation}
Combined with the Virasoro-type constraint
\begin{equation}
\partial_a X^-= \dot{X}^i \partial_a X^i,
\end{equation}
we find that, just as in the string theory case, membranes must have
Neumann boundary conditions in the light-front directions.

These considerations would seem to lead to the conclusion that
transverse M5-branes simply cannot be constructed in matrix theory.
On the other hand, it was argued in Ganor, Ramgoolam, and Taylor (1997)
that a transverse M5-brane may be constructed using S-duality when the
theory has been compactified on a 3-torus.  To construct an infinite
extended transverse M5-brane in this fashion would require performing
S-duality on $(3+1)$-dimensional ${\cal N} = 4$ supersymmetric
Yang-Mills theory with gauge group $U(\infty)$, which is a poorly
understood procedure.  In Taylor and Van Raamsdonk (2000), however, a
finite size transverse M5-brane with geometry $T^3 \times S^2$ was
constructed using S-duality of the four-dimensional $U(N)$ with finite
$N$.  Furthermore, it was shown that this object couples correctly to
the supergravity fields even in the absence of an explicit transverse
M5-brane charge; similar results had been found earlier by Lifschytz
(1997).  Taken together, all these results for transverse M5-branes
seem to indicate that transverse M5-branes in matrix theory can be
constructed as quantum states in the theory, but that they are
essentially solitonic objects and do not carry a classically conserved
charge.

We now turn to the wrapped, or longitudinal, M5-brane which we
will refer to as the ``L5-brane''.  This object appears as a D4-brane
in the IIA theory.  An infinite flat D4-brane was considered as a matrix
theory background in Berkooz and Douglas (1997) by including extra fields
corresponding to strings stretching between the D0-branes of matrix
theory and the background D4-brane.  As in the case of the membrane,
however, we would like to find a way to explicitly describe a
dynamical L5-brane using the matrix degrees of freedom.
From the L5-brane charge $z^{ijkl} \sim\imm^{+ -ijkl}$ discussed in
Section \ref{sec:extended}, we know that the charge measuring the
L5-brane four-volume in the $ijkl$ plane is given by
\begin{equation}
2 \pi^2 {\rm Tr}\;\epsilon_{ijkl} X^i X^j X^k X^l \,.
\label{eq:L5-brane-volume}
\end{equation}
Another way to motivate this charge is that it is the T-dual of the
instanton number in a four-dimensional gauge theory, which measures
D0-brane charge on D4-branes.
Just as for the membrane charge, higher multipole moments of the
L5-brane charge are constructed by inserting factors of $X^m$ into
Eq.~(\ref{eq:matrix-m}) and performing a symmetrized trace. 

Unlike the matrix membrane, there is no general theory describing
an arbitrary L5-brane geometry in matrix theory language.  In fact,
the only L5-brane configurations which have been explicitly
constructed to date are those corresponding to the highly symmetric
geometries $S^4, {\bc P}^2$, and ${\br}^4$.  We now make a few brief
comments about these configurations.

The L5-brane with isotropic $S^4$ geometry is similar in many ways to
the membrane with $S^2$ geometry discussed in section
\ref{sec:compact-membranes}.  There are a number of unusual features
of the $S^4$  system, however, which deserve mention.  For full
details of the construction see Castelino, Lee, and Taylor (1998); a
related construction from the noncommutative geometry point of view
is given in Grosse, Klim\v{c}\'{i}k, and Pre\v{s}najder (1996).
A rotationally invariant spherical L5-brane can only be constructed
for those values of $N$ which are of the form
$
N = (n + 1) (n + 2) (n + 3)/6
$ where $n$ is integral.  For $N$ of this form we define the
configuration by setting $ X_i = r G_i/n, i \in \{1, \ldots, 5\},$
where $G_i$ are the generators of the $n$-fold symmetric tensor
product representation of the five four-dimensional Euclidean gamma
matrices $\Gamma_i$.  For any $n$ this configuration has the
geometrical properties expected of $n$ superimposed L5-branes
contained in the locus of points describing a 4-sphere.  As for the
spherical membrane discussed in \ref{sec:compact-membranes} the
configuration is confined to the appropriate spherical locus.  The
configuration is symmetric under $SO(5)$ and has the correct spectrum
and the L5-brane dipole moment of $n$ spherical branes.  The energy
and equations of motion of this system agree with those expected from
M-theory.  Unlike the $S^2$ membrane, there is no obvious way of
including small fluctuations of the membrane geometry around the
perfectly isotropic 4-sphere L5-brane in a systematic way.  In the
case of the membrane, we know that for any particular geometry the
fluctuations around that geometry can be encoded into matrices which
form an arbitrarily good approximation to a smooth fluctuation, through
the procedure of replacing functions described in terms of an
orthonormal basis by appropriate matrix analogues.  In the case of the
L5-brane no such procedure is known.   

The infinite flat L5-brane was  constructed in Ganor, Ramgoolam
and Taylor (1997) and Banks, Seiberg, and Shenker (1997).
Like the infinite membrane,
the infinite L5-brane with geometry of a flat ${\br}^4 \subset{\br}^9$ can
be viewed as a local limit of a large spherical geometry or it can be
constructed directly.  We need to find a set of operators $X^{1-4}$ on
some Hilbert space
satisfying
\begin{equation}
\epsilon_{ijkl} X^i X^j X^k X^l  = \frac{\epsilon^4}{2 \pi^2}  \identity.
\end{equation}
Such a configuration can be constructed using matrices which are
tensor products of the form $\identity \otimes Q, P$ and $Q, P \otimes
\identity$.  This gives a ``stack of D2-branes'' solution with
D2-brane charge as well as D4-brane charge.  It is also
possible to construct a configuration with no D2-brane charge by
identifying $X^a$ with the components of the covariant derivative
operator for an instanton on $S^4$,
$
X^i = i \partial^i + A_i
$. This construction is known as the Banks-Casher instanton (Banks and
Casher, 1980).  Just as for the spherical L5-brane, it is not known
how to construct small fluctuations of the L5-brane geometry around
any of these flat solutions.

The only other known configuration of an L5-brane in matrix theory
corresponds to a brane with geometry $ {\bc P}^2$.  This configuration
was constructed by Nair and Randjbar-Daemi (1998) as a particular
example of a coset space $G/H$ with $G = SU(3)$ and $H = U(2)$.  They
chose the matrices
$X_i =rt_i/\sqrt{N}$,
where $t_i$ are generators spanning ${\bf g}/{\bf h}$ in a particular
representation of $SU(3)$.  The geometry defined in this fashion seems
to be in some ways better behaved than the $S^4$ geometry.  For one
thing, configurations of a single brane can be constructed with
arbitrarily large $N$.  Furthermore, it seems to be possible to
include local fluctuations as symmetric functions of the matrices
$t_i$.  This configuration extends in only four spatial dimensions,
however, which makes the geometrical interpretation less clear.

Clearly, there are many aspects of the L5-brane in matrix theory which
are not understood.  The principal outstanding problem is to find a
systematic way of describing an arbitrary L5-brane geometry including
its fluctuations.  One approach to this might be to find a
way of regularizing the world-volume theory of an M5-brane in a
fashion similar to the matrix regularization of the supermembrane.
Just as for the construction of a covariant version of the matrix
membrane theory, a generalization of the Nambu bracket may be helpful
in finding such a matrix M5-brane theory.
It is also possible that understanding the structure of noncommutative
4-manifolds might help clarify this question.  This is one of many
places where noncommutative geometry  ties in closely with
matrix theory.  We will discuss other such connections with
noncommutative geometry in Section~\ref{sec:matrix-summary}.

In this and the previous subsection we have discussed the construction
of membrane and 5-brane configurations from the matrix degrees of
freedom.  Additional features which appear when multiple membranes of
different kinds are included in the configuration were discussed in
Ohta, Shimizu, and Zhou (1998) and de Roo, Panda,  and Van der Schaar (1998).

\section{Matrix theory in a general background}
\label{sec:matrix-general-background}

So far we have only discussed matrix theory as a description of
M-theory in infinite flat space.
In this section we consider the possibility of extending the theory
to compact and curved spaces.  
We discuss the
compactification of the theory on tori in subsection \ref{sec:tori}.
In subsection \ref{sec:curved-background}  we discuss
the problem of using matrix theory methods to describe M-theory in a
curved background space-time.

\subsection{Matrix theory on tori}
\label{sec:tori}

The compactification of matrix theory on a toroidallly compactified
space-time is most easily understood using an explicit representation
of T-duality in type IIA string theory.
In string theory, T-duality is a symmetry which relates the type IIA
theory compactified on a circle of radius $R$ with type IIB theory
compactified on a circle with dual radius $\hat{R} = \alpha'/R$.
In the perturbative type II string theory, T-duality exchanges winding
and momentum modes of the closed string around the compact direction.
For open strings, Dirichlet and Neumann boundary conditions are
exchanged by T-duality, so that Dirichlet $p$-branes are mapped under
T-duality to Dirichlet $(p \pm 1)$-branes (Dai, Leigh,
and Polchinski, 1989).
It was shown by Witten (1996) that
the low-energy theory
describing a system of $N$  parallel D$p$-branes in flat space is the
dimensional reduction of ${\cal N} = 1$, $(9 + 1)$-dimensional super
Yang-Mills theory to $p + 1$ dimensions.  In the case of $N$
D0-branes, this gives the Lagrangian Eq.~(\ref{eq:D0-Lagrangian}).  
In terms of these low-energy field theories describing D$p$-brane
dynamics, T-duality has the effect of exchanging transverse scalars
and gauge fields associated with the compact direction in the
$p$-brane and $(p + 1)$-brane world-volume theories through
\begin{equation}
X^j \rightarrow (2 \pi \alpha') (i \partial_j + A_j)\,.
\label{eq:T-duality}
\end{equation}
With this identification, the low-energy action describing $N$
D0-branes on a $d$-torus is precisely identifiable with the
dimensional reduction of 10D super Yang-Mills to a $(d +
1)$-dimensional theory on the dual torus (Taylor, 1997a).  This allows
us to identify the matrix model of M-theory compactified on a torus
$T^d$ as a $(d + 1)$-dimensional supersymmetric Yang-Mills theory.
The argument of Seiberg and Sen reviewed in section \ref{sec:proof} is
valid in this situation, so that $U(N)$ maximally supersymmetric
Yang-Mills theory on $(T^d)^*$ should describe DLCQ M-theory compactified
on $T^d$.  When $d \leq 3$, the quantum super Yang-Mills theory is
renormalizable so this is a sensible way to approach the theory.  As
the dimension of the torus increases, however, the matrix description
of the theory develops more and more complications.  In general, the
super Yang-Mills theory on the $d$-torus encodes the full U-duality
symmetry group of M-theory on $T^d$ in a rather nontrivial fashion.

We will discuss the compactification of the theory on a circle $S^1$
in Section \ref{sec:matrix-string}.
Compactification of the theory on a two-torus was discussed by Sethi
and Susskind (1997).  They pointed out that as the $T^2$ shrinks, a
new dimension appears whose quantized momentum modes correspond to
magnetic flux on the $T^2$.  In the limit where the area of the torus
goes to 0, an $O (8)$ symmetry appears.  This corresponds with the
fact that IIB string theory appears as a limit of M-theory on a small
2-torus (Schwarz, 1995; Aspinwall, 1996).

Compactification of the theory on a three-torus was discussed in
Susskind (1996) and Ganor, Ramgoolam, and Taylor (1997).  In this
case, M-theory on $T^3$ is equivalent to $(3+1)$-dimensional super
Yang-Mills theory on a torus.  This theory is conformal and finite.
M-theory on $T^3$ has a special type of T-duality symmetry under which
all three dimensions of the torus are inverted.  In the matrix
description this is encoded in the Montanen-Olive S-duality of the 4D
super Yang-Mills theory.

When compactified on $T^4$, the manifest symmetry group of the theory
is $SL(4,Z)$.  The expected U-duality group of M-theory compactified
on $T^4$ is $SL(5,Z)$, however.  It was pointed out by Rozali
(1997) that the U-duality group can be completed by
interpreting instantons on $T^4$ as momentum states in a fifth compact
dimension.  This means that Matrix theory on $T^4$ is most naturally
described in terms of a (5 + 1)-dimensional theory with a chiral $(2,
0)$ supersymmetry.  This $(2, 0)$ theory with 16
supersymmetries (see for example Seiberg, 1998) appears to play a
crucial role in numerous aspects of the physics of M-theory and
5-branes, and has been studied extensively in recent years.

Compactification on $T^5$ was discussed in Berkooz, Rozali, and Seiberg
(1977) and Seiberg (1997b).
Compactification on tori of higher dimensions continues to lead to
more complicated situations, particularly when one gets to $T^6$, when
the matrix theory description seems to be as complicated as the
original M-theory (Elitzur, Giveon, Kutasov, and Rabinovici, 1998;
Losev, Moore, and Shatashvili, 1998; Brunner and Karch, 1998; Hanany
and Lifschytz, 1998).  Despite the complexity of $T^6$
compactification, however, it was suggested by Kachru, Lawrence, and
Silverstein (1998) that compactification of Matrix theory on a more
general Calabi-Yau 3-fold might actually lead to a simpler theory than
that resulting from compactification on $T^6$.  If this speculation is
correct and a more explicit description of the theory on a Calabi-Yau
compactification could be found, it might make matrix theory a
possible approach for studying realistic 4D phenomenology.

A significant amount of literature has been produced on the subject of
compactification of matrix theory on tori and orbifolds, of which we
have only mentioned a few aspects.  One particularly interesting
orbifold compactification of M-theory is the Ho\v{r}ava-Witten (1996)
compactification leading to heterotic string theory.  The construction
of a matrix heterotic string theory was considered in Danielsson and
Ferretti (1997), Kachru and Silverstein (1997), Motl (1996), Lowe
(1997a, 1997b), Banks and Motl (1997), Rey (1997a), Ho\v{r}ava (1997),
Motl and Susskind (1997),
and Krogh (1999a, 1999b).  The reader interested in more details regarding
toroidal or orbifold compactifications of matrix theory is referred to
Fischler, Halyo, Rajaraman, and Susskind (1997), Banks (1999), and
Obers and Pioline (1999) for reviews and further references.

\subsection{Matrix theory in curved backgrounds}
\label{sec:curved-background}

In the previous section we discussed matrix theory compactifications
on tori, which have nontrivial topology but are locally flat.
We now briefly discuss the problem of formulating matrix theory in a
space which has the topology of ${\br}^9$  but which may be
curved or have other nontrivial background fields.  We would like to
generalize the matrix theory action to one which includes a general
supergravity background given by a metric tensor, 3-form field, and
gravitino field which together satisfy the equations of motion of 11D
supergravity.  This issue has been discussed 
by many authors, although limited progress has been made in this
direction so far.

In Seiberg (1997b) it was argued that light-front M-theory on an
arbitrary compact or non-compact manifold should be reproduced by the
low-energy D0-brane action on the same compact manifold, although no
explicit description of this low-energy theory was given.  Douglas
(1998a, 1998b) proposed that any formulation of matrix theory in a
curved background should satisfy a number of axioms.  The most
restrictive of these axioms is a condition stating that for a pair
of D0-branes at points $x^i$ and $y^i$, corresponding to diagonal $2
\times 2$ matrices, the masses of the off-diagonal fields should be
given by the geodesic distance between the points $x^i$ and $y^i$ in
the given background metric.  It was shown by Douglas, Kato, and Ooguri
(1998) that the first few terms in
a weak field expansion of the multiple D0-brane action on
a Ricci-flat K\"ahler manifold can be constructed in a fashion which
is consistent with the geodesic length condition as well as Douglas'
other axioms. These authors also found, however, that these conditions
do not uniquely determine most of the terms in the action, so that a
more general principle is still needed to construct the action to all
orders.

In Taylor and Van Raamsdonk (1999a), the matrix theory representation
of the supercurrent components reviewed in \ref{sec:two-body-general}
was used to construct the terms in the matrix theory action describing
linear couplings to a general supergravity background (see also
Lifschytz, 1998b).  A related
construction from the membrane point of view was carried out in
Dasgupta, Nicolai, and Plefka (2000).  One interesting feature of this
construction is that the combinatorics of the symmetrized trace
prescription is necessary for Douglas' geodesic length condition to be
satisfied.  This proposal can in principle be generalized to described
$m$th order couplings to the the supergravity background fields, where
matrix expressions are needed for quantities which can be determined
from an $m$-loop matrix theory calculation.  Whether these terms can
be calculated and sensibly organized into higher-order couplings of
matrix theory to background fields depends on whether higher-loop
matrix theory results are protected by supersymmetric
nonrenormalization theorems.

In Douglas, Ooguri, and Shenker (1997) and Douglas and Ooguri (1998)
two-graviton scattering for matrix theory on a large K3 surface was
considered.  These authors concluded that no finite  $N$ matrix theory
action could reproduce gravitational physics in such a curved
background.  The difficulty in this situation first arises
from terms quadratic in the background curvature tensor.  This is
compatible with the observations mentioned in Section
\ref{sec:interaction-summary} that supersymmetric nonrenormalization
theorems first fail  for four graviton interactions.  These combined
pieces of evidence make it quite plausible that classical supergravity
on curved (or flat) spaces will not be describable by any finite $N$
matrix theory, but that the large $N$ limit must be understood for
further progress to be made.

\section{Related models}
\label{sec:related-models}

The BFSS conjecture stating that matrix quantum mechanics is a
complete description of flat space M-theory in light front coordinates
was the first of a series of related conjectures that M-theory and
other string theories can be described in certain regimes or with
certain backgrounds by quantum mechanical or quantum field theoretical
models.  In this section we briefly review several of these other
conjectures and discuss their relationship to the matrix model of
M-theory theory which we have focused on in the rest of this review.

\subsection{The IKKT matrix model of IIB string theory}
\label{sec:IKKT}

Shortly after the original BFSS paper, it was proposed by Ishibashi,
Kawai, Kitazawa, and Tsuchiya (1996) that a (0+0)-dimensional matrix
model should give a Poincar\'e invariant description of type IIB string
theory in a flat space background.  The argument given for this
conjecture follows a similar line of reasoning to the derivation of
matrix theory as a regularized light-front membrane theory reviewed in
Section \ref{sec:quantized-membrane}.  Ishibashi {\it et al.}
started with the Green-Schwarz form of the IIB string action, written
following Schild (1977) as
\begin{equation}
S = \int d^2 \sigma \left[ \sqrt{g} \alpha \left( \frac{1}{4} 
\{X^\mu, X^\nu\}^2 -\frac{i}{2}  \bar{\psi} \Gamma_\mu\{ X^\mu, \psi\}
\right) + \beta \sqrt{g} \right],
\label{eq:Schild}
\end{equation}
where $\{\cdot, \cdot\}$ is a canonical Poisson bracket on the string
world-volume and $\alpha, \beta$ are constants.  Performing the matrix
regularization of this theory {\it a la} Goldstone and Hoppe leads to
the 0-dimensional matrix model arising from the dimensional reduction
in all ten dimensions of 10D ${\cal N} = 1$ super Yang-Mills
\begin{equation}
S = \alpha \left( -\frac{1}{4}  {\rm Tr}\;[A_\mu, A_\nu]^2
-\frac{1}{2}{\rm Tr}\; \left( \bar{\psi} \Gamma^\mu
[A_\mu, \psi] \right) \right) + \beta {\rm Tr}\; {\bf 1}.
\label{eq:IKKT}
\end{equation}
This action is then integrated over all $N \times N$ matrices $X^\mu,
\psi$, giving a finite-dimensional integral for finite $N$.
The integral of Eq.~(\ref{eq:Schild}) over all metrics $g$ was interpreted
in this model as leading to a sum over all values of $N$ in the
partition function of the theory.  This (0+0)-dimensional matrix
model of type IIB string theory is often referred to as the ``IKKT''
model.  Other related matrix formulations of type IIB string theory
have been discussed in Periwal (1997), Yoneya (1997), Fayyazuddin,
Makeenko, Oleson, Smith, and Zarembo (1997), Kitsunezaki and Nishimura
(1998), Hirano and Kato (1997), Tada and Tsuchiya (1999).  A related
matrix formulation of type I string theory was investigated by Tokura
and Itoyama (1998); see Itoyama and Tsuchiya (1999) for a review.

Since the initial formulation of this model by Ishibashi, Kawai,
Kitazawa, and Tsuchiya, many further extensions of this model have
been carried out.  For a review of some of this work, see Aoki, Iso,
Kawai, Kitazawa, Tsuchiya, and Tada (1999).  Because the partition
function for this model is simply a finite-dimensional integral for
finite $N$, this is in principle the simplest of the matrix models in
which to carry out explicit calculations.  Since this model
furthermore has the virtue of manifest Poincar\'e invariance, it is
potentially a more powerful framework than the matrix model of
M-theory, which as we have discussed here is restricted to a
light-front description of the full eleven-dimensional theory.  In
some sense this matrix model can be thought of in terms of the
low-energy theory of $N$ D-instantons, although there does not seem to
be an argument analogous to the Seiberg/Sen limiting argument which
justifies the dropping of higher-order terms in the Born-Infeld theory
for this model.  There is a separate argument for the validity of this
model, which comes from relating the Schwinger-Dyson loop equations
for Wilson loops in the IKKT model to the type IIB string field theory
in light-cone gauge (Fukuma, Kawai, Kitazawa, and Tsuchiya, 1998).
The role of the light-cone and its relationship with space-time
causality in the IIB matrix model, however, is not yet clearly
understood.  One very intriguing suggestion which has been made for
the IKKT model is that the dimension (four) of observable space-time
arises as the natural fractal dimension of a branched polymer which
describes the dynamics of the model (Aoki, Iso, Kawai, Kitazawa, and
Tada, 1999).  While evidence for this speculation is not yet
conclusive (see for instance, Ambjorn, Anagnostopoulos, Bietenholz,
Hotta, and Nishimura, 2000), it is clearly important to develop a
deeper understanding of this model.

\subsection{The matrix model of light-front IIA string theory}
\label{sec:matrix-string}

Another matrix formulation of string theory arises from acting with
T-duality on the matrix model of M-theory we have been discussing.
The resulting matrix string theory asserts that a light-cone
description of type IIA string theory in flat space is given by (1+
1)-dimensional maximally supersymmetric Yang-Mills theory.  This
matrix string theory was first described by Motl (1997), and was
further refined by Banks and Seiberg (1997) and Dijkgraaf, Verlinde,
and Verlinde (1997, 1998).  The model can be derived from the matrix
model of M-theory in the following fashion: Consider Matrix theory
compactified on a circle $S^1$ in dimension 9.  As discussed in
\ref{sec:tori}, under T-duality on the circle this theory can be
described by super Yang-Mills theory in $(1+1)$-D on the dual circle
$\hat{S}^1$.  In the BFSS formulation of Matrix theory, this
corresponds to M-theory compactified on a 2-torus.  If we now think of
dimension 9 rather than dimension 11 as the dimension which has been
compactified to get a IIA theory, we see immediately that this super
Yang-Mills theory should provide a light-front description of type IIA
string theory.  Because we are now interpreting dimension 9 as the
dimension of M-theory which is compactified to give type IIA string
theory, the fundamental objects which carry momentum $p^{+}$ are no
longer D0-branes, but rather strings with longitudinal momentum.
Thus, it is natural to interpret $N/R$ in this super Yang-Mills theory
as the longitudinal string momentum.

To to be slightly more explicit about this matrix string theory
conjecture, consider the Matrix theory Hamiltonian (working in Planck
units and dropping factors of order unity)
\begin{equation}
H = R_{11} {\rm Tr}\; \left[
  P_a P_a -  [X^a, X^b]^2  +
\theta^T
\gamma_a[X^a, \theta] \right]\ .
\end{equation}
After compactification on $R_9$ we identify $X^9 \rightarrow R^9
D_\sigma$, $P_9 \rightarrow R_9 \dot{A}_9\sim E_9/R_9$, where
$\sigma \in[0, 2 \pi]$ is the coordinate on the dual circle.  With
these identifications, and using $g \sim R_9^{3/2}$, the Hamiltonian
can be rewritten in the form
\begin{eqnarray}
H  & = &  \frac{R_{11}}{2 \pi}  \int d \sigma \; {\rm Tr}\; \left[
 P_a P_a + (D_\sigma X^a)^2 + 
\theta^TD_\sigma \theta
\right.\nonumber\\
& &\hspace{1in} \left.
+
\frac{1}{g^2}  \left( E^2  - [X^a, X^b]^2 \right)
+ \frac{1}{g} \theta^T
\gamma_a[X^a, \theta]\right]\ .
\end{eqnarray}
This is essentially the form of the Green-Schwarz light-front string
Hamiltonian, with the modification that the fields are now $N \times
N$ matrices which do not necessarily commute.  This means that  the
theory automatically contains multi-string objects living in a second
quantized Hilbert space.  Furthermore, it is possible to construct
extended string theory objects in terms of the noncommuting matrix
variables, by a simple translation from the original Matrix theory
language.  For example, the type IIA D0-brane charge in this model is
given by the electric flux $F_{09}$ along the compact direction in the
(1+1)-dimensional super Yang-Mills theory.  A complete list of the
charges and their couplings to background supergravity fields is given
in Schiappa (2000).

A particularly nice feature of the matrix IIA string theory is the way
in which the individual string bits carrying a single unit of
longitudinal momentum combine to form long strings, as shown in Motl
(1997).  As the string coupling becomes small $g \rightarrow 0$, the
coefficient of the term $[X^a, X^b]^2$ in the Hamiltonian becomes very
large.  This forces the matrices to become simultaneously
diagonalizable.  Because the string configuration is defined over
$S^1$, however, the matrix configuration need not be periodic in
$\sigma$.  The matrices $X^a (0)$ and $X^a (2 \pi)$ can be related by
an arbitrary permutation.  The lengths of the cycles of this
permutation determine the numbers of string bits which combine into
long strings whose longitudinal momentum $N/R_{11}$ can become large
in the large $N$ limit.  As the coupling becomes very small, the
theory therefore essentially becomes a sigma model on
$({\br}^8)^N/S^N$.  The twisted sectors of this theory correspond to
the sectors where the string bits are combined in different
permutations.  In this picture, string interactions appear as vertex
operators in the conformal field theory arising as the infrared limit
of the sigma model, as discussed in Dijkgraaf, Verlinde, and Verlinde
(1997).  Further details regarding string interactions in matrix
string theory can be found in Wynter (1997, 1998, 2000), Giddings,
Hacquebord, and Verlinde (1999), Bonelli, Bonora, and Nesti (1998,
1999), Bonelli, Bonora, Nesti, and Tomasiello (1999), Grignani and
Semenoff (1999), Hacquebord (1999), Brax (2000), and Grignani, Orland,
Paniak, and Semenoff (2000).  Other aspects of matrix string theory
were discussed in Verlinde (1997), Bonora and Chu (1997), Brax and
Wynter (1999), Bill\'o, Caselle, D'Adda, and Provero (1999), Kostov
and Vanhove (1998), Sugino (1999), and Balieu and Laroche (1999).

\subsection{The AdS/CFT correspondence}
\label{sec:CFT}

From the point of view taken in \ref{sec:BFSS} the essential
connection between matrix quantum mechanics and M-theory arises
because the same limit which gives the nonrelativistic Yang-Mills
theory for D0-branes can be interpreted as corresponding to a limit of
lightlike compactification of M-theory.  Following the BFSS matrix
theory conjecture, it was found that there are numerous other
situations in which an appropriate field theory limit of a system of
multiple branes can be related to M-theory and string theory in
certain limiting backgrounds.  The simplest and best studied example
of this for higher-dimensional branes is the case of many D3-branes.
The first clue that a similar correspondence might exist for D3-branes
was the demonstration by Klebanov (1997) that the leading term in a
semiclassical calculation of the absorption cross-section of a dilaton
s-wave by a system of many 3-branes is precisely reproduced by the
(3+1)-dimensional super Yang-Mills theory describing the low-energy
dynamics of the system.  This and other evidence led Maldacena (1998c)
to conjecture that the large $N$ limit of $U(N)$ maximally
supersymmetric Yang-Mills theory in (3+1) dimensions should precisely
reproduce the physics of type IIB string theory in the near-horizon
limit of the D3-brane supergravity solution.  This near-horizon
geometry is a manifold of the form ${\rm AdS}_5 \times S^5$.
Maldacena motivated his conjecture by observing that by taking the
limit $\alpha' \rightarrow 0$ and taking the distance scale $r$ on the
supergravity side to zero such that $r/\alpha'$ remains constant, the
physics on the D3-brane side is the Yang-Mills limit of the
Born-Infeld theory, while the physics on the supergravity side is
precisely that of IIB string theory in the near-horizon ${\rm AdS}_5
\times S^5$ geometry.  An enormous amount of work has been done to
extend and verify this conjecture in many different situations,
including those with reduced supersymmetry.  Further development of
this subject is beyond the scope of this review, and we refer the
reader to the comprehensive review by Aharony, Gubser, Maldacena,
Ooguri, and Oz (2000) for further details.  We will restrict ourselves
here to a few brief comments about the connection between this AdS/CFT
conjecture for D3-branes and the matrix description of M-theory.  Just
as the matrix string theory described in the previous subsection can
be related to the matrix of M-theory through T-duality on a circle
$S^1$, it is tempting to imagine that there is a connection between
matrix theory and the D3-brane AdS/CFT conjecture which may be made
precise by considering T-duality on a three-torus.  This duality
replaces matrix quantum mechanics with the same 4D Yang-Mills theory
which appears in the AdS/CFT correspondence.  One difficulty in making
such a connection precise is that the connection between the theories
is described very differently in the two cases.  In the matrix model
case, we expect to be able to explicitly describe the quantum gravity
S-matrix in terms of scattering of localized D0-brane wavefunctions.
In the AdS/CFT picture, on the other hand, correlation functions in
the Yang-Mills theory correspond to interactions between supergravity
fields in the bulk of the AdS space with sources on the boundary
(Gubser, Klebanov, and Polyakov, 1998; Witten, 1998).  While the very
different nature of these two correspondences makes it difficult to
relate them in a precise fashion, connections between the matrix
theory and AdS/CFT conjectures were discussed in Balasubramanian,
Gopakumar, and Larsen (1998), Hyun (1998), Itzhaki, Maldacena,
Sonnenschein, and Yankielowicz (1998), Jevicki and Yoneya (1999), Hyun
and Kiem (1999), de Alwis (1999), Silva (1998), Martinec and Sahakian
(1999), Chepelev (1999), Sekino and Yoneya (1999), and Yoneya (2000).
The connections between these points of view, and the regions of
overlap between the various limits associated with matrix theory and
the AdS/CFT conjecture for D0-branes are discussed in Polchinski
(1999).

\section{Conclusions}
\label{sec:matrix-summary}

In this review we have focused on some basic aspects of
matrix theory.  We have described two complementary ways of thinking
about matrix theory: first as a quantized regularized theory of a
supermembrane, which can be interpreted as a second-quantized theory
of objects moving in an eleven-dimensional target space, and second as
the DLCQ of M-theory, which is equivalent to a simple limit of type IIA
string theory through the Seiberg-Sen limiting argument.  We have
reviewed perturbative matrix theory calculations which correspond
precisely with linearized eleven-dimensional supergravity at the
one-loop level, and with nonlinear interactions between three gravitons at
the two-loop level, but which seem to disagree with higher-order
nonlinearities in gravity at the three-loop level.  As we have
discussed, showing that matrix theory agrees with classical
supergravity to all orders probably requires new insight into the
nature of the large $N$ limit and the structure of quantum states in
the theory.  We have shown that using matrix degrees of freedom, it is
possible to describe pointlike objects which have many of the physical
properties of supergravitons, as well as extended objects which behave
like the supermembrane and 5-brane of M-theory.  For supergravitons
and membranes this story is fairly complete, at least classically;
for M5-branes, however, only a few very special geometries have been
described in matrix language, and a systematic description of
M5-branes, even at the classical level, is still lacking.
We have reviewed progress on generalizing matrix theory to backgrounds
other than flat eleven-dimensional Minkowski space.  Finding a
description of the theory when the background is curved seems to
involve resolving many of the same issues which arise in comparing
with nonlinear classical supergravity.  Finally, we discussed 
related models which describe other M-theory or string theory
backgrounds in terms of higher-dimensional field theories.

There are many aspects of matrix theory which we have covered only
briefly, or not at all, in this review.  These include matrix theory
black holes\footnote{See, {\it e.g.}, Banks, Fischler, Klebanov, and
Susskind (1998), Kabat and Lifschytz (2000) and references therein.},
orbifold compactifications of matrix theory, matrix models of the
six-dimensional $(0, 2)$ theory and little string
theory\footnote{These matrix models were first developed in Aharony,
Berkooz, Kachru, Seiberg, Silverstein (1998), Aharony, Berkooz, and
Seiberg (1998), Witten (1997), Ganor and Sethi (1998); for a review of
this work and further developments in this direction see Banks
(1999).}, and the matrix models of string theory briefly mentioned in
the previous section.

Matrix theory has given us a remarkable new perspective on M-theory
and string theory, by giving us a well-defined, in principle
calculable, model for a quantum theory of supergravity.  While this
model has given us many new insights, at this point it seems clear
that for further progress in directly using this model to better
understand the physics of M-theory, some new ideas about how to
understand the quantum theory and the large $N$ limit are probably
needed.  Resolving the outstanding issues surrounding both the
connection of the model with classical nonlinear supergravity and the
formation of the model in a general space-time background is clearly
necessary if we ever wish to use this model to make new statements
about corrections to classical supergravity in phenomenologically
interesting models such as M-theory on compact 7-manifolds or
orbifolds.

While direct progress on matrix theory seems at this point to be
slowing, the development of this model over the last several years
has led to a number of ideas which have fueled interesting
developments in many other related areas.  We conclude this review
with a brief mention of some of these areas.

One important area of research on which matrix theory has had significant
impact is the ongoing study of supersymmetric nonrenormalization
theorems in quantum field theories.  Motivated in part by the BFSS conjecture,
Dine and Seiberg (1997) proved a nonrenormalization theorem for the
$F^4$ terms in the effective action of (3+1)-dimensional super
Yang-Mills theory.  As we discussed in
Section~\ref{sec:matrix-interactions}, the 
matrix theory conjecture motivated a great deal of effort towards
proving such nonrenormalization theorems for the matrix quantum
mechanics theory.  This work has already improved our understanding of
the role of supersymmetry in field theories of various dimensions.
Finding some general principles which explain why certain terms in the
effective action are renormalized and others are not would be a
great step forward in the study of supersymmetric field theories.

Another direction which recent work has taken which was motivated, at
least in part, by results from matrix theory is the study of
noncommutative field theory and noncommutative geometry in string
theory.  A review of noncommutative geometry in the context of matrix
theory is given in Konechny and Schwarz
(2000).  It was pointed out by Connes, Douglas, and Schwarz (1998) and
Douglas and Hull (1998) that the T-duality construction of Taylor
(1997a) relating D0-branes on a torus to D$p$-branes on the dual torus
can be generalized by considering boundary conditions giving a
noncommutative gauge theory on the dual torus.  They showed that this
construction was equivalent to thinking of the D0-branes in a constant
$B$-field background.  The connection between string theory in a
constant $B$ field and noncommutative geometry was studied further by
Seiberg and Witten (1998), leading to a flurry of activity in this
area.  Throughout this recent work, one theme is the idea that for a
D$p$-brane in a constant $B$-field, a gauge transformation removes the
$B$ field in the bulk and produces a magnetic or electric flux $F$ on
the D$p$-brane world-volume.  For $p = 2$, the resulting system is
simply a D2-brane bound to multiple D0-branes, which is described
equivalently through the matrix theory language and the language of
fuzzy geometry on the D2-brane, using the Moyal (1949) product.  The
connection between these points of view is discussed in Cornalba and
Schiappa (1999), Alekseev, Recknagel, and Schomerus (1999), Floratos
and Leontaris (1999), Castro (1999), Cornalba (1999), and many other
papers.

Another aspect of matrix theory which has wide-ranging applications is
the explicit construction reviewed in
Section~\ref{sec:matrix-interactions} of the multipole moments of the
matrix theory stress tensor, membrane current and M5-brane current.
This higher moment structure, which describes higher-dimensional
extended objects in terms of the degrees of freedom of
lower-dimensional objects, is very general, and has a precise analogue
in type II string theory, where it is possible to describe the
supercurrents and charges of both higher- and lower-dimensional
Dirichlet- and NS-branes in terms of the degrees of freedom living in
the world-volume theory of a system of D$p$-branes (Taylor and Van
Raamsdonk, 1999a, 1999b; Myers, 1999).  This structure has many
possible applications to D-brane physics.  It was pointed out by Myers
(1999) that putting a system of D$p$-branes in a constant background
$(p +4)$-form flux will produce a dielectric effect in which spherical
bubbles of D$(p + 2)$-branes will be formed with dipole moments which
screen the background field.  This dielectric effect has been used in
the work of Polchinski and Strassler (2000) on string duals of super
Yang-Mills theories with reduced supersymmetry, and in the work of
McGreevy, Susskind, and Toumbas (2000) on giant gravitons in AdS
space.  The fact that extended objects can be constructed from the
matrices describing pointlike D0-branes seems to be one of the
fundamental lessons of matrix theory.  The fundamental problem of
M-theory at this point is finding a background-independent formulation
in terms of fundamental degrees of freedom from which all extended
objects in the theory can be built.  It seems likely that the insights
learned from matrix theory may be useful in finding such a set of
fundamental degrees of freedom and understanding how they can be used
to build the strings and branes in the theory and describe their
interactions.

\section*{Acknowledgments}

This work was supported in part by the A.\ P.\ Sloan Foundation, in
part by the U.S.\ Department of Energy (DOE) through contract
\#DE-FC02-94ER40818, and in part through the National Science
Foundation (NSF) under grant No.\ PHY94-07194.  The author would like
to thank the Institute for Theoretical Physics in Santa Barbara and
the University of Tokyo, Komaba, for hospitality during the completion
of this work.  Thanks to Joe Polchinski, Sav Sethi, Steve Shenker,
Lenny Susskind, Mark Van Raamsdonk, and Tamiaki Yoneya for reading a
preliminary version of this review and making numerous helpful
suggestions.

Some of the material in this review article appeared previously in
transcripts of pedagogical lectures which were presented at MIT, at
the Korean Institute for Advanced Study, at the Komaba '99 workshop in
Tokyo, 1999, and at the NATO Advanced Study Institute, Akureyri,
Iceland, August 1999.

\bibliographystyle{plain}


\end{document}